# Mechanizing Gödel's Incompleteness Theorems and Provability Logic

Shogo Saitou[1] and Mashu Noguchi[2]

[1] Tohoku University, Sendai, Japan
palalansouki@gmail.com
[2] Kobe University, Kobe, Japan
me@sno2wman.net
ORCID: 0009-0000-8653-3403

**Abstract.** We mechanized proofs of Gödel's first and second incompleteness theorems, Solovay's arithmetical completeness theorem for GL, and related results in the Lean 4 theorem prover.

**Remark** Since the word *formalize* is used in two different senses, which may cause confusion, we strictly distinguish between the words *formalize* and *mechanize*.

***formalize*** the formalization of mathematics as a technique in the context of mathematical logic and metamathematics.

***mechanize*** the formalization of mathematics in an interactive theorem prover, verifiable on an actual computer.

Following this convention, what we have done can be stated succinctly: *mechanizing formalized mathematics*.

## 1 Introduction

*Gödel's incompleteness theorems* are among the most significant results in mathematical logic. In his seminal paper [46], he proved what is now known as the first incompleteness theorem (G1); moreover, he outlined the second incompleteness theorem (G2) in the last short section. G2 was later proved rigorously by Hilbert and Bernays [65]. We state the theorems in modern terms: G1, with Rosser's improvement [118], states that for any consistent axiomatic system with sufficient expressive power to execute arithmetic, there exists a proposition that can neither be proved nor disproved within the system. G2 states that, for any consistent reasonable axiomatic system as in G1, the proposition formally representing the system's own consistency cannot be proved within the system itself.

Gödel also made another important observation: that provability can be regarded as a modality. In his early work [47], he observed that the provability (*Beweisbar*, $\mathfrak{B}$) of intuitionistic logic can be treated similarly to the modal operator $\Box$ in the modal logic now called S4. However, it follows from G2 that abstracting the behavior of the provability predicate, the most central notion of the incompleteness theorems, does not yield S4. Solovay [142] showed that the modal logic called GL precisely captures the behavior of the standard provability predicate. This fact, known as *Solovay's arithmetical completeness theorem*, was a most fundamental result in the subfield of modal logic called *provability logic*.

On the other hand, recently, there has been much active work on mechanizing mathematics using interactive theorem provers, which guarantees the validity of existing and new results and enables AI/LLM-assisted or automated proving. There are many well-known interactive theorem provers such as Rocq [117], Isabelle [67], HOL Light [61, 62], Agda [1], and Lean [103], and mathematics has been mechanized in each of them, including in the field of mathematical logic[1]. In particular, for mechanizing Gödel's incompleteness theorems, this line of work began with Shankar in 1986 [130, 131], and continues with O'Connor [106, 107], Harrison [60], Paulson [110], and

[1] Some of these mechanizations are summarized in [39].

Popescu and Traytel [114, 115], Kirst et al. [78, 80]. As for provability logic, modal-logical properties of `GL`, such as its semantical completeness and automated solvers, have been mechanized by Harrison [61:Chapter 20][2], Goré and Kelly [50], Goré, Ramanayake and Shillito [53], Maggesi and Perini Brogi [95, 96], Gignoux [45]. However, the existing mechanizations of the incompleteness theorems are either abstract or not carried out fully within arithmetic. For instance, O'Connor's implementation assumes several facts needed for the proof of G2 as axioms, and Paulson's mechanization of G2 uses hereditarily finite sets, not arithmetic [109]. To the best of our knowledge, no full mechanization of the incompleteness theorems entirely within arithmetic has been reported, and neither has any mechanization of the arithmetical side of provability logic, such as Solovay's arithmetical completeness theorem.

We report our project: *Formalized Formal Logic* (abbreviated as *FFL* below), for mechanizing mathematical logic. The main contributions of this paper are completely `sorry`-free mechanizations of Gödel's first (Theorem 2.3) and second (Theorem 2.18) incompleteness theorems and of Solovay's arithmetical completeness theorem (Theorem 3.24). We have also mechanized several results related to incompleteness and provability logic; for these, we refer the reader to the respective sections.

Our work is carried out in Lean 4, an interactive theorem prover, together with mathlib4 [98], its community-developed mathematics library. Lean 4 is based on the Calculus of Inductive Constructions (CIC) [103], and features dependent types, quotient types, and support for noncomputable definitions, making it highly expressive. In addition, its powerful metaprogramming infrastructure like `aesop` [88] and `grind` [102] enables efficient proof automation and extensibility.

### 1.1 Paper structure

This report is organized as follows.

- Section 2 describes the mechanization of the incompleteness theorems and their corollaries, together with the technical details.
- Section 3 describes provability logic, in particular Solovay's arithmetical completeness theorem and the classification theorem.
- Section 4 describes the future development and direction of FFL, with reference to related work.
- Section 5 is an appendix that briefly describes the use of AI in our mechanization.

Since it is not our purpose to state all of the mathematical definitions and facts here, we state them somewhat informally and generally assume that their proofs are known to the reader; see the references given at the beginning of each section. Moreover, we also assume that the reader is familiar with the notation, syntax, and functionalities of Lean 4, both as a programming language and as an interactive theorem prover. Readers unfamiliar with Lean may consult a standard textbook such as [13].

### 1.2 Repositories

Our mechanization is currently hosted in several repositories on GitHub. We note that our mechanization is still under development at the time of writing this paper, so the statements and definitions described in this paper may have been revised in the latest versions of these repositories. This report is based on the following fixed versions.

- Section 2, the mechanization of the incompleteness theorems: https://github.com/FormalizedFormalLogic/Foundation/tree/v1.
- Section 3, the mechanization of provability logic: https://github.com/FormalizedFormalLogic/ProvabilityLogic/tree/v1.

Each excerpted code snippet is annotated with the URL of its source as a reference. These repositories are licensed under the Apache License 2.0.

### 1.3 Declaration of AI usage

Our main mechanizations of the three results, the first and second incompleteness theorems and Solovay's arithmetical completeness theorem were done between 2023 and 2025, and up to that point they contained no AI-generated code. This can be verified from the following commits, at which each result first became `sorry`-free.[3]

- Gödel's first incompleteness theorem: e9325d82 (2024/09/04).
- Gödel's second incompleteness theorem: 2da7151e (2024/09/04).
- Solovay's arithmetical completeness theorem: 4a34d75c (2025/04/06).

[2]We do not know when Harrison's mechanization of modal logic was carried out.

[3]The first two commits were made in FormalizedFormalLogic/Arithmetization, later merged into Foundation as a subtree.

Some proofs in modal logic and provability logic make use of AI-assisted mechanizations. This is discussed in detail in Appendix: Section 5.

### 1.4 Acknowledgement

First, we thank Taishi Kurahashi, Yuta Sato, C7X and Malvin Gattinger for reading an early draft of this report and providing us with valuable comments and reviews. Second, during the development of FFL, we thank Hunter Monroe (@hmonroe), C7X (@indiscernibles) and Trevor Morris (@gotrevor), who were mainly engaged in active discussion and experimentation. We also received financial support; this work was partially supported by JST CREST JPMJCR25I5 and JST BOOST JPMJBY24E2. In addition, we received financial support from individuals and companies[4]. We gratefully acknowledge all of this support here.

## 2 Mechanization of the incompleteness theorems

We mechanized the following two results. Here, arithmetic sentence/theory means a sentence/theory in the language $\mathcal{L}_{\mathrm{OR}} = \{0, 1, +, \cdot, <, =\}$.

**Theorem 2.3 (Gödel's first incompleteness theorem)** Let $T$ be a $\Delta_1$-definable, $\Sigma_1$-sound arithmetic theory stronger than $\mathsf{R}_0$. Then $T$ is incomplete.

**Theorem 2.18 (Gödel's second incompleteness theorem)** Let $T$ be a $\Delta_1$-definable, consistent arithmetic theory stronger than $\mathsf{I}\Sigma_1$. Then $T \nvdash \mathsf{Con}_T$.

The proofs largely follow the standard approach using derivability conditions in the literature (see, for example, [57]). We therefore omit the details and instead comment on several technical and methodological aspects of the formalization.

### 2.1 Syntax

We use a locally nameless representation for terms and formulas of first-order logic. A similar approach is adopted in [59].

In standard logical terminology, this amounts to using *semiterms* and *semiformulas*, which generalize terms and formulas, respectively [32]. Variable symbols are divided into two classes: infinitely many *free variables* and finitely many *bound variables*. A semiterm is a term generated using these two kinds of variables. A semiformula is generated from semiterms in the usual way, but may contain bound variables that are not bound by any quantifier. We mechanized the type of semiformulas that may contain free variables of type $\xi$ and $n$ bound variables as `Semiformula L ξ n`:

```
inductive Semiformula (L : Language) (ξ : Type*) : ℕ → Type _ where
|  verum : Semiformula L ξ n
| falsum : Semiformula L ξ n
|    rel : {arity : ℕ} → L.Rel arity → (Fin arity → Semiterm L ξ n) → Semiformula L ξ n
|   nrel : {arity : ℕ} → L.Rel arity → (Fin arity → Semiterm L ξ n) → Semiformula L ξ n
|    and : Semiformula L ξ n → Semiformula L ξ n → Semiformula L ξ n
|     or : Semiformula L ξ n → Semiformula L ξ n → Semiformula L ξ n
|    all : Semiformula L ξ (n + 1) → Semiformula L ξ n
|    exs : Semiformula L ξ (n + 1) → Semiformula L ξ n
```

NOTE: We write `Formula L ξ` for `Semiformula L ξ 0`, and `Sentence L` for sentences, namely `Formula L Empty`.

SOURCE: 1

Mechanization using semiterm/semiformulas is more than a technical device to deal with quantifiers; it also offers practical advantages. For example, a frequently encountered situation in proof theory and model theory, such as a formula $\varphi(x, y, z)$ with parameters from $M$, can be expressed by the single type `Semiformula L M 3`.

[4]See: https://formalizedformallogic.github.io#financial-supports

### 2.2 On internal argument

In proofs of the incompleteness theorems, especially G2, the principal obstacle is often the internalization of metamathematics—terms, formulas, provability, elementary proof theory, and so forth—a process commonly called *arithmetization* or *bootstrapping*. In other words, these notions must be formally defined and their properties proved *within* the formalized deductive system itself, which in our case is $\mathsf{I}\Sigma_1$. A naïve, purely syntactic approach to this task encounters the following difficulties[5].

**Bureaucracy of the deductive system** When sufficiently complex formulas are involved (which is most often the case in practice), the deductive system can become unmanageably intricate. A task that is already difficult to formalize in Lean becomes exceedingly burdensome when it must instead be carried out within a still more restrictive formal system that is itself defined inside a formal system (Lean). Moreover, tackling G2 requires climbing yet another level. One would have to work with a formal system inside a formal system inside a formal system, which is hardly practical.

**Non-canonicity of bootstrapping** Bootstrapping is a formalization of metamathematics. This requires encoding of the metamathematical notions, the *Gödel numbering*. Unfortunately, there is neither a canonical choice of encoding nor a unique mathematically natural construction. One must instead develop a large body of intrinsically complicated combinatorics, often involving numerous ad-hoc constructions. This complicates the proofs and, for the reasons just discussed, makes mechanization difficult.

Our solution is to avoid syntactic bureaucracy by employing a model-theoretic argument via the completeness theorem of first-order logic. That is, we show $T \vDash \varphi$ instead of $T \vdash \varphi$. This largely resolves the first problem. Although it does not eliminate the second, it mitigates its complexity to feasible levels.

In the weak mathematics considered here, one frequently needs to track restrictions on formula complexity, e.g. $\Sigma_i, \Pi_i$. With a model-theoretic argument, however, it is unnecessary to exhibit an actual formula; it suffices to establish that the predicate in question is *definable* in appropriate complexity. As discussed below, we designed this part of the development so that it can be handled almost automatically using Aesop [88]. Thus, the bureaucratic overhead of syntax, especially that associated with (external) formulas, can be *almost* eliminated. There nevertheless remain situations in which a concrete formula must be supplied. For example, the second incompleteness theorem asserts $T \nvdash \mathsf{Con}_T$, and stating this result requires an explicit, model-independent formula $\mathsf{Con}_T$.

In addition, to avoid directly handling formalized statements, such as $\mathsf{Pr}_T(x)$, as much as possible, we use an abstract characterization of provability predicates when discussing the derivability conditions. We discuss this in detail in Section 2.4.

### 2.3 First incompleteness theorem

It is known that the first incompleteness theorem holds even for extremely weak arithmetic theories. Among these, we use the arithmetic theory $\mathsf{R}_0$ due to Cobham (cf. [152]).

**Definition 2.1** The theory $\mathsf{R}_0$ consists of the equality axioms for $\mathcal{L}_{\mathrm{OR}}$, together with the following variable-free atomic formulas in $\mathcal{L}_{\mathrm{OR}}$,

$$\begin{aligned} \overline{n} + \overline{m} &= \overline{n+m} && \text{for all } n, m \in \mathbb{N} \\ \overline{n} \cdot \overline{m} &= \overline{n \cdot m} && \text{for all } n, m \in \mathbb{N} \\ \overline{n} &\neq \overline{m} && \text{for all } n, m \in \mathbb{N} \text{ such that } n \neq m \end{aligned}$$

together with the following axiom scheme:

$$\forall x \left[ x < \overline{n} \leftrightarrow \bigvee_{i<n} (x = \overline{i}) \right]$$

```
inductive R0 : ArithmeticTheory
| equal : ∀ φ ∈ EQ ℒₒᵣ, R0 φ
| Ω₁ (n m : ℕ) : R0 “↑n + ↑m = ↑(n + m)”
| Ω₂ (n m : ℕ) : R0 “↑n * ↑m = ↑(n * m)”
| Ω₃ (n m : ℕ) : n ≠ m → R0 “↑n ≠ ↑m”
```

[5] Nevertheless, carrying out such a construction is worthwhile. These syntactic operations are constructive and can be developed over very weak base theories, such as $\mathsf{S}^1_2$.

```
| Ω₄ (n : ℕ) : R0 “∀ x, x < ↑n ↔ ⋁ i < n, x = ↑i”

notation "R₀" => R0
```

SOURCE: 1

The key theorem is that every $\Sigma_1$-sound theory extending $\mathsf{R}_0$ weakly represents every recursively enumerable (r.e.) set [72, 152]. More precisely:

**Theorem 2.2** Let $T \supseteq \mathsf{R}_0$ be a $\Sigma_1$-sound theory and let $S$ be an r.e. set. Then there is a $\mathcal{L}_{\mathrm{OR}}$-formula $\mathsf{Rep}_S(x)$ such that

$$n \in S \iff T \vdash \mathsf{Rep}_S(\overline{n})$$

Let $\ulcorner \bullet \urcorner$ denote a Gödel coding of formulas. Define the set $D$ by $\ulcorner \varphi \urcorner \in D \iff T \vdash \neg\varphi(\ulcorner \varphi \urcorner)$. As shown below, since $T$ is $\Delta_1$-definable, there is a provability predicate $\mathsf{Pr}_T(x)$, definable by a $\Sigma_1$-formula, such that $\mathbb{N} \models \mathsf{Pr}_T(\ulcorner \psi \urcorner) \iff T \vdash \psi$; hence $D$ is r.e. Theorem 2.3 now follows from Theorem 2.2 by the standard diagonal argument.

**Theorem 2.3 (Gödel's first incompleteness theorem [46, 72, 152])** Let $T$ be a $\Delta_1$-definable, $\Sigma_1$-sound arithmetic theory stronger than $\mathsf{R}_0$. Then $T$ is incomplete; that is, there exists an arithmetic sentence $\varphi$ such that $T \nvdash \varphi$ and $T \nvdash \neg\varphi$.

```
theorem incomplete (T : ArithmeticTheory) [T.Δ₁] [R₀ ⪯ T] [T.SoundOnHierarchy Σ 1] : Incomplete T
```

NOTE: Here `Incomplete T` is an abbreviation of `∃ φ, T ⊬ φ ∧ T ⊬ ∼φ`.

SOURCE: 1

As a corollary of Theorem 2.3, we can also prove Gödel's theorem in the following form.

**Corollary 2.4** Let $T$ be a $\Delta_1$-definable, $\Sigma_1$-sound arithmetic theory stronger than $\mathsf{R}_0$. Then there exists a sentence $\sigma$ that is true but unprovable in $T$; that is, $\mathbb{N} \models \sigma$ but $T \nvdash \sigma$.

```
theorem exists_true_but_unprovable_sentence_of_sigma1sound
  (T : ArithmeticTheory) [T.Δ₁] [R₀ ⪯ T] [T.SoundOnHierarchy Σ 1] :
  ∃ δ : ArithmeticSentence, ℕ↓[ℒₒᵣ] ⊧ δ ∧ T ⊬ δ
```

SOURCE: 1

### 2.4 Provability abstraction

Before proving G2, we introduce a theory of the provability predicate, called *provability abstraction*, because working directly with a raw provability predicate is technically cumbersome. This notion is closely related to provability logic, which treats provability as a modality (see Section 3). With these abstractions, the incompleteness theorems can be mechanized abstractly, by purely syntactic manipulations. Concretely constructing a "provability" satisfying the abstract derivability conditions then immediately yields the concrete statements of the incompleteness theorems. Mechanizing the incompleteness theorems via such an abstract provability has previously been studied by Popescu and Traytel [114, 115].

**Definition 2.5 (Provability predicate)** Suppose that $\mathcal{L}$-sentences admit a Gödel numbering in language $\mathcal{L}_0$. For an $\mathcal{L}_0$-theory $T_0$ and an $\mathcal{L}$-theory $T$, a unary $\mathcal{L}_0$-semisentence $\mathfrak{B}(x)$ is called a *$T$-provability predicate over $T_0$*, if the following holds for every $\mathcal{L}$-sentence $\sigma$.

$$\textbf{D1} \quad T \vdash \sigma \text{ implies } T_0 \vdash \mathfrak{B}(\ulcorner \sigma \urcorner)$$

That is, $\mathfrak{B}(x)$ is required to satisfy at least the derivability condition **D1**. In what follows, we simply write $\mathfrak{B}\sigma$ for $\mathfrak{B}(\ulcorner \sigma \urcorner)$. We further define the following properties, where $\sigma$ and $\pi$ range over $\mathcal{L}$-sentences. The conditions **D3** and **Kre** are defined only when $T_0$ and $T$ are theories in the same language, i.e., when $\mathcal{L}_0 = \mathcal{L}$.

$$\textbf{D2} : T_0 \vdash \mathfrak{B}(\sigma \to \pi) \to \mathfrak{B}\sigma \to \mathfrak{B}\pi \qquad \textbf{Kre} : T \vdash \mathfrak{B}\sigma \text{ implies } T \vdash \sigma$$

$$\textbf{D3} : T_0 \vdash \mathfrak{B}\sigma \to \mathfrak{B}\mathfrak{B}\sigma \qquad \textbf{Ros} : T \vdash \neg\sigma \text{ implies } T_0 \vdash \neg\mathfrak{B}\sigma$$

```
structure Provability [L.ReferenceableBy L₀] (T₀ : Theory L₀) (T : Theory L) where
  prov : Semisentence L₀ 1
  bew_def {σ : Sentence L} : T ⊢ σ → T₀ ⊢ prov/[⌜σ⌝]

variable {L₀ L : Language} [L.ReferenceableBy L₀] {T₀ : Theory L₀} {T : Theory L}

@[coe] def pr (𝔅 : Provability T₀ T) (σ : Sentence L) : Sentence L₀ := 𝔅.prov/[⌜σ⌝]
instance : CoeFun (Provability T₀ T) (fun _ ↦ Sentence L → Sentence L₀) := ⟨pr⟩

class HBL2 [L.ReferenceableBy L₀] {T₀ : Theory L₀} {T : Theory L} (𝔅 : Provability T₀ T) where
  D2 {σ τ : Sentence L} : T₀ ⊢ 𝔅 (σ ➝ τ) ➝ 𝔅 σ ➝ 𝔅 τ

class HBL3 [L.ReferenceableBy L] {T₀ T : Theory L} (𝔅 : Provability T₀ T) where
  D3 {σ : Sentence L} : T₀ ⊢ 𝔅 σ ➝ 𝔅 (𝔅 σ)

class Kreisel [L.ReferenceableBy L] {T₀ T : Theory L} (𝔅 : Provability T₀ T) where
  KR {σ : Sentence L} : T ⊢ 𝔅 σ → T ⊢ σ

class Rosser [L.ReferenceableBy L₀] {T₀ : Theory L₀} {T : Theory L} (𝔅 : Provability T₀ T) where
  Ros {σ : Sentence L} : T ⊢ ∼σ → T₀ ⊢ ∼𝔅 σ
```

Source: 1

The standard provability predicate $\mathsf{Pr}_T$ satisfies **D2**, **D3** and **Kre**. In provability logic discussed in Section 3, we mainly assume those conditions on the provability predicate. The condition **Kre** is a derivability condition introduced by Visser [160] under the name *Kreisel's condition*[6]. Anticipating the later construction, the standard provability predicate is a $\Sigma_1$-predicate, so the condition **Kre** can be regarded as a purely syntactic counterpart of the $\Sigma_1$-soundness of $T$; the converse implication, which corresponds to $\Sigma_1$-completeness, is the content of **D1**.

In fact, abstracting provability alone does not suffice to mechanize the incompleteness theorems: we also need to abstract the diagonalization.

**Definition 2.6 (Diagonalization abstraction)** Suppose that $\mathcal{L}$-sentences admit a Gödel numbering in $\mathcal{L}$. An $\mathcal{L}$-theory $T$ is called *diagonalizable* if one can construct a map $\mathsf{fixedpoint}_\bullet$, sending a unary $\mathcal{L}$-formula to an $\mathcal{L}$-sentence, such that

$$T \vdash \mathsf{fixedpoint}_\theta \leftrightarrow \theta(\ulcorner \mathsf{fixedpoint}_\theta \urcorner)$$

for any $\theta(x)$. We call $\mathsf{fixedpoint}_\theta$ the *fixpoint* of $\theta$.

Let $T_0, T$ be $\mathcal{L}$-theories such that $T_0$ is diagonalizable, and let $\mathfrak{B}$ be a provability of $T_0, T$. Then the fixpoint of $\neg\mathfrak{B}(x)$ is called the *Gödel sentence* and is denoted by $\mathsf{G}_{\mathfrak{B}}$.

```
class Diagonalization [L.ReferenceableBy L] (T : Theory L) where
  fixedpoint : Semisentence L 1 → Sentence L
  diag (θ) : T ⊢ fixedpoint θ ⭤ θ/[⌜fixedpoint θ⌝]
```

[6] Visser attributes the origin of this condition to [81]. To be precise, Visser required both directions.

```
variable {L : Language} [L.ReferenceableBy L] {T₀ T : Theory L} [Diagonalization T₀]

def gödel (𝔅 : Provability T₀ T) : Sentence L := fixedpoint T₀ “x. ¬!𝔅.prov x”
```

SOURCE: 1

For the arithmetic theory stronger than $\mathsf{I}\Sigma_1$, we have a *standard diagonalization*, which is obtained by the standard construction of diagonalization.

```
theorem diagonal {T : ArithmeticTheory} [𝐈𝚺₁ ⪯ T] (θ : ArithmeticSemisentence 1) :
    T ⊢ fixedpoint θ ⭤ θ/[⌜fixedpoint θ⌝]
```

SOURCE: 1

With these tools at hand, the incompleteness theorems and their corollaries can be proved by syntactic manipulations alone. In what follows, let $T_0 \subseteq T$ be $\mathcal{L}$-theories such that $T_0$ is diagonalizable and $T$ is consistent, and let $\mathfrak{B}$ be a $T$-provability predicate over $T_0$. The first incompleteness theorem is proved as follows.

**Proposition 2.7 (Abstract version of G1)**
1. $T \nvdash \mathsf{G}_{\mathfrak{B}}$.
2. If $\mathfrak{B}$ satisfies **Kre**, then $T \nvdash \neg\mathsf{G}_{\mathfrak{B}}$.

Hence $\mathsf{G}_{\mathfrak{B}}$ is independent of $T$, and therefore $T$ is incomplete.

```
variable {L : Language} [L.ReferenceableBy L] [L.DecidableEq]
variable {T₀ T : Theory L} [Diagonalization T₀] [T₀ ⪯ T] [Consistent T]
variable {𝔅 : Provability T₀ T}

theorem unprovable_gödel : T ⊬ gödel 𝔅

theorem unrefutable_gödel [𝔅.Kreisel] : T ⊬ ∼gödel 𝔅

theorem gödel_independent [𝔅.Kreisel] : Independent T (gödel 𝔅)

theorem first_incompleteness [𝔅.Kreisel] : Incomplete T
```

SOURCE: 1

The second incompleteness theorem can likewise be mechanized.

**Proposition 2.8 (Abstract version of G2)** Assume that $\mathfrak{B}$ satisfies **D2** and **D3**. The sentence $\neg\mathfrak{B}\bot$ is a natural expression of consistency; we denote it by $\mathsf{Con}_{\mathfrak{B}}$. Then the following hold.
1. $T \nvdash \mathsf{Con}_{\mathfrak{B}}$.
2. If $\mathfrak{B}$ satisfies **Kre**, then $T \nvdash \neg\mathsf{Con}_{\mathfrak{B}}$. Hence $\mathsf{Con}_{\mathfrak{B}}$ is also independent of $T$.

```
variable {L₀ L : Language} [L.ReferenceableBy L₀] {T₀ : Theory L₀} {T : Theory L}

def con (𝔅 : Provability T₀ T) : Sentence L₀ := ∼𝔅 ⊥

variable {L : Language} [L.ReferenceableBy L] [L.DecidableEq]
variable {T₀ T : Theory L} [Diagonalization T₀] [T₀ ⪯ T]
variable {𝔅 : Provability T₀ T} [𝔅.HBL]

theorem con_unprovable [Consistent T] : T ⊬ 𝔅.con

theorem con_unrefutable [Consistent T] [𝔅.Kreisel] : T ⊬ ∼𝔅.con

theorem con_independent [Consistent T] [𝔅.Kreisel] : Independent T 𝔅.con
```

SOURCE: 1

There is, however, a view that $\mathsf{Con}_{\mathfrak{B}}$ is not the only natural expression of consistency. Variants of G2 arising from this view are discussed later.

As further results, we can also mechanize Löb's theorem and the formalized Löb's theorem.

**Proposition 2.9 (Abstract version of Löb's theorem)** Assume that $\mathfrak{B}$ satisfies **D2** and **D3**. Then the following hold, where $\sigma$ is an arbitrary $\mathcal{L}$-sentence.

$$\text{Löb's theorem} \quad T \vdash \mathfrak{B}\sigma \to \sigma \text{ implies } T \vdash \sigma.$$
$$\text{Formalized Löb's theorem} \quad T_0 \vdash \mathfrak{B}(\mathfrak{B}\sigma \to \sigma) \to \mathfrak{B}\sigma.$$

```
variable {L : Language} [L.ReferenceableBy L] [L.DecidableEq]
variable {T₀ T : Theory L} [Diagonalization T₀] [T₀ ⪯ T]
variable {𝔅 : Provability T₀ T} [𝔅.HBL]

theorem löb_theorem {σ : Sentence L} (H : T ⊢ 𝔅 σ ➝ σ) : T ⊢ σ

theorem formalized_löb_theorem {σ : Sentence L} : T₀ ⊢ 𝔅 (𝔅 σ ➝ σ) ➝ 𝔅 σ
```

Source: 1

Note that **D1**, **D2**, **D3** and the formalized Löb's theorem correspond roughly to the necessitation rule and the axioms K, 4, and L of modal logic, respectively. This yields the observation that arithmetical soundness holds for the standard provability predicate.

In view of the reason we gave for introducing **Kre** into the abstraction, requiring **Kre** in the abstract G1 of Proposition 2.7 corresponds to requiring the $\Sigma_1$-soundness of $T$. If we instead impose on $\mathfrak{B}$ the condition **Ros**, then the abstract G1 can be proved assuming only that $T$ is consistent. This is precisely an abstraction of the incompleteness theorem as improved by Rosser [118].

**Proposition 2.10 (Abstract version of Gödel–Rosser theorem)** Let $\mathfrak{R}$ be a $T$-provability predicate over $T_0$ satisfying **Ros**. In this case, the Gödel sentence for $\mathfrak{R}$ is called the *Rosser sentence*. Then we have $T \nvdash \mathsf{G}_{\mathfrak{R}}$ and $T \nvdash \neg\mathsf{G}_{\mathfrak{R}}$. That is, $\mathsf{G}_{\mathfrak{R}}$ is independent of $T$; note in particular that **Kre** is not required. On the other hand, for the consistency statement $\mathsf{Con}_{\mathfrak{R}}$ defined above, we have $T \vdash \mathsf{Con}_{\mathfrak{R}}$[7].

```
variable {L : Language} [L.ReferenceableBy L]
variable {T₀ T : Theory L} [Diagonalization T₀] [T₀ ⪯ T] [Consistent T]
variable {𝔅 : Provability T₀ T} [𝔅.Rosser]

theorem unrefutable_rosser : T ⊬ ∼(gödel 𝔅)

theorem rosser_independent : Independent T (gödel 𝔅)

theorem rosser_first_incompleteness (𝔅 : Provability T₀ T) : Incomplete T

theorem kreisel_remark : T ⊢ 𝔅.con
```

Source: 1

Indeed, the concrete statement of the Gödel–Rosser theorem given later, obtained by instantiating this abstraction, does not require $\Sigma_1$-soundness.

We next describe refutability (*Widerlegbar*) $\mathfrak{W}$. Concerning G2, formal consistency can be expressed in ways other than the formalized consistency $\neg\mathfrak{B}\bot$ introduced in Proposition 2.8. For instance, the statement "no sentence is both provable and refutable" may also be regarded as a natural expression of consistency. The version of G2 obtained by formalizing this consistency is usually attributed to Jeroslow [71][8]. A naive attempt to formalize

[7] In the Japanese mathematical logic community, this is often called *Kreisel's remark*.

[8] See, e.g., Kurahashi [84] for how subtle differences in the required conditions, and various versions of the statement of G2, arise depending on how consistency is formalized.

Jeroslow's G2 on top of the provability abstraction, however, becomes slightly cumbersome if only $\mathfrak{B}$ is available. The reason is that $\mathfrak{B}$ is in fact an arithmetical predicate taking the Gödel number of a formula: dealing with refutability, that is, with negated sentences, would force us to handle a "function" computing the Gödel number of $\neg\sigma$ from that of $\sigma$, and such a function is awkward to accommodate within the provability abstraction. We therefore abstract refutability itself, rather than a function computing the Gödel number of a negation. This allows us to formalize Jeroslow's G2 concisely.

**Definition 2.11 (Refutability abstraction)** For an $\mathcal{L}_0$-theory $T_0$ and an $\mathcal{L}$-theory $T$, a unary $\mathcal{L}_0$-semisentence $\mathfrak{W}(x)$ is called a *$T$-refutability predicate over $T_0$*, if the following holds for every $\mathcal{L}$-sentence $\sigma$.

$$T \vdash \neg\sigma \Longrightarrow T_0 \vdash \mathfrak{W}(\ulcorner\sigma\urcorner)$$

As with $\mathfrak{B}$, we abbreviate $\mathfrak{W}(\ulcorner\sigma\urcorner)$ as $\mathfrak{W}\sigma$. We say that $\mathfrak{W}$ is *sound on* an $\mathcal{L}$-sentence $\sigma$ if $T \vdash \mathfrak{W}\sigma \Longrightarrow T \vdash \neg\sigma$.

Let $\mathfrak{W}$ be a $T$-refutability predicate over $T_0$ and suppose that $T_0$ is diagonalizable. Then the fixpoint of $\mathfrak{W}(x)$ is called the *Jeroslow sentence* and is denoted by $\mathsf{J}_{\mathfrak{W}}$.

```
structure Refutability [L.ReferenceableBy L₀] (T₀ : Theory L₀) (T : Theory L) where
  refu : Semisentence L₀ 1
  refu_def {σ : Sentence L} : T ⊢ ∼σ → T₀ ⊢ refu/[⌜σ⌝]

@[coe] def Refutability.rf (𝔚 : Refutability T₀ T) (σ : Sentence L) : Sentence L₀ := 𝔚.refu/[⌜σ⌝]
instance : CoeFun (Refutability T₀ T) (fun _ ↦ Sentence L → Sentence L₀) := ⟨Refutability.rf⟩

variable {L : Language} [L.ReferenceableBy L] {T₀ T : Theory L} [Diagonalization T₀]

class Refutability.SoundOn (𝔚 : Refutability T₀ T) (σ : Sentence L) where
  sound_on : T ⊢ 𝔚 σ → T ⊢ ∼σ

def jeroslow (𝔚 : Refutability T₀ T) : Sentence L := fixedpoint T₀ 𝔚.refu
```

Source: 1

The following is immediate for the Jeroslow sentence.

**Proposition 2.12** If $T$ is consistent and $\mathfrak{W}$ is sound on $\mathsf{J}_{\mathfrak{W}}$, then $T \nvdash \mathsf{J}_{\mathfrak{W}}$.

```
lemma unprovable_jeroslow [T₀ ⪯ T] [Consistent T] [𝔚.SoundOn (jeroslow 𝔚)] : T ⊬ jeroslow 𝔚
```

Source: 1

We now state Jeroslow's incompleteness theorem.

**Proposition 2.13 (Abstract version of Jeroslow's G2 [71])** Let $\mathsf{Safe}_{\mathfrak{B},\mathfrak{W}}(x) \equiv \neg(\mathfrak{B}x \wedge \mathfrak{W}x)$ be the unary formula stating that a sentence is not both provable and refutable (*safe*), and let $\mathsf{FLoN}_{\mathfrak{B},\mathfrak{W}} \equiv \forall x, \mathsf{Safe}_{\mathfrak{B},\mathfrak{W}}(x)$ be the sentence expressing consistency in the sense that every sentence is safe (the *formalized law of non-contradiction*).

If $T$ is consistent and $T_0 \vdash \mathsf{J}_{\mathfrak{W}} \to \mathfrak{B}\mathsf{J}_{\mathfrak{W}}$, then $T \nvdash \mathsf{FLoN}_{\mathfrak{B},\mathfrak{W}}$.

```
variable [L.DecidableEq] [L.ReferenceableBy L] {T₀ T : Theory L}
variable [Diagonalization T₀] [T₀ ⪯ T] {𝔅 : Provability T₀ T} {𝔚 : Refutability T₀ T}

def safe (𝔅 : Provability T₀ T) (𝔚 : Refutability T₀ T) : Semisentence L 1 :=
  "x. ¬(!𝔅.prov x ∧ !𝔚.refu x)"

def flon (𝔅 : Provability T₀ T) (𝔚 : Refutability T₀ T) : Sentence L := "∀ x, !(safe 𝔅 𝔚) x"

class FormalizedCompleteOn (𝔅 : Provability T₀ T) (σ) where
  formalized_complete_on : T₀ ⊢ σ ➝ 𝔅 σ
```

```
lemma unprovable_flon [Consistent T] [𝔅.FormalizedCompleteOn (jeroslow 𝔚)] : T ⊬ flon 𝔅 𝔚
```

NOTE: The hypothesis $T_0 \vdash \mathsf{J}_{\mathfrak{W}} \to \mathfrak{B}\mathsf{J}_{\mathfrak{W}}$ is mechanized as the class `FormalizedCompleteOn`, which is an abstract version of formalized Γ-completeness for $\mathfrak{B}$. For instance, formalized $\Sigma_1$-completeness is expressed as `[∀ σ ∈ Σ₁, 𝔅.FormalizedCompleteOn σ]`.

SOURCE: 1 2

Making this abstraction concrete, that is, actually constructing the desired provability $\mathfrak{B}$ and refutability $\mathfrak{W}$, is the goal of the following sections.

### 2.5 Second incompleteness theorem

The goal of this section is to construct a *standard* provability predicate, thereby making the abstraction Proposition 2.8 introduced in the previous section concrete. We first take $\mathsf{I}\Sigma_1$ as the base theory for our proof of G2.

**Definition 2.14** We call $\mathsf{PA}^-$ (the theory of discrete ordered semirings) the finite axiom system consisting of universal $\mathcal{L}_{\mathrm{OR}}$-sentences describing basic properties. For a unary arithmetical formula $\varphi(x)$ (which may contain parameters), we define the formula $\mathsf{I}\varphi$ expressing the universal closure of the following instance of mathematical induction:

$$\varphi(0) \to \forall x\, [\varphi(x) \to \varphi(x+1)] \to \forall x\, \varphi(x)$$

For a class of formulas Γ, we define $\mathsf{I}\Gamma$ as the union of $\mathsf{PA}^-$ with $\mathsf{I}\varphi$ for every formula $\varphi$ belonging to Γ. We then let $\mathsf{I}\Sigma_1$ be the theory in which mathematical induction is available for all $\Sigma_1$-formulas, and $\mathsf{PA}$ the theory in which it is available for all formulas.

```
abbrev       addZero : ArithmeticSentence := “∀ x, x + 0 = x”
abbrev      addAssoc : ArithmeticSentence := “∀ x y z, (x + y) + z = x + (y + z)”
abbrev       addComm : ArithmeticSentence := “∀ x y, x + y = y + x”
abbrev     addEqOfLt : ArithmeticSentence := “∀ x y, x < y → ∃ z, x + z = y”
abbrev        zeroLe : ArithmeticSentence := “∀ x, 0 ≤ x”
abbrev     zeroLtOne : ArithmeticSentence := “0 < 1”
abbrev oneLeOfZeroLt : ArithmeticSentence := “∀ x, 0 < x → 1 ≤ x”
abbrev      addLtAdd : ArithmeticSentence := “∀ x y z, x < y → x + z < y + z”
abbrev       mulZero : ArithmeticSentence := “∀ x, x * 0 = 0”
abbrev        mulOne : ArithmeticSentence := “∀ x, x * 1 = x”
abbrev      mulAssoc : ArithmeticSentence := “∀ x y z, (x * y) * z = x * (y * z)”
abbrev       mulComm : ArithmeticSentence := “∀ x y, x * y = y * x”
abbrev      mulLtMul : ArithmeticSentence := “∀ x y z, x < y ∧ 0 < z → x * z < y * z”
abbrev         distr : ArithmeticSentence := “∀ x y z, x * (y + z) = x * y + x * z”
abbrev      ltIrrefl : ArithmeticSentence := “∀ x, x ≮ x”
abbrev       ltTrans : ArithmeticSentence := “∀ x y z, x < y ∧ y < z → x < z”
abbrev         ltTri : ArithmeticSentence := “∀ x y, x < y ∨ x = y ∨ x > y”

inductive PeanoMinus : ArithmeticTheory
  | equal        : ∀ φ ∈ EQ ℒₒᵣ, PeanoMinus φ
  | addZero      : PeanoMinus PeanoMinus.Axiom.addZero
  | addAssoc     : PeanoMinus PeanoMinus.Axiom.addAssoc
  | ...
notation "PA⁻" => PeanoMinus

def succInd {ξ} (φ : Semiformula L ξ 1) : Formula L ξ :=
  “!φ 0 → (∀ x, !φ x → !φ (x + 1)) → ∀ x, !φ x”

def InductionScheme (Γ : Semiformula L ℕ 1 → Prop) : Theory L :=
  { ψ | ∃ φ : Semiformula L ℕ 1, Γ φ ∧ ψ = .univCl (succInd φ) }

abbrev InductionOnHierarchy (Γ : Polarity) (k : ℕ) : ArithmeticTheory := PA⁻ ∪ InductionScheme ℒₒᵣ
(Arithmetic.Hierarchy Γ k)
prefix:max "IND " => InductionOnHierarchy

abbrev ISigma (k : ℕ) : ArithmeticTheory := IND Σ k
notation "IΣ₁" => ISigma 1
```

```
abbrev Peano : ArithmeticTheory := PA⁻ ∪ InductionScheme ℒₒᵣ Set.univ
notation "PA" ⇒ Peano
```

SOURCE: 1 2

$\mathsf{I}\Sigma_1$ is in fact unnecessarily strong. For a sharper result, one could weaken the base theory to Buss's theory $\mathsf{S}^1_2$ [31], over which the standard proof can be carried out with few changes[9]. Moreover, Nelson's interpretation $\mathsf{Q} \rhd \mathsf{S}^1_2$ extends the second incompleteness theorem to a broad class of theories that interpret Robinson arithmetic $\mathsf{Q}$ [159]. Although this is an appealing direction, we do not pursue it because it would make the mechanization prohibitively complex (see Section 4.2 for future work). Working in $\mathsf{I}\Sigma_1$ makes recursive definitions of predicates and functions easier to handle, since Theorem 2.15 is available.

As noted above, our internal arithmetical arguments are carried out in an arbitrarily fixed model of $\mathsf{I}\Sigma_1$, which we henceforth denote by $\mathbf{V}$. A *class* is a subset of $\mathbf{V}$.

Because only induction restricted to $\Sigma_1$-formulas is available over $\mathbf{V}$, the ($\Sigma_i$-, $\Pi_i$-, and $\Delta_i$-) definability of relations and functions on $\mathbf{V}$ is important. In practice, this can often be inferred automatically from the stated definition. To automate the substantial amount of such reasoning, we make extensive use of Aesop [88] whenever no explicit defining formula is needed.

Let $\mathrm{Bit}(x, y)$ be the predicate asserting that the $x$-th digit in the binary expansion of $y$ is 1. Ackermann coding, obtained from the membership relation defined below, provides a means of representing hereditarily finite sets within arithmetic [112].

$$x \in y \iff \mathrm{Bit}(x, y)$$

To work with $\mathrm{Bit}(x, y)$ in weak arithmetic, we also mechanized the well-known fact due to Gaifman and Dimitracopoulos [41], that the graph of exponentiation is representable by a $\Delta_0$-formula and that its inductive properties are provable in $\mathsf{I}\Delta_0$.

$$\begin{array}{l}
\mathrm{Exp}(a_0, a_1) \equiv [[a_0 = 0] \land [a_1 = 1]] \lor [(\exists x_0 < ((((a_1 \cdot a_1) \cdot a_1) \cdot a_1) + 1))[(\exists x_1 < ((((a_1 \cdot a_1) \cdot a_1) \cdot a_1) + 1))[[[(\exists x_2 < (x_0 + 1))[[[[0 \not< (((1 + 1) + 1) + 1)] \lor [(\exists x_3 < (((1 + 1) + 1) + 1))[x_0 = \\
(((((1+1)+1)+1)\cdot x_2)+x_3)]]] \land [[((((1+1)+1)+1) \neq 0] \lor [x_2 = 0]]] \land [(\exists x_3 < (x_2+1))[[[[0 \not< (((1+1)+1)+1)] \lor [(\exists x_4 < (((1+1)+1)+1))[x_2 = (((((1+1)+1)+1)\cdot x_3)+x_4)]]] \land [[((((1+1)+1)+1) \neq 0] \lor [x_3 = \\
0]]] \land [[[[(((((1+1)+1)+1)\cdot x_3) \neq x_2] \land [(((((1+1)+1)+1)\cdot x_3) \not< x_2]] \lor [x_2 = (((((1+1)+1)+1)\cdot x_3)+1)]] \land [[x_2 \not< ((((1+1)+1)+1)\cdot x_3)] \lor [1 = 0]]]]]]] \land [(\exists x_2 < (x_1+1))[[[[0 \not< (((1+1)+1)+1)] \lor [(\exists x_3 < \\
(((1+1)+1)+1))[x_1 = (((((1+1)+1)+1)\cdot x_2)+x_3)]]] \land [[((((1+1)+1)+1) \neq 0] \lor [x_2 = 0]]] \land [(\exists x_3 < (x_2+1))[[[[0 \not< (((1+1)+1)+1)] \lor [(\exists x_4 < (((1+1)+1)+1))[x_2 = (((((1+1)+1)+1)\cdot x_3)+x_4)]]] \land [[((((1+1)+1)+1) \neq \\
0] \lor [x_3 = 0]]] \land [[[[(((((1+1)+1)+1)\cdot x_3) \neq x_2] \land [(((((1+1)+1)+1)\cdot x_3) \not< x_2]] \lor [x_2 = (((((1+1)+1)+1)\cdot x_3)+(1+1))]] \land [[x_2 \not< ((((1+1)+1)+1)\cdot x_3)] \lor [(1+1) = 0]]]]]]] \land [[(\forall x_2 < (a_1+1))[[x_2 = (1+1)] \lor [[[[0 \not< \\
x_2] \lor [(\exists x_3 < (x_2+1))[[1 < x_3] \land [[(\exists x_4 < (x_2+1))[x_2 = (x_3\cdot x_4)]] \land [(\forall x_4 < (x_3+1))[x_3 \neq ((1+1)\cdot x_4)]]]]]] \lor [(\forall x_3 < ((1+1)\cdot x_2))[[[(\exists x_4 < (x_3+1))[[[[0 \not< 1] \lor [(\exists x_5 < 1)[x_3 = ((1\cdot x_4)+x_5)]]] \land [[1 \neq 0] \lor [x_4 = \\
0]]] \land [(\forall x_5 < (x_4+1))[x_4 \neq ((1+1)\cdot x_5)]]]] \lor [[(\forall x_4 < (x_3+1))[[[[0 < (1+1)] \land [(\forall x_5 < (1+1))[x_3 \neq (((1+1)\cdot x_4)+x_5)]]] \lor [[(1+1) = 0] \land [x_4 \neq 0]]] \lor [(\exists x_5 < (x_4+1))[x_4 = ((1+1)\cdot x_5)]]]] \lor [(\exists x_4 < (x_3+1))[[[0 < \\
x_4] \land [(\forall x_5 < (x_4+1))[[1 \not< x_5] \lor [[(\forall x_6 < (x_4+1))[x_4 \neq (x_5\cdot x_6)]] \lor [(\exists x_6 < (x_5+1))[x_5 = ((1+1)\cdot x_6)]]]]]] \land [[(1+1) < x_4] \land [[[(\exists x_5 < (x_3+1))[[[[0 \not< x_4] \lor [(\exists x_6 < x_4)[x_3 = ((x_4\cdot x_5)+x_6)]]] \land [[x_4 \neq 0] \lor [x_5 = \\
0]]] \land [(\forall x_6 < (x_5+1))[x_5 \neq ((1+1)\cdot x_6)]]]] \land [(\forall x_5 < (x_4+1))[[[[(x_5\cdot x_5) \neq x_4] \land [(x_5\cdot x_5) \not< x_4]] \lor [x_4 \not< ((x_5+1)\cdot(x_5+1))]] \lor [[(x_5\cdot x_5) \neq x_4] \lor [(\forall x_6 < (x_3+1))[[[[0 < x_5] \land [(\forall x_7 < x_5)[x_3 \neq ((x_5\cdot x_6)+x_7)]]] \lor [[x_5 = \\
0] \land [x_6 \neq 0]]] \lor [(\exists x_7 < (x_6+1))[x_6 = ((1+1)\cdot x_7)]]]]]]]] \lor [[(\exists x_5 < (x_4+1))[[[[(x_5\cdot x_5) = x_4] \lor [(x_5\cdot x_5) < x_4]] \land [x_4 < ((x_5+1)\cdot(x_5+1))]] \land [[(x_5\cdot x_5) = x_4] \land [(\exists x_6 < (x_3+1))[[[[0 \not< x_5] \lor [(\exists x_7 < x_5)[x_3 = \\
((x_5\cdot x_6)+x_7)]]] \land [[x_5 \neq 0] \lor [x_6 = 0]]] \land [(\forall x_7 < (x_6+1))[x_6 \neq ((1+1)\cdot x_7)]]]]]]] \land [(\forall x_5 < (x_3+1))[[[[0 < x_4] \land [(\forall x_6 < x_4)[x_3 \neq ((x_4\cdot x_5)+x_6)]]] \lor [[x_4 = 0] \land [x_5 \neq 0]]] \lor [(\exists x_6 < (x_5+1))[x_5 = ((1+1)\cdot x_6)]]]]]]]]]] \lor [(\forall x_4 < \\
(x_3+1))[[[[0 < x_2] \land [(\forall x_5 < x_2)[x_3 \neq ((x_2\cdot x_4)+x_5)]]] \lor [[x_2 = 0] \land [x_4 \neq 0]]] \lor [(\exists x_5 < (x_4+1))[x_4 = ((1+1)\cdot x_5)]]]]]]] \lor [[[(\exists x_3 < (x_0+1))[[(\exists x_4 < (x_0+1))[[[[0 \not< x_2] \lor [(\exists x_5 < x_2)[x_0 = ((x_2\cdot x_4)+x_5)]]] \land [[x_2 \neq \\
0] \lor [x_4 = 0]]] \land [(\exists x_5 < (x_4+1))[[[[0 \not< x_2] \lor [(\exists x_6 < x_2)[x_4 = ((x_2\cdot x_5)+x_6)]]] \land [[x_2 \neq 0] \lor [x_5 = 0]]] \land [[[[(x_2\cdot x_5) \neq x_4] \land [(x_2\cdot x_5) \not< x_4]] \lor [x_4 = ((x_2\cdot x_5)+x_3)]] \land [[x_4 \not< (x_2\cdot x_5)] \lor [x_3 = 0]]]]]]] \land [(\exists x_4 < (x_0+1))[[[[0 \not< \\
(x_2\cdot x_2)] \lor [(\exists x_5 < (x_2\cdot x_2))[x_0 = (((x_2\cdot x_2)\cdot x_4)+x_5)]]] \land [[(x_2\cdot x_2) \neq 0] \lor [x_4 = 0]]] \land [(\exists x_5 < (x_4+1))[[[[0 \not< (x_2\cdot x_2)] \lor [(\exists x_6 < (x_2\cdot x_2))[x_4 = (((x_2\cdot x_2)\cdot x_5)+x_6)]]] \land [[(x_2\cdot x_2) \neq 0] \lor [x_5 = 0]]] \land [[[[((x_2\cdot x_2)\cdot x_5) \neq \\
x_4] \land [((x_2\cdot x_2)\cdot x_5) \not< x_4]] \lor [x_4 = (((x_2\cdot x_2)\cdot x_5)+((1+1)\cdot x_3))]] \land [[x_4 \not< ((x_2\cdot x_2)\cdot x_5)] \lor [((1+1)\cdot x_3) = 0]]]]]]]]] \land [(\exists x_3 < (x_1+1))[[(\exists x_4 < (x_1+1))[[[[0 \not< x_2] \lor [(\exists x_5 < x_2)[x_1 = ((x_2\cdot x_4)+x_5)]]] \land [[x_2 \neq 0] \lor [x_4 = \\
0]]] \land [(\exists x_5 < (x_4+1))[[[[0 \not< x_2] \lor [(\exists x_6 < x_2)[x_4 = ((x_2\cdot x_5)+x_6)]]] \land [[x_2 \neq 0] \lor [x_5 = 0]]] \land [[[[(x_2\cdot x_5) \neq x_4] \land [(x_2\cdot x_5) \not< x_4]] \lor [x_4 = ((x_2\cdot x_5)+x_3)]] \land [[x_4 \not< (x_2\cdot x_5)] \lor [x_3 = 0]]]]]]] \land [(\exists x_4 < (x_1+1))[[[[0 \not< \\
(x_2\cdot x_2)] \lor [(\exists x_5 < (x_2\cdot x_2))[x_1 = (((x_2\cdot x_2)\cdot x_4)+x_5)]]] \land [[(x_2\cdot x_2) \neq 0] \lor [x_4 = 0]]] \land [(\exists x_5 < (x_4+1))[[[[0 \not< (x_2\cdot x_2)] \lor [(\exists x_6 < (x_2\cdot x_2))[x_4 = (((x_2\cdot x_2)\cdot x_5)+x_6)]]] \land [[(x_2\cdot x_2) \neq 0] \lor [x_5 = 0]]] \land [[[[((x_2\cdot x_2)\cdot x_5) \neq \\
x_4] \land [((x_2\cdot x_2)\cdot x_5) \not< x_4]] \lor [x_4 = (((x_2\cdot x_2)\cdot x_5)+(x_3\cdot x_3))]] \land [[x_4 \not< ((x_2\cdot x_2)\cdot x_5)] \lor [(x_3\cdot x_3) = 0]]]]]]]]]] \lor [[(\exists x_3 < (x_0+1))[[(\exists x_4 < (x_0+1))[[[[0 \not< x_2] \lor [(\exists x_5 < x_2)[x_0 = ((x_2\cdot x_4)+x_5)]]] \land [[x_2 \neq 0] \lor [x_4 = \\
0]]] \land [(\exists x_5 < (x_4+1))[[[[0 \not< x_2] \lor [(\exists x_6 < x_2)[x_4 = ((x_2\cdot x_5)+x_6)]]] \land [[x_2 \neq 0] \lor [x_5 = 0]]] \land [[[[(x_2\cdot x_5) \neq x_4] \land [(x_2\cdot x_5) \not< x_4]] \lor [x_4 = ((x_2\cdot x_5)+x_3)]] \land [[x_4 \not< (x_2\cdot x_5)] \lor [x_3 = 0]]]]]]] \land [(\exists x_4 < (x_0+1))[[[[0 \not< \\
(x_2\cdot x_2)] \lor [(\exists x_5 < (x_2\cdot x_2))[x_0 = (((x_2\cdot x_2)\cdot x_4)+x_5)]]] \land [[(x_2\cdot x_2) \neq 0] \lor [x_4 = 0]]] \land [(\exists x_5 < (x_4+1))[[[[0 \not< (x_2\cdot x_2)] \lor [(\exists x_6 < (x_2\cdot x_2))[x_4 = (((x_2\cdot x_2)\cdot x_5)+x_6)]]] \land [[(x_2\cdot x_2) \neq 0] \lor [x_5 = 0]]] \land [[[[((x_2\cdot x_2)\cdot x_5) \neq \\
x_4] \land [((x_2\cdot x_2)\cdot x_5) \not< x_4]] \lor [x_4 = (((x_2\cdot x_2)\cdot x_5)+(((1+1)\cdot x_3)+1))]] \land [[x_4 \not< ((x_2\cdot x_2)\cdot x_5)] \lor [(((1+1)\cdot x_3)+1) = 0]]]]]]]]] \land [(\exists x_3 < (x_1+1))[[(\exists x_4 < (x_1+1))[[[[0 \not< x_2] \lor [(\exists x_5 < x_2)[x_1 = ((x_2\cdot x_4)+x_5)]]] \land [[x_2 \neq \\
0] \lor [x_4 = 0]]] \land [(\exists x_5 < (x_4+1))[[[[0 \not< x_2] \lor [(\exists x_6 < x_2)[x_4 = ((x_2\cdot x_5)+x_6)]]] \land [[x_2 \neq 0] \lor [x_5 = 0]]] \land [[[[(x_2\cdot x_5) \neq x_4] \land [(x_2\cdot x_5) \not< x_4]] \lor [x_4 = ((x_2\cdot x_5)+x_3)]] \land [[x_4 \not< (x_2\cdot x_5)] \lor [x_3 = 0]]]]]]] \land [(\exists x_4 < (x_1+1))[[[[0 \not< \\
(x_2\cdot x_2)] \lor [(\exists x_5 < (x_2\cdot x_2))[x_1 = (((x_2\cdot x_2)\cdot x_4)+x_5)]]] \land [[(x_2\cdot x_2) \neq 0] \lor [x_4 = 0]]] \land [(\exists x_5 < (x_4+1))[[[[0 \not< (x_2\cdot x_2)] \lor [(\exists x_6 < (x_2\cdot x_2))[x_4 = (((x_2\cdot x_2)\cdot x_5)+x_6)]]] \land [[(x_2\cdot x_2) \neq 0] \lor [x_5 = 0]]] \land [[[[((x_2\cdot x_2)\cdot x_5) \neq \\
x_4] \land [((x_2\cdot x_2)\cdot x_5) \not< x_4]] \lor [x_4 = (((x_2\cdot x_2)\cdot x_5)+((1+1)\cdot(x_3\cdot x_3)))]] \land [[x_4 \not< ((x_2\cdot x_2)\cdot x_5)] \lor [((1+1)\cdot(x_3\cdot x_3)) = 0]]]]]]]]]]]]]] \land [(\exists x_2 < ((a_1\cdot a_1)+1))[[x_2 \neq (1+1)] \land [[[[0 < x_2] \land [(\forall x_3 < (x_2+1))[[1 \not< \\
x_3] \lor [[(\forall x_4 < (x_2+1))[x_2 \neq (x_3\cdot x_4)]] \lor [(\exists x_4 < (x_3+1))[x_3 = ((1+1)\cdot x_4)]]]]]] \land [(\exists x_3 < ((1+1)\cdot x_2))[[[(\forall x_4 < (x_3+1))[[[[0 < 1] \land [(\forall x_5 < 1)[x_3 \neq ((1\cdot x_4)+x_5)]]] \lor [[1 = 0] \land [x_4 \neq 0]]] \lor [(\exists x_5 < (x_4+1))[x_4 = \\
((1+1)\cdot x_5)]]]] \land [[(\exists x_4 < (x_3+1))[[[[0 \not< (1+1)] \lor [(\exists x_5 < (1+1))[x_3 = (((1+1)\cdot x_4)+x_5)]]] \land [[(1+1) \neq 0] \lor [x_4 = 0]]] \land [(\forall x_5 < (x_4+1))[x_4 \neq ((1+1)\cdot x_5)]]]] \land [(\forall x_4 < (x_3+1))[[[0 \not< x_4] \lor [(\exists x_5 < (x_4+1))[[1 < \\
x_5] \land [[(\exists x_6 < (x_4+1))[x_4 = (x_5\cdot x_6)]] \land [(\forall x_6 < (x_5+1))[x_5 \neq ((1+1)\cdot x_6)]]]]]] \lor [[(1+1) \not< x_4] \lor [[[(\forall x_5 < (x_3+1))[[[[0 < x_4] \land [(\forall x_6 < x_4)[x_3 \neq ((x_4\cdot x_5)+x_6)]]] \lor [[x_4 = 0] \land [x_5 \neq 0]]] \lor [(\exists x_6 < (x_5+1))[x_5 = \\
((1+1)\cdot x_6)]]]] \lor [(\exists x_5 < (x_4+1))[[[[(x_5\cdot x_5) = x_4] \lor [(x_5\cdot x_5) < x_4]] \land [x_4 < ((x_5+1)\cdot(x_5+1))]] \land [[(x_5\cdot x_5) = x_4] \land [(\exists x_6 < (x_3+1))[[[[0 \not< x_5] \lor [(\exists x_7 < x_5)[x_3 = ((x_5\cdot x_6)+x_7)]]] \land [[x_5 \neq 0] \lor [x_6 = 0]]] \land [(\forall x_7 < \\
(x_6+1))[x_6 \neq ((1+1)\cdot x_7)]]]]]]]] \land [[(\forall x_5 < (x_4+1))[[[[(x_5\cdot x_5) \neq x_4] \land [(x_5\cdot x_5) \not< x_4]] \lor [x_4 \not< ((x_5+1)\cdot(x_5+1))]] \lor [[(x_5\cdot x_5) \neq x_4] \lor [(\forall x_6 < (x_3+1))[[[[0 < x_5] \land [(\forall x_7 < x_5)[x_3 \neq ((x_5\cdot x_6)+x_7)]]] \lor [[x_5 = 0] \land [x_6 \neq \\
0]]] \lor [(\exists x_7 < (x_6+1))[x_6 = ((1+1)\cdot x_7)]]]]]]] \lor [(\exists x_5 < (x_3+1))[[[[0 \not< x_4] \lor [(\exists x_6 < x_4)[x_3 = ((x_4\cdot x_5)+x_6)]]] \land [[x_4 \neq 0] \lor [x_5 = 0]]] \land [(\forall x_6 < (x_5+1))[x_5 \neq ((1+1)\cdot x_6)]]]]]]]]]]] \land [(\exists x_4 < (x_3+1))[[[[0 \not< x_2] \lor [(\exists x_5 < \\
x_2)[x_3 = ((x_2\cdot x_4)+x_5)]]] \land [[x_2 \neq 0] \lor [x_4 = 0]]] \land [(\forall x_5 < (x_4+1))[x_4 \neq ((1+1)\cdot x_5)]]]]]]] \land [[(\exists x_3 < (x_0+1))[[[[0 \not< x_2] \lor [(\exists x_4 < x_2)[x_0 = ((x_2\cdot x_3)+x_4)]]] \land [[x_2 \neq 0] \lor [x_3 = 0]]] \land [(\exists x_4 < (x_3+1))[[[[0 \not< x_2] \lor [(\exists x_5 < \\
x_2)[x_3 = ((x_2\cdot x_4)+x_5)]]] \land [[x_2 \neq 0] \lor [x_4 = 0]]] \land [[[[(x_2\cdot x_4) \neq x_3] \land [(x_2\cdot x_4) \not< x_3]] \lor [x_3 = ((x_2\cdot x_4)+a_0)]] \land [[x_3 \not< (x_2\cdot x_4)] \lor [a_0 = 0]]]]]]] \land [(\exists x_3 < (x_1+1))[[[[0 \not< x_2] \lor [(\exists x_4 < x_2)[x_1 = ((x_2\cdot x_3)+x_4)]]] \land [[x_2 \neq \\
0] \lor [x_3 = 0]]] \land [(\exists x_4 < (x_3+1))[[[[0 \not< x_2] \lor [(\exists x_5 < x_2)[x_3 = ((x_2\cdot x_4)+x_5)]]] \land [[x_2 \neq 0] \lor [x_4 = 0]]] \land [[[[(x_2\cdot x_4) \neq x_3] \land [(x_2\cdot x_4) \not< x_3]] \lor [x_3 = ((x_2\cdot x_4)+a_1)]] \land [[x_3 \not< (x_2\cdot x_4)] \lor [a_1 = 0]]]]]]]]]]]]]]]
\end{array}$$

**Fig. 1.** Explicit presentation of a $\Delta_0$-graph of exponential function of our construction.

The $\mathsf{I}\Sigma_1$ version of the Knaster–Tarski theorem, stated below, is useful for defining recursively defined structures over $\mathbf{V}$ with appropriate complexity.

[9] In many proofs, including ours, derivability condition **D3** is established using formalized $\Sigma_1$-completeness. Whether this principle holds in $\mathsf{S}^1_2$ remains an open problem [20]. One must therefore prove the sharper formalized $\Sigma^{\mathrm{b}}_1$-completeness theorem.

**Theorem 2.15 (Version of the Knaster–Tarski theorem)** Let $\Phi : \mathcal{P}(\mathbf{V}) \to \mathcal{P}(\mathbf{V})$ be a class-valued function. Assume that $\Phi$ satisfies the following conditions.

**Definability** A predicate $P(x, c) :\equiv x \in \Phi(\{z \mid z \in c\})$ is $\Delta_1$-definable with parameters.
**Monotonicity** $\boldsymbol{C} \subseteq \boldsymbol{C}'$ implies $\Phi(\boldsymbol{C}) \subseteq \Phi(\boldsymbol{C}')$.
**Finiteness** If $x \in \Phi(\boldsymbol{C})$ holds, then $x \in \Phi(\{z \in \boldsymbol{C} \mid z < m\})$ holds for some $m \in \mathbf{V}$.

Then we have a $\Sigma_1$ class $\mathbf{Fix}_\Phi$ such that

$$\Phi(\mathbf{Fix}_\Phi) = \mathbf{Fix}_\Phi$$

Additionally, if it satisfies the following condition, $\mathbf{Fix}_\Phi$ is $\Delta_1$.

**Strong finiteness** If $x \in \Phi(\boldsymbol{C})$ holds, then $x \in \Phi(\{z \in \boldsymbol{C} \mid z < x\})$ holds.

$\mathbf{Fix}_\Phi$ satisfies the following structural induction principle.

**Theorem 2.16 (Structural induction)** Assume that $\Phi$ satisfies the strong finiteness property. The predicate $\mathbf{Fix}_\Phi$ above satisfies the induction principle of the following form. Let $\psi$ be a $\Sigma_1$ or $\Pi_1$-predicate (which may contain parameters from $\mathbf{V}$):

$$\forall \boldsymbol{C} \subseteq \mathbf{Fix}_\Phi\, [\forall x \in \boldsymbol{C}\, \psi(x) \to \forall x \in \Phi(\boldsymbol{C})\, \psi(x)] \quad \text{implies} \quad \forall x \in \mathbf{Fix}_\Phi\, \psi(x)$$

A practically important point is that the defining formulas of $\mathbf{Fix}_\Phi$ can be explicitly constructed from the defining formula of $P(x, c)$. In the mechanization, we first call such a formula a `Blueprint k` ( `k` is the number of parameters). Obviously it is purely syntactic and independent of any model. We then define `Construction V φ`, the model-theoretic realization of a `φ : Blueprint k`. Our mechanization of Theorem 2.15 and Theorem 2.16 is stated with these two parameters.

```
structure Blueprint (k : ℕ) where
  core : Δ₁.Semisentence (k + 2)

structure Construction {k : ℕ} (φ : Blueprint k) where
  Φ : (Fin k → V) → Set V → V → Prop
  defined : Δ₁.Defined (fun v ↦ Φ (v ·.succ.succ) {x | x ∈ v 1} (v 0)) φ.core
  monotone {C C' : Set V} (h : C ⊆ C') {v x} : Φ v C x → Φ v C' x

class Construction.Finite {k : ℕ} {φ : Blueprint k} (c : Construction V φ) where
  finite {C : Set V} {v x} : c.Φ v C x → ∃ m, c.Φ v {y ∈ C | y < m} x

class Construction.StrongFinite {k : ℕ} {φ : Blueprint k} (c : Construction V φ) where
  strong_finite {C : Set V} {v x} : c.Φ v C x → c.Φ v {y ∈ C | y < x} x

variable (c : Construction V φ)

def Construction.Fixpoint (v) (x : V) : Prop

theorem Construction.case [c.Finite] : c.Fixpoint v x ↔ c.Φ v {z | c.Fixpoint v z} x

theorem Construction.induction [c.StrongFinite] {P : V → Prop} (hP : Γ-[1]-Predicate P)
    (H : ∀ C : Set V, (∀ x ∈ C, c.Fixpoint v x ∧ P x) → ∀ x, c.Φ v C x → P x) :
    ∀ x, c.Fixpoint v x → P x
```

Source: 1

Syntactic structures such as terms, formulas, and proofs are all recursively generated and can therefore be constructed using `Blueprint` and `Construction`. Moreover, because these structures are generated well-foundedly, they satisfy the strong finiteness property. It follows uniformly that the corresponding predicates are $\Delta_1$-definable and satisfy the structural induction principles. These facts immediately yield definitions over $\mathbf{V}$ of basic syntactic operations such as substitution.

```
variable {L : Language} [L.Encodable] [L.LORDefinable]

instance IsSemiformula.definable : Δ₁-Relation[V] (IsSemiformula L)

instance Proof.definable {T : Theory L} [T.Δ₁] : Δ₁-Relation[V] (Proof T)

def Provable (φ : V) : Prop := ∃ d, Proof T d φ

instance Provable.definable : Σ₁-Predicate[V] Provable T
```

SOURCE: 1 2

We can routinely verify that the predicate `provabilityPred` is a provability predicate in the sense of Definition 2.5, and moreover that it satisfies the derivability conditions **D1**, **D2**, **D3**, and **Kre**.

```
variable {L : Language} [L.Encodable] [L.LORDefinable]

/-- Hilbert-Bernays provability condition D1 -/
theorem internalize_provability {φ} : T ⊢ φ → Provable T (⌜φ⌝ : V)

/-- Hilbert-Bernays provability condition D2 -/
theorem modus_ponens {φ ψ : Proposition L}
    (hφψ : Provable T (⌜φ → ψ⌝ : V)) (hφ : Provable T (⌜φ⌝ : V)) :
    Provable T (⌜ψ⌝ : V)

/-- Hilbert-Bernays provability condition D3 -/
theorem sigma_one_complete {σ : ArithmeticSentence} (hσ : Hierarchy Σ 1 σ) :
    V↓[ℒₒᵣ] ⊨ σ → Provable T (⌜σ⌝ : V)

theorem provable_internalize {σ : ArithmeticSentence} :
    Provable T (⌜σ⌝ : V) → Provable T (⌜provabilityPred T σ⌝ : V)

noncomputable abbrev Theory.standardProvability : Provability IΣ₁ T where
  prov := provable T
  bew_def := provable_D1

instance : T.standardProvability.HBL2 := ⟨provable_D2⟩

instance [PA⁻ ⪯ T] : T.standardProvability.HBL3 := ⟨provable_D3⟩
```

SOURCE: 1 2 3 4

On the other hand, making Proposition 2.8 concrete requires the theory to be diagonalizable (Definition 2.6). Since $T \supseteq \mathsf{I}\Sigma_1$, the fixpoint theorem holds.

**Theorem 2.17 (Fixpoint theorem)** Suppose $T \supseteq \mathsf{I}\Sigma_1$. For any unary arithmetical formula $\theta(x)$, one can construct an arithmetic sentence $\mathsf{fixedpoint}_\theta$ such that

$$T \vdash \mathsf{fixedpoint}_\theta \leftrightarrow \theta(\ulcorner \mathsf{fixedpoint}_\theta \urcorner)$$

Hence the theory $T$ is diagonalizable.

```
noncomputable def diag (θ : ArithmeticSemisentence 1) : ArithmeticSemisentence 1 :=
  “x. ∀ y, !ssnum y x x → !θ y”

noncomputable def fixedpoint (θ : ArithmeticSemisentence 1) : ArithmeticSentence :=
  (diag θ)/[⌜diag θ⌝]

theorem diagonal (θ : ArithmeticSemisentence 1) : T ⊢ fixedpoint θ ↔ θ/[⌜fixedpoint θ⌝]

noncomputable instance : Diagonalization IΣ₁ where
  fixedpoint := fixedpoint
  diag θ := diagonal θ
```

SOURCE: 1 2

Combining the results above, Proposition 2.8 immediately yields our final result: a mechanization of the second incompleteness theorem.

**Theorem 2.18 (Gödel's second incompleteness theorem [46])** Let $T$ be a $\Delta_1$-definable arithmetic theory stronger than $\mathsf{I}\Sigma_1$, and let $\mathsf{Con}_T$ be the consistency statement $\neg\mathsf{Pr}_T(\ulcorner\bot\urcorner)$ of $T$, where $\mathsf{Pr}_T(x)$ is the standard provability predicate of $T$ constructed above.
1. If $T$ is consistent, then $T \nvdash \mathsf{Con}_T$.
2. If $T$ is $\Sigma_1$-sound, then $T \nvdash \neg\mathsf{Con}_T$. Hence $\mathsf{Con}_T$ is independent of $T$.

```
/-- Gödel's second incompleteness theorem -/
theorem consistent_unprovable [Consistent T] : T ⊬ T.consistent.val

theorem inconsistent_unprovable [ArithmeticTheory.SoundOnHierarchy T 𝚺 1] : T ⊬ ∼T.consistent.val
```

Source: 1

## 2.6 Some further results related to the incompleteness theorems

Using the tools developed so far, we have also proved several theorems related to Gödel's incompleteness theorems.

### 2.6.1 Variants of fixedpoint lemma

The following fixpoint theorems, which generalize Theorem 2.17, also hold; see [27] for the proofs. Although we omit the details, they are needed when we establish arithmetical completeness in Section 3. Throughout this subsection, we assume $T \supseteq \mathsf{I}\Sigma_1$.

**Theorem 2.19** For any family $\{\theta_i\}_{i<k}$ of $k$-ary arithmetical formulas $\theta_i(x_0, ..., x_{k-1})$, one can construct arithmetic sentences $\mathsf{fixedpoint}_0, ..., \mathsf{fixedpoint}_{k-1}$ such that, for every $i < k$,

$$T \vdash \mathsf{fixedpoint}_i \leftrightarrow \theta_i(\ulcorner \mathsf{fixedpoint}_0 \urcorner, ..., \ulcorner \mathsf{fixedpoint}_{k-1} \urcorner)$$

```
noncomputable def multifixedpoint (θ : Fin k → ArithmeticSemisentence k) (i : Fin k)
  : ArithmeticSentence := ...

theorem multidiagonal (θ : Fin k → ArithmeticSemisentence k)
  : T ⊢ multifixedpoint θ i ⭤ (Rew.subst fun j ↦ ⌜multifixedpoint θ j⌝) ▹ (θ i)
```

Source: 1

**Theorem 2.20** For any $(k+1)$-ary arithmetical formula $\theta(x, \vec{y})$, one can construct a $k$-ary arithmetical formula $\mathsf{fixedpoint}_\theta(\vec{y})$ such that

$$T \vdash \forall \vec{y}, (\mathsf{fixedpoint}_\theta(\vec{y}) \leftrightarrow \theta(\ulcorner \mathsf{fixedpoint}_\theta \urcorner, \vec{y}))$$

```
noncomputable def parameterizedFixedpoint (θ : ArithmeticSemisentence (k + 1))
  : ArithmeticSemisentence k := ...

theorem parameterized_diagonal (θ : ArithmeticSemisentence (k + 1))
  : T ⊢ ∀¹* (parameterizedFixedpoint θ ⭤ “!θ !!(⌜parameterizedFixedpoint θ⌝) ⋯”)
```

Source: 1

**2.6.2 Löb's theorem** Instantiating the abstract version stated in Proposition 2.9, we immediately obtain the concrete Löb's theorem. As related work, Bailitis has mechanized Löb's theorem both in Isabelle, on top of Paulson's mechanization of the incompleteness theorems (see [111:Chapter 13]), and in Rocq [15].

**Theorem 2.21 (Löb's theorem and formalized Löb's theorem [91])** Let $T \supseteq \mathsf{I}\Sigma_1$ be a $\Delta_1$-definable theory and let $\sigma$ be any sentence.

**Löb's theorem** If $T \vdash \mathsf{Pr}_T(\ulcorner \sigma \urcorner) \to \sigma$, then $T \vdash \sigma$.
**Formalized Löb's theorem** $\mathsf{I}\Sigma_1 \vdash \mathsf{Pr}_T(\ulcorner \mathsf{Pr}_T(\ulcorner \sigma \urcorner) \to \sigma \urcorner) \to \mathsf{Pr}_T(\ulcorner \sigma \urcorner)$.

```
variable {T : ArithmeticTheory} [T.Δ₁] [IΣ₁ ⪯ T] {σ : ArithmeticSentence}

theorem löb_theorem : T ⊢ provabilityPred T σ ➝ σ → T ⊢ σ

theorem formalized_löb_theorem : IΣ₁ ⊢ provabilityPred T (provabilityPred T σ ➝ σ) ➝
provabilityPred T σ
```

Source: 1

**2.6.3 Tarski's undefinability theorem** As a corollary of the fixpoint theorem, we can prove Tarski's theorem on the undefinability of truth. First, we prove the following lemma.

**Lemma 2.22** Let $T \supseteq \mathsf{I}\Sigma_1$ be a consistent theory. Then there is no unary formula $\tau(x)$ such that $T \vdash \sigma \leftrightarrow \tau(\ulcorner \sigma \urcorner)$ for every sentence $\sigma$.

```
lemma not_exists_tarski_predicate {T : ArithmeticTheory} [IΣ₁ ⪯ T] [Consistent T] : ¬∃ τ :
ArithmeticSemisentence 1, ∀ σ, T ⊢ σ ⭤ τ/[⌜σ⌝]
```

Source: 1

Taking as $T$ the *true arithmetic* $\mathsf{TA}$, the theory of all sentences true in $\mathbb{N}$, we immediately obtain the desired theorem.

**Theorem 2.23 (Tarski's undefinability theorem [146])** There is no truth predicate $\mathsf{True}(x)$ such that $\mathbb{N} \models \sigma$ if and only if $\mathbb{N} \models \mathsf{True}(\ulcorner \sigma \urcorner)$ for every sentence $\sigma$.

```
theorem undefinability_of_truth : ¬∃ τ : ArithmeticSemisentence 1, ∀ σ : ArithmeticSentence,
ℕ↓[ℒₒᵣ] ⊧ σ ⭤ ℕ↓[ℒₒᵣ] ⊧ τ/[⌜σ⌝]
```

Source: 1

In contrast to this theorem, it is known that for a complexity class $\Gamma$ of formulas, there is a partial truth predicate $\mathsf{True}_\Gamma(x)$, obtained by replacing "for any sentence" with "for any $\Gamma$-sentence" in the definition of $\mathsf{True}(x)$, which is itself definable by a $\Gamma$-formula (cf. [57]). This fact has not been mechanized yet; consequently, several statements of provability logic proved via partial truth predicates remain unmechanized, as we discuss further in Section 3.6.

**2.6.4 Church's theorem and undecidability of first-order logic** In Mathlib, computability of a predicate (at the meta level of Lean) is defined by `ComputablePred` (cf. [33]), which requires the types of the domain and the range to be `Primcodable`[10]. Since the type of formulas of our arithmetic is `Primcodable` via a suitable encoding, we can ask whether the set $\mathrm{Thm}(T)$ of sentences provable from a theory $T$ is computable. This yields the following theorem, commonly known as Church's theorem.

**Theorem 2.24 (Church's theorem (for $\Sigma_1$-sound theories))** For a $\Sigma_1$-sound theory $T \supseteq \mathsf{R}_0$, $\mathrm{Thm}(T)$ is not computable.

```
theorem uncomputable_theory_of_sigma1Sound {T : ArithmeticTheory} [R₀ ⪯ T] [T.SoundOnHierarchy 𝚺 1]
: ¬ComputablePred T.theory
```

Source: 1

Now take as $T$ the theory $\mathsf{PA}^-$, a finitely axiomatized $\Sigma_1$-sound fragment of $\mathsf{PA}$ stronger than $\mathsf{R}_0$. Since its axioms can be conjoined into a single sentence, the deduction theorem applies, and the undecidability of first-order logic over the language of arithmetic follows.

**Theorem 2.25 (Undecidability of first-order logic)** First-order logic over the language $\mathcal{L}_{\mathrm{OR}}$ is not computable. That is, there is no algorithm that decides, for a given $\mathcal{L}_{\mathrm{OR}}$-sentence $\sigma$, whether $\varnothing \vdash \sigma$ or $\varnothing \nvdash \sigma$.

```
theorem undecidability_first_order_logic : ¬ComputablePred ((∅ : ArithmeticTheory).theory)
```

Source: 1

For the speed-up theorem (Theorem 2.38) below, we have also mechanized a version that weakens $\Sigma_1$-soundness to mere consistency, at the cost of strengthening the base theory from $\mathsf{R}_0$ to $\mathsf{I}\Sigma_1$.

**Theorem 2.26 (Church's theorem (for consistent theories))** For a consistent theory $T \supseteq \mathsf{I}\Sigma_1$, $\mathrm{Thm}(T)$ is not computable.

```
theorem uncomputable_theory_of_consistent {T : ArithmeticTheory} [𝐈𝚺₁ ⪯ T] [Entailment.Consistent
T]
: ¬ComputablePred T.theory
```

Source: 1

Note that this version does not directly apply to the above proof of the undecidability of first-order logic: that proof requires the theory to be given by a single sentence, whereas $\mathsf{I}\Sigma_1$, naïvely presented by an infinite axiom scheme, is not finitely axiomatized as it stands (although it is in fact finitely axiomatizable).

**2.6.5 Gödel–Rosser first incompleteness theorem** In the setting of Theorem 2.3, the theory $T$ was required to be $\Sigma_1$-sound. By instantiating Proposition 2.10, we can prove the Gödel–Rosser incompleteness theorem [118], which weakens this requirement to mere consistency.

**Theorem 2.27 (Gödel–Rosser first incompleteness theorem [118])** Let $T \supseteq \mathsf{I}\Sigma_1$ be a $\Delta_1$-definable and consistent theory. Then $T$ is incomplete.

```
variable {T : Theory L} [T.Δ₁] [Entailment.Consistent T]

noncomputable abbrev Theory.rosserProvability : Provability 𝐈𝚺₁ T where
```

[10] That is, encoding into and decoding from natural numbers are primitive recursive.

```
  prov := T.rosserProvable
  bew_def := rosserProvable_D1

instance : T.rosserProvability.Rosser := ⟨rosserProvable_rosser⟩

theorem incomplete_GR (T : ArithmeticTheory) [T.Δ₁] [𝐈𝚺₁ ≼ T] [Entailment.Consistent T] :
Entailment.Incomplete T
```

NOTE: The assumption `[T.SoundOnHierarchy 𝚺 1]` in the mechanization of Theorem 2.3 is replaced by `[Entailment.Consistent T]`.

SOURCE: 1

The required provability predicate satisfying **Ros** is constructed by so-called *witness comparison* (see [57, 89]); we omit the details here. Note also that this proof internally uses the fixpoint theorem (Theorem 2.17); hence, unlike Theorem 2.3, we assume that $T$ extends $\mathsf{I}\Sigma_1$, which is stronger than $\mathsf{R}_0$. Similarly, Corollary 2.4 can also be strengthened; we omit the statement.

### 2.6.6 Craig's trick, soundness and definability

Moreover, by the technique known as *Craig's trick* below, the requirement of $\Delta_1$-definability can also be weakened to being r.e. As explained in connection with Church's theorem, a natural encoding of formulas and the like into $\mathbb{N}$ is implemented in Lean; hence we may simply define a theory $T$ to be r.e. when the predicate $\sigma \in T$ is r.e.

**Theorem 2.28 (Craig's trick)** For an r.e. theory $T$, one can construct a primitive recursive theory $T^{\mathrm{C}}$ equivalent to $T$; that is, $T \vdash \sigma \iff T^{\mathrm{C}} \vdash \sigma$ for every sentence $\sigma$. In particular, $T^{\mathrm{C}}$ is consistent if $T$ is, and $T^{\mathrm{C}}$ is incomplete if $T$ is.

```
class Theory.RE (T : Theory L) : Prop where
  re : REPred (· ∈ T)

class Theory.Primrec (T : Theory L) : Prop where
  primrec : PrimrecPred (· ∈ T)

instance {T : Theory L} [T.Primrec] : T.RE

def Theory.craig (T : Theory L) [T.RE] : Theory L

instance : T.craig.Primrec

noncomputable instance : (T.craig).Δ₁

instance : T ≊ T.craig

instance [Consistent T] : Consistent T.craig
```

NOTE: Here `T ≊ T.craig` expresses that the two theories are equivalent.

SOURCE: 1 2

By this trick, Theorem 2.27 and Corollary 2.4 can be strengthened further. This is one of the strongest (i.e., with the weakest assumptions) statements of G1 that we have mechanized[11].

**Theorem 2.29 (G1 for r.e. theories)** Let $T \supseteq \mathsf{I}\Sigma_1$ be an r.e. and consistent theory. Then $T$ is incomplete. Moreover, there also exists a sentence that is true but unprovable in $T$.

[11] Since the assumption is changed to extending $\mathsf{I}\Sigma_1$, it is not comparable with Theorem 2.3, which holds for any extension of $\mathsf{R}_0$.

```
theorem incomplete_GR_of_RE (T : ArithmeticTheory) [T.RE] [𝐈𝚺₁ ⪯ T] [Consistent T] : Incomplete T

theorem exists_true_but_unprovable_sentence_of_RE_of_consistent
  (T : ArithmeticTheory) [T.RE] [𝐈𝚺₁ ⪯ T] [Consistent T] :
  ∃ δ : ArithmeticSentence, ℕ↓[ℒₒᵣ] ⊧ δ ∧ T ⊬ δ
```

Source: 1

As for G2, at present it can only be stated in the following form, because of an issue with coding in the arithmetization.

**Theorem 2.30 (G2 for r.e. theories)** Let $T$ be an r.e., consistent arithmetic theory stronger than $\mathsf{I}\Sigma_1$. Then $T \nvdash \mathsf{Con}_{T^{\mathrm{C}}}$, i.e. $T$ cannot prove the consistency statement of $T^{\mathrm{C}}$.

```
theorem craig_consistent_unprovable_of_RE (T : ArithmeticTheory) [T.RE] [𝐈𝚺₁ ⪯ T] [Consistent T]
: T ⊬ T.craig.consistent.val
```

Source: 1

**Remark 2.31** To restate this in terms of the consistency of $T$ itself, one would have to mechanize $T \vdash \forall x\, [\mathsf{Pr}_T(x) \leftrightarrow \mathsf{Pr}_{T^{\mathrm{C}}}(x)]$. This, however, requires delicate adjustments, such as making $\mathsf{Pr}_T(x)$ codable for theories that are merely r.e. rather than $\Delta_1$-definable, as well as the cumbersome task of formalizing Craig's trick itself within arithmetic and carrying it out there; we have therefore not mechanized it yet.

To instantiate the theorems and corollaries stated in this paper with a concrete theory $T$ such as $\mathsf{I}\Sigma_1$ or $\mathsf{PA}$, the $\Sigma_1$-soundness (and hence consistency) and the recursive enumerability of these theories must themselves be mechanized. We have done this as well.

**Proposition 2.32** $\mathsf{I}\Sigma_1$ and $\mathsf{PA}$ are $\Sigma_1$-sound, hence consistent.

```
instance sigmaOneSound_ISigmaOne : 𝐈𝚺₁.SoundOnHierarchy 𝚺 1

instance sigmaOneSound_Peano : 𝐏𝐀.SoundOnHierarchy 𝚺 1

instance (T : ArithmeticTheory) [T.SoundOnHierarchy 𝚺 1] : Entailment.Consistent T
```

Source: 1 2

The following also holds. Here $\Delta_1$-definability is stated only because of the issue pointed out in Remark 2.31, and is not essential; in ordinary mathematics one may always replace an r.e. theory by an equivalent $\Delta_1$-definable one via Craig's trick.

**Proposition 2.33** $\mathsf{I}\Sigma_1$ and $\mathsf{PA}$ are r.e. and $\Delta_1$-definable.

```
instance : 𝐏𝐀.RE

instance : 𝐈𝚺₁.RE

noncomputable instance : 𝐏𝐀.Δ₁

noncomputable instance : 𝐈𝚺₁.Δ₁
```

Source: 1

Hence all the theorems mechanized here can indeed be instantiated with concrete theories such as $\mathsf{I\Sigma}_1$ and $\mathsf{PA}$.

#### 2.6.7 Jeroslow's second incompleteness theorem

Similarly, from Proposition 2.13, we can also concretely mechanize Jeroslow's second incompleteness theorem [71]. We mention that Popescu and Traytel [115:Theorem 30] mechanized Jeroslow's theorem only at the abstract level.

**Theorem 2.34 (Jeroslow's second incompleteness theorem [71])** Let $T \supseteq \mathsf{I\Sigma}_1$ be a $\Delta_1$-definable and consistent theory. Then $T \nvdash \forall x.\neg(\mathsf{Pr}_T(x) \land \mathsf{Pr}_T(\dot\neg x))$, where $\dot\neg$ denotes the function taking the Gödel number of a sentence to that of its negation, i.e., $\dot\neg\ulcorner\sigma\urcorner = \ulcorner\neg\sigma\urcorner$.

```
theorem unprovable_formalized_law_of_noncontradiction {T : ArithmeticTheory} [T.Δ₁] [𝐈𝚺₁ ⪯ T]
[Entailment.Consistent T]
: T ⊬ (∀¹ ∼(provable T ⋏ T.refutable))
```

Source: 1

#### 2.6.8 FGH Theorem

Using machinery similar to the witness comparison used in the proof of the Gödel–Rosser theorem, the following theorem can be mechanized easily. The FGH theorem, due to Friedman–Goldfarb–Harrington (see: [157:Section 3]), states that over $\mathsf{I\Sigma}_1 + \mathsf{Con}_T$ every $\Sigma_1$-sentence is equivalent to one of the form $\mathsf{Pr}_T(\ulcorner\pi\urcorner)$.

**Theorem 2.35 (FGH Theorem)** Let $T \supseteq \mathsf{I\Sigma}_1$ be $\Delta_1$-definable theory. For any $\Sigma_1$-sentence $\sigma$, there exists a $\Sigma_1$-sentence $\pi$ satisfying the following:

$$\mathsf{I\Sigma}_1 + \mathsf{Con}_T \vdash \sigma \leftrightarrow \mathsf{Pr}_T(\ulcorner\pi\urcorner)$$

```
theorem fgh_theorem_con (hσ : Hierarchy 𝚺 1 σ) :
  ∃ π : 𝚺₁.Sentence, 𝐈𝚺₁ ∪ T.Con ⊢ σ ⭤ provabilityPred T π.val := by
```

Source: 1

#### 2.6.9 On proof size

Formalization also allows us to discuss provability by a proof of *feasible* length or complexity in a certain sense.

**Theorem 2.36** Let $T \supseteq \mathsf{I\Sigma}_1$ be a $\Delta_1$-definable and $\Sigma_1$-sound theory, let $f$ be a $\Sigma_1$-definable function, and let $e$ be an arbitrary natural number. Then we can construct the *restricted provability predicate* $\mathsf{Pr}_T^{\langle f,e\rangle}(x)$, a further restriction of the provability predicate expressing that "provable by a $T$-proof whose Gödel number is less than $f(e)$". As with the usual Gödel sentence, let $\mathsf{G}_T^{\langle f,e\rangle}$ be a fixpoint of $\neg\mathsf{Pr}_T^{\langle f,e\rangle}(x)$.

Then $\mathbb{N} \models \mathsf{G}_T^{\langle f,e\rangle}$ and $T \vdash \mathsf{G}_T^{\langle f,e\rangle}$, but every $T$-proof of $\mathsf{G}_T^{\langle f,e\rangle}$ has code at least $f(e)$.

```
variable {T : ArithmeticTheory} [T.Δ₁] [𝐈𝚺₁ ⪯ T] [T.SoundOnHierarchy 𝚺 1]
         {fDef : 𝚺₁.Semisentence 2} {e : ℕ}

theorem true_restrictedGödel (f : ℕ → ℕ) [𝚺₁-Function₁ f via fDef] :
  ℕ↓[ℒₒᵣ] ⊧ T.restrictedGödel fDef e

theorem provable_restrictedGödel (f : ℕ → ℕ) [𝚺₁-Function₁ f via fDef] :
  T ⊢ T.restrictedGödel fDef e
```

```
theorem lower_bound_gödelNumber_proof_restrictedGödel (f : ℕ → ℕ) [Σ₁-Function₁ f via fDef] :
  ∀ b : T ⊢! T.restrictedGödel fDef e, f (ORingStructure.numeral e) ≤ ⌜b⌝
```

NOTE: To use these theorems, one must supply both a concrete function `f` and a $\Sigma_1$-formula `fDef` representing it; the instance `[Σ₁-Function₁ f via fDef]` states that `fDef` actually defines `f`. `T ⊢! T.restrictedGödel fDef e` denotes the `Type` of "$T$-proofs of $\mathsf{G}_T^{(f,e)}$" (not the `Prop` of provability).

SOURCE: 1

As a concrete example of such an $f$, we can take the superexponential function $\mathsf{supexp}$[12], formalized over $\mathsf{I}\Sigma_1$, which yields the following corollary.

**Corollary 2.37** Let $T \supseteq \mathsf{I}\Sigma_1$ be a $\Delta_1$-definable and $\Sigma_1$-sound theory, and let $e$ be an arbitrary natural number. Then $T \vdash \mathsf{G}_T^{(\mathsf{supexp},e)}$, but every $T$-proof of $\mathsf{G}_T^{(\mathsf{supexp},e)}$ has code at least $\mathsf{supexp}(e)$.

```
variable {T : ArithmeticTheory} [T.Δ₁] [𝐈𝚺₁ ⪯ T] [T.SoundOnHierarchy 𝚺 1] {e : ℕ}

theorem provable_restrictedGödel_superexp : T ⊢ T.restrictedGödel superexpDef e

theorem lower_bound_gödelNumber_proof_restrictedGödel_superexp :
  ∀ b : T ⊢! T.restrictedGödel superexpDef e, Superexp.superexp e ≤ ⌜b⌝
```

SOURCE: 1 2

In our mechanization, coding an $n$-character formula or proof yields a Gödel number roughly of the order of $2^n$. Hence, taking as $f$ the far faster-growing $\mathsf{supexp}$ and taking $e$ extremely large, say $10^9$, this corollary suggests that there are true sentences which are provable in $T$ but admit no $T$-proof that a human could ever *practically* write down, or even read.

Furthermore, we have also mechanized the speed-up theorem due to Ehrenfeucht–Mycielski [37][13].

**Theorem 2.38 (Ehrenfeucht–Mycielski speed-up theorem [37])** Let $\min_T(\sigma)$ be the least Gödel number of a $T$-proof of $\sigma$ if $T \vdash \sigma$, and 0 if $T \nvdash \sigma$.

Let $T \supseteq \mathsf{I}\Sigma_1$ be a $\Delta_1$-definable theory and take a sentence $\sigma$ with $T \nvdash \sigma$. Then, for any computable function $f$, there exists a sentence $\pi$ such that $T \vdash \pi$ and $f(\min_{T+\sigma}(\pi)) < \min_T(\pi)$.

For example, taking $f(x) = 2^{x+1}$, there is a sentence $\pi$ with $\min_{T+\sigma}(\pi) < \log_2 \min_T(\pi)$: adding $\sigma$ as an axiom shrinks its proof to logarithmic order.

```
noncomputable def Theory.minProof (T : Theory L) [T.Δ₁] (σ : Sentence L) : ℕ
  := sInf (Set.range λ d : T ⊢!₂! (σ : Proposition L) ↦ (⌜d⌝ : ℕ))

theorem ehrenfeucht_mycielski_speedup {T : Theory L} [T.Δ₁] {σ : Sentence L} [L.Primcodable]
  (hU : ¬ComputablePred (insert (∼σ) T).theory) (f : ℕ → ℕ) (hf : Computable f) :
  ∃ π : Sentence L, T ⊢ π ∧ f ((insert σ T).minProof π) < T.minProof π

variable {T : ArithmeticTheory} [T.Δ₁] [𝐈𝚺₁ ⪯ T] {σ : ArithmeticSentence} (hσ : T ⊬ σ)

theorem ehrenfeucht_mycielski_speedup_arithmetic (f : ℕ → ℕ) (hf : Computable f) :
  ∃ π : ArithmeticSentence, T ⊢ π ∧ f ((insert σ T).minProof π) < T.minProof π

example : ∃ π : ArithmeticSentence, T ⊢ π ∧ (insert σ T).minProof π < Nat.log 2 (T.minProof π)
```

SOURCE: 1

[12]Define $2^y_x$ by $2^0_x = x$ and $2^{y+1}_x = 2^{2^y_x}$, and let $\mathsf{supexp}(x) = 2^x_x$. For example, $\mathsf{supexp}(2) = 16$ and $\mathsf{supexp}(3) = 2^{256}$; also $2^x \leq \mathsf{supexp}(x)$ holds for $x \geq 1$.

[13]Observations of this kind go back to Gödel [48].

**Remark 2.39** For a general, not necessarily arithmetic, theory $T$, the same conclusion follows assuming that provability in $T + \neg\sigma$ is not computable ( `ehrenfeucht_mycielski_speedup` above). In the arithmetic case, this assumption follows from Theorem 2.26, since $T \nvdash \sigma$ makes $T + \neg\sigma$ a consistent theory containing $\mathsf{I}\Sigma_1$.

Although one may question measuring the complexity of a proof simply by its Gödel number, such discussions of restricted provability are closely related to Parikh's feasibility [108] and to bounded arithmetic [31]. We consider these mechanizations to be a first step in that direction.

**2.6.10 Lindenbaum algebra** The *Lindenbaum algebra* $\mathfrak{A}_T$ of a theory $T$ is obtained by the usual construction quotienting sentences by the equivalence relation given by $T \vdash \sigma \leftrightarrow \pi$. We have also mechanized some results on these algebras: in particular, for a theory $T$ for which the Gödel–Rosser first incompleteness theorem holds, $\mathfrak{A}_T$ is a dense Boolean algebra.

**Theorem 2.40** Let $T \supseteq \mathsf{I}\Sigma_1$ be a $\Delta_1$-definable theory. Then the Lindenbaum algebra $\mathfrak{A}_T$ of $T$ is densely ordered: whenever $\varphi < \psi$ in $\mathfrak{A}_T$, there exists $\xi$ such that $\varphi < \xi < \psi$.

```
lemma dense (T : ArithmeticTheory) [IΣ₁ ≼ T] [T.Δ₁] {φ ψ : LindenbaumAlgebra T} :
    φ < ψ → ∃ ξ, φ < ξ ∧ ξ < ψ

instance (T : ArithmeticTheory) [IΣ₁ ≼ T] [T.Δ₁] : DenselyOrdered (LindenbaumAlgebra T)
```

Source: 1

Now, it is clear that the Lindenbaum algebra of an arithmetic theory is countable. Combined with the well-known fact that any two countable, dense, and nontrivial Boolean algebras are order isomorphic (cf. [58:Chapter 16][14]), we immediately obtain the following result.

**Theorem 2.41** For any $\Delta_1$-definable consistent theories $T, U \supseteq \mathsf{I}\Sigma_1$, the Lindenbaum algebras $\mathfrak{A}_T$ and $\mathfrak{A}_U$ are isomorphic.

```
theorem iso_of_countable_atomless {α β : Type*}
    [BooleanAlgebra α] [Countable α] [Nontrivial α] [DenselyOrdered α]
    [BooleanAlgebra β] [Countable β] [Nontrivial β] [DenselyOrdered β] :
    Nonempty (α ≃o β)

theorem lindenbaum_iso (T U : ArithmeticTheory)
    [IΣ₁ ≼ T] [T.Δ₁] [Consistent T] [IΣ₁ ≼ U] [U.Δ₁] [Consistent U] :
    Nonempty (LindenbaumAlgebra T ≃o LindenbaumAlgebra U)
```

Source: 1 2

That is, the Lindenbaum algebras of $\mathsf{I}\Sigma_1$, $\mathsf{PA}$, and even $\mathsf{ZF}$ (if consistent and although not mechanized) are all isomorphic; in this sense, the Lindenbaum algebra of a reasonable theory does not have a rich structure. By a theorem of Pour-El and Kripke [116], this isomorphism can moreover be taken to be recursive, but such a refinement has not been mechanized at present.

On the other hand, the algebras obtained by extending the Lindenbaum algebra with provability as an explicit unary operator are called *diagonalizable algebras* or *Magari algebras* (cf. [94, 132]). It is known, for example, that the diagonalizable algebras of $\mathsf{PA}$ and $\mathsf{ZF}$ are not isomorphic [133], and these algebras are deeply related to provability logic, which we discuss in Section 3. No mechanization of these algebras has been carried out at present.

[14] This fact is usually stated in terms of atomlessness, but our mechanization states it in terms of density, following Mathlib's `DenselyOrdered` class.

**2.6.11 Arithmetic Zoo** For visualization of our mechanized progress, we show the *Arithmetic Zoo* (Figure 2), which displays the relative strength of the theories established in our mechanization[15].

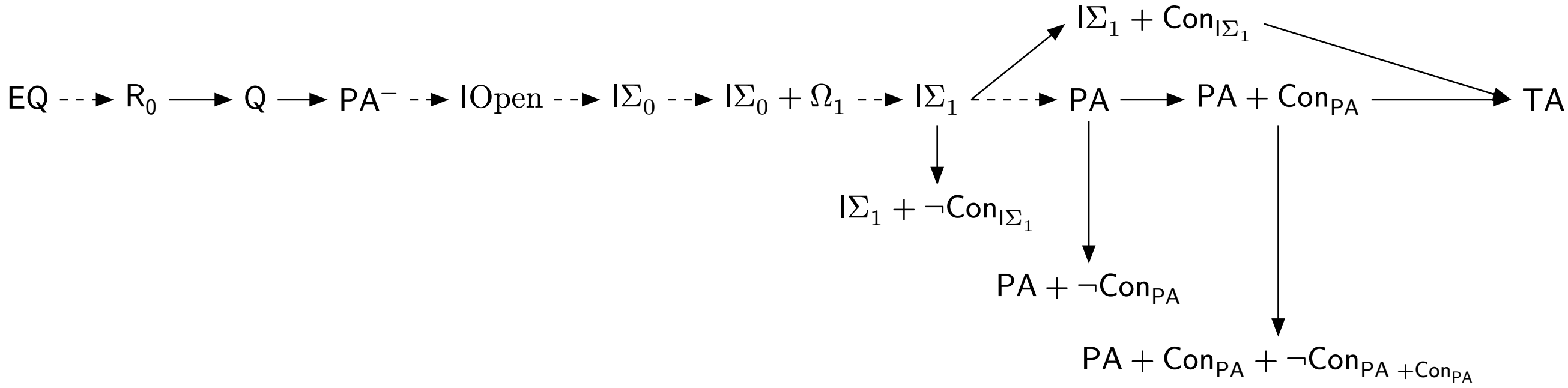

**Figure 2.** *Arithmetic theory zoo*: a dashed arrow ⇢ denotes inclusion, and → denotes proper inclusion.

In fact all of these inclusions are proper and some edges are missing; for instance, $\mathsf{I}\Sigma_1 + \mathsf{Con}_{\mathsf{I}\Sigma_1}$ is a subtheory of $\mathsf{PA}$. Mechanizing this properness and these missing edges is left for future work. Moreover, there are many subsystems of arithmetic, such as $\mathsf{B}\Gamma$ and $\mathsf{L}\Gamma$, obtained by adding to $\mathsf{I}\Sigma_0$ the collection principle or the least number principle for a class $\Gamma$ of formulas, which lie for instance between $\mathsf{I}\Sigma_0$ and $\mathsf{PA}$ (see: [57]). Adding these is also of interest, and their mechanization is currently under working.

# 3 Provability logic

In this section, we describe our mechanization of modal logic, in particular of provability logic. We have mechanized Solovay's arithmetical completeness theorem [142], the most fundamental and important result in the field of provability logic. We have also mechanized the classification theorem of provability logics due to Beklemishev [19]. As in the previous section, we keep the introduction of definitions and facts brief. For the details of modal logic and provability logic, we refer the reader to the standard textbooks [27, 36, 141] and the surveys [10, 20, 70, 153].

## 3.1 Basics of modal logic

We first set up the basic framework of modal logic.

**Definition 3.1** Formulas of modal logic are built from propositional variables (denoted by Var), the primitive logical connectives $\bot$ and $\to$, and the modal operator $\Box$. The remaining operators $\top, \neg, \land, \lor, \leftrightarrow, \Diamond$ are introduced as the usual abbreviations. We also abbreviate $\boxdot A \equiv A \land \Box A$. A *substitution* is a map $s$ assigning a formula to each propositional variable, and $A[s]$ denotes the formula obtained from $A$ by replacing every occurrence of each propositional variable $p$ with $s(p)$. For a finite set of formulas $\Gamma$, we write $\Box\Gamma = \{\Box B \mid B \in \Gamma\}$. The set of subformulas of $A$ is denoted by $\mathrm{Sub}(A)$. A set of formulas is called a *logic* when we want to emphasize this viewpoint, and we write $L \vdash A$ for $A \in L$ when $L$ is a logic.

```
inductive Formula (α : Type*)
| atom : α → Formula α
| bot  : Formula α
| imp  : Formula α → Formula α → Formula α
| box  : Formula α → Formula α

abbrev neg (A : Formula α) : Formula α := A ➝ ⊥
abbrev top : Formula α := ∼⊥
abbrev or (A B : Formula α) : Formula α := ∼A ➝ B
abbrev and (A B : Formula α) : Formula α := ∼(A ➝ ∼B)
abbrev dia (A : Formula α) : Formula α := ∼□(∼A)
abbrev boxdot (A : Formula α) : Formula α := A ⋏ □A
```

[15]For the mechanization itself, see for example Foundation/FirstOrder/Incompleteness/Examples.lean . A auto-generated zoo from Lean is also available at https://formalizedformallogic.github.io/Foundation/zoo/arithmetic.png.

```
abbrev Substitution (α β) := α → Formula β

def subst (s : Substitution α β) : Formula α → Formula β
  | atom a  ⇒ (s a)
  | ⊥       ⇒ ⊥
  | □A      ⇒ □(A.subst s)
  | A ➝ B   ⇒ A.subst s ➝ B.subst s
notation:95 A "⟦" s "⟧" ⇒ Formula.subst s A

abbrev Logic (α) := Set (Formula α)
```

NOTE: In the mechanization, the propositional variables are parameterized by a type `α`. For instance, taking `α` to be `Empty` yields the formulas containing no propositional variables (called closed or letterless formulas).

SOURCE: 1 2 3

In the present paper, we mainly characterize the logic GL in three ways: by a Gentzen-style sequent calculus, by Kripke semantics, and by a Hilbert-style proof system. Although GL is usually defined in the Hilbert style, when proving the Kripke completeness, introducing a sequent calculus makes both the mathematical proofs and the implementation of the mechanization simpler. Moreover, as applications, the interpolation theorem and the fixpoint theorem can be derived easily via the sequent calculus (we will discuss this in Section 3.2). Hence, in our mechanization we first define the Gentzen-style sequent calculus, and eventually prove the equivalence of all these characterizations (Theorem 3.7).

We first introduce the Gentzen-style sequent calculus. Our sequent calculus for GL is due to Sambin and Valentini [122].

**Definition 3.2** A *sequent* $\Gamma \Rightarrow \Delta$ is a pair of finite sets of formulas. The sequent calculus $\mathcal{G}_{\mathsf{GL}}$ for GL consists of the following rules, where in the weakening rules (WL) and (WR) we assume $\Gamma \subseteq \Gamma'$ and $\Delta \subseteq \Delta'$ respectively. We write $\mathcal{G}_{\mathsf{GL}} \vdash \Gamma \Rightarrow \Delta$ if the sequent $\Gamma \Rightarrow \Delta$ is provable in $\mathcal{G}_{\mathsf{GL}}$.

$$\frac{}{A \Rightarrow A}\,(\mathrm{Ax}) \qquad \frac{}{\bot \Rightarrow}\,(\bot\mathrm{L})$$

$$\frac{\Gamma \Rightarrow \Delta}{\Gamma' \Rightarrow \Delta}\,(\mathrm{WL}) \qquad \frac{\Gamma \Rightarrow \Delta}{\Gamma \Rightarrow \Delta'}\,(\mathrm{WR})$$

$$\frac{\Gamma \Rightarrow A, \Delta \qquad B, \Gamma \Rightarrow \Delta}{A \to B, \Gamma \Rightarrow \Delta}\,(\to\mathrm{L}) \qquad \frac{A, \Gamma \Rightarrow B, \Delta}{\Gamma \Rightarrow A \to B, \Delta}\,(\to\mathrm{R})$$

$$\frac{\Box A, \Gamma, \Box\Gamma \Rightarrow A}{\Box\Gamma \Rightarrow \Box A}\,(\Box_{\mathsf{GL}})$$

Moreover, the sequent calculus $\mathcal{G}_{\mathsf{GL}} + (\mathrm{Cut})$ is obtained from $\mathcal{G}_{\mathsf{GL}}$ by adding the following cut rule.

$$\frac{\Gamma_1 \Rightarrow A, \Delta_1 \qquad A, \Gamma_2 \Rightarrow \Delta_2}{\Gamma_1, \Gamma_2 \Rightarrow \Delta_1, \Delta_2}\,(\mathrm{Cut})$$

```
structure Sequent (α : Type u) where
  ant : FormulaFinset α
  suc : FormulaFinset α
infix:50 " ⇒ " ⇒ Sequent.mk

inductive LogicGL.ProofGentzen : Sequent α → Type u
| axm (A) : ProofGentzen ({A} ⇒ {A})
| botL : ProofGentzen ({⊥} ⇒ ∅)
| wkL  {Γ Γ' Δ}  : ProofGentzen (Γ ⇒ Δ) → Γ ⊆ Γ' → ProofGentzen (Γ' ⇒ Δ)
| wkR  {Γ Δ Δ'}  : ProofGentzen (Γ ⇒ Δ) → Δ ⊆ Δ' → ProofGentzen (Γ ⇒ Δ')
| impL {Γ Δ A B} : ProofGentzen (Γ ⇒ (insert A Δ)) → ProofGentzen (insert B Γ ⇒ Δ) →
```

```
                   ProofGentzen ((insert (A ➝ B) Γ) ⇒ Δ)
  | impR {Γ Δ A B} : ProofGentzen ((insert A Γ) ⇒ (insert B Δ)) →
                   ProofGentzen (Γ ⇒ (insert (A ➝ B) Δ))
  | boxGL {Γ A} : ProofGentzen ((insert (□A) (Γ ∪ Γ.box)) ⇒ {A}) → ProofGentzen (Γ.box ⇒ {□A})
notation:120 "⊢ᵍ[GL]! " S:121 => LogicGL.ProofGentzen S

abbrev LogicGL.ProvableGentzen (S : Sequent α) : Prop := Nonempty (⊢ᵍ[GL]! S)
notation:120 "⊢ᵍ[GL] " S:121 => LogicGL.ProvableGentzen S

inductive LogicGL.GentzenWithCutProof : Sequent α → Type u
  | ...
  | cut {Γ₁ Γ₂ Δ₁ Δ₂ A} : GentzenWithCutProof (Γ₁ ⇒ insert A Δ₁) → GentzenWithCutProof (insert A Γ₂ ⇒
Δ₂) →
                        GentzenWithCutProof (Γ₁ ∪ Γ₂ ⇒ Δ₁ ∪ Δ₂)
notation:120 "⊢ᵍᶜ[GL]! " S:121 => LogicGL.GentzenWithCutProof S

abbrev LogicGL.GentzenWithCutProvable (S : Sequent α) : Prop := Nonempty (⊢ᵍᶜ[GL]! S)
notation:120 "⊢ᵍᶜ[GL] " S:121 => LogicGL.GentzenWithCutProvable S
```

SOURCE: 1 2

Note that $\mathcal{G}_{\mathsf{GL}}$ itself contains no cut rule. The cut-elimination theorem for $\mathcal{G}_{\mathsf{GL}} + (\mathrm{Cut})$ is also mechanized.

**Theorem 3.3 (Cut elimination for $\mathcal{G}_{\mathsf{GL}}$ [14, 122])** If $\mathcal{G}_{\mathsf{GL}} + (\mathrm{Cut}) \vdash \Gamma \Rightarrow \Delta$, then $\mathcal{G}_{\mathsf{GL}} \vdash \Gamma \Rightarrow \Delta$.

```
theorem LogicGL.ProvableGentzen.of_with_cut {S : Sequent α} : ⊢ᵍᶜ[GL] S → ⊢ᵍ[GL] S
```

SOURCE: 1

Here we note that this cut-elimination theorem is mechanized as a semantical cut elimination, via the Kripke semantics explained below. In other words, we do not present a deterministic/computable/syntactic cut-elimination algorithm ( `def cutEliminationAlgorithm : ⊢ᵍᶜ[GL]! S → ⊢ᵍ[GL]! S` ), such as the ones repeatedly discussed in [52, 122]. For the purpose of our mechanization, the cut rule is introduced to show the equivalence with the Hilbert-style system, i.e., for modus ponens, and it suffices that it can be eliminated; hence we put off a rigorous mechanization of such an algorithm. For a syntactic cut-elimination algorithm for the sequent calculus of $\mathsf{GL}$, see, e.g., the mechanization in Rocq by Goré, Ramanayake, and Shillito [53].

Next, we introduce Kripke semantics. Since we are not concerned with modal logic in general, we omit the notion of frames and work only with models.

**Definition 3.4** Let $W$ be a nonempty set, whose elements are called *worlds* or *points*. A *Kripke model* is a triple $M = \langle W, R, V \rangle$, where $R \subseteq W \times W$ (the *accessibility relation*) and $V : W \times \mathrm{Var} \to \{0, 1\}$ (the *valuation*). When there is no danger of confusion, we write $x \prec y$ for $xRy$.

We use the following terminology for models.

- A model is *finite* if $W$ is a finite set.
- A model is *transitive* (resp. *irreflexive*) if $R$ is transitive (resp. irreflexive).
- A point $r \in W$ is a *root* if $rRx$ for every $x \in W$ with $x \neq r$. A model with a designated root is called a *rooted model*.
- A model is a $\mathsf{GL}$-*model* if $R$ is transitive and conversely well-founded. In particular, a finite, transitive, and irreflexive model is called a *finite* $\mathsf{GL}$-*model*; every such model is a $\mathsf{GL}$-model.

For a model $M$, a point $x$ of $M$, and a formula $A$, the *forcing relation* $M, x \Vdash A$ is defined as follows.

- $M, x \Vdash p$ iff $V(x, p) = 1$.
- $M, x \nVdash \bot$.
- $M, x \Vdash A \to B$ iff $M, x \Vdash A$ implies $M, x \Vdash B$.
- $M, x \Vdash \Box A$ iff $M, y \Vdash A$ for every $y \in W$ with $xRy$.

```
structure Model (κ : Type u) [Nonempty κ] (α : Type v) where
  Rel' : κ → κ → Prop
```

```
  Val' : κ → α → Prop

def Model.World.Forces (M : Model κ α) (x : M.World) : Formula α → Prop
| #a     ⇒ M x a
| ⊥      ⇒ False
| A ➝ B ⇒ Forces M x A → Forces M x B
| □A     ⇒ ∀ y, x ≺ y → Forces M y A
notation:55 x:56 " ⊩[" M "] " A:56 ⇒ Model.World.Forces M x A

class IsGL (M : Model κ α) extends IsTrans _ M.Rel, IsConverseWellFounded _ M.Rel

class IsFiniteGL (M : Model κ α) extends IsTrans _ M.Rel, Std.Irrefl M.Rel where
  [finite : Finite M.World]

abbrev Model.Root (M : Model κ α) := { r : M.World // ∀ x, x ≠ r → r ≺ x }

structure RootedModel (κ) [Nonempty κ] (α) extends Model κ α where
  root : toModel.Root
```

NOTE: $W$ is given as an arbitrary nonempty type `κ`, and a model is implemented as a pair of a relation and a valuation. An advantage of taking the model `M` as an explicit argument of the forcing relation, as in `x ⊩[M] A`, is that the type of `x` (namely `M.World`) can be inferred from the notation. Conversely, if `x` is already inferred to be a world of `M`, then `M` is determined by unification, and hence can be omitted as in `x ⊩[_] A`.

SOURCE: 1 2

For later use, we also introduce the notions of the *rank* of a point and the *height* of a finite GL-model.

**Definition 3.5** Let $M$ be a finite GL-model.

- The *rank* $\mathrm{rank}(x) < \omega$ of a point $x$ is the maximal length $n$ of the $R$-chains $x \prec y_1 \prec \cdots \prec y_n$ starting from $x$.
- The *height* of a rooted finite GL-model $M$ is the rank of its root.

These notions are well-defined since $M$ is conversely well-founded.

```
noncomputable def World.rank {M : Model κ α} [Fintype M.World] [M.IsGL] (x : M.World) : ℕ :=
  cwfHeight (· ≺ ·) x

noncomputable def height (M : RootedModel κ α) [Fintype M.World] [M.IsGL] : ℕ := M.root.1.rank
```

SOURCE: 1

Finally, we introduce the Hilbert-style proof system.

**Definition 3.6** The Hilbert-style proof system $\mathcal{H}_{\mathsf{GL}}$ for GL consists of the following axioms and inference rules. We write $\mathcal{H}_{\mathsf{GL}} \vdash A$ if $A$ is provable in $\mathcal{H}_{\mathsf{GL}}$, and define the logic $\mathsf{GL} := \{A : \mathcal{H}_{\mathsf{GL}} \vdash A\}$.

1. Tautologies of classical propositional logic (cf. [36])
2. Axiom K: $\Box(A \to B) \to (\Box A \to \Box B)$
3. Axiom 4: $\Box A \to \Box\Box A$
4. Axiom Löb: $\Box(\Box A \to A) \to \Box A$
5. Inference rules: modus ponens (MP) and the necessitation rule (Nec).

```
inductive LogicGL.ProofHilbert : Formula α → Type u
| implyK   {A B}   : ProofHilbert $ A ➝ B ➝ A
| implyS   {A B C} : ProofHilbert $ (A ➝ B ➝ C) ➝ (A ➝ B) ➝ (A ➝ C)
| dne      {A}     : ProofHilbert $ ~~A ➝ A
| andElimL {A B}   : ProofHilbert $ (A ⋏ B) ➝ A
| andElimR {A B}   : ProofHilbert $ (A ⋏ B) ➝ B
| andIntro {A B}   : ProofHilbert $ A ➝ B ➝ (A ⋏ B)
| orIntroL {A B}   : ProofHilbert $ A ➝ (A ⋎ B)
| orIntroR {A B}   : ProofHilbert $ B ➝ (A ⋎ B)
| orElim   {A B C} : ProofHilbert $ (A ➝ C) ➝ (B ➝ C) ➝ ((A ⋎ B) ➝ C)
```

```
  | modalK   {A B}   : ProofHilbert $ □(A ➝ B) ➝ (□A ➝ □B)
  | modal4   {A}     : ProofHilbert $ □A ➝ □□A
  | modalL   {A}     : ProofHilbert $ □(□A ➝ A) ➝ □A
  | mdp      {A B}   : ProofHilbert (A ➝ B) → ProofHilbert A → ProofHilbert B
  | nec      {A}     : ProofHilbert A → ProofHilbert (□A)
notation:50 "⊢ʰ[GL]! " A:51 => LogicGL.ProofHilbert A

abbrev LogicGL.ProvableHilbert (A : Formula α) := Nonempty (⊢ʰ[GL]! A)
notation:50 "⊢ʰ[GL] " A:51 => LogicGL.ProvableHilbert A

abbrev LogicGL {α} : Logic α := { A | ⊢ʰ[GL] A }
```

NOTE: Łukasiewicz's three axioms would suffice to prove all tautologies of classical propositional logic, but then the axioms listed here would have to be proved syntactically, which is extremely tedious; so we adopt all of them as axioms. Axiom 4 is syntactically derivable from the others (cf. [27:Chapter 1, Theorem 18]), but its proof is a tedious puzzle, so our mechanization adopts it as an axiom.

SOURCE: 1 2

As the equivalence of these characterizations, we mechanized the following.

**Theorem 3.7 (Characterization of GL)** The following are equivalent.

1. $\mathsf{GL} \vdash A$.
2. $\mathcal{H}_{\mathsf{GL}} \vdash A$.
3. $\mathcal{G}_{\mathsf{GL}} \vdash\Rightarrow A$.
4. $\mathcal{G}_{\mathsf{GL}} + (\mathrm{Cut}) \vdash\Rightarrow A$.
5. $\Rightarrow 0 : A$ is provable in the labelled sequent calculus (see Section 3.3).
6. $A$ is forced at every point of every finite $\mathsf{GL}$-model.
7. $A$ is forced at the root of every rooted finite $\mathsf{GL}$-model.
8. $A$ is forced at the root of every rooted finite $\mathsf{GL}$-model that is a tree.
9. For every $n \geq 1$, $A$ is forced at every point of every finite $\mathsf{GL}$-model whose set of points is $\{0, 1, \cdots, n-1\}$.
10. For every $n \geq 1$, $A$ is forced at the root of every rooted finite $\mathsf{GL}$-model whose set of points is $\{0, 1, \cdots, n-1\}$.

```
theorem LogicGL.provability_TFAE {α : Type u} [DecidableEq α] {A : Formula α} : [
  A ∈ LogicGL,
  ⊢ʰ[GL] A,
  ⊢ᵍ[GL] (∅ ⇒ {A}),
  ⊢ᵍᶜ[GL] (∅ ⇒ {A}),
  ⊢ˡᵍ[GL] (∅ , ∅ ⇒ˡ {(0 : Label) : A}),
  ∀ {κ : Type u}, [Nonempty κ] → ∀ M : Model κ α, [M.IsFiniteGL] → M ⊧ A,
  ∀ {κ : Type u}, [Nonempty κ] → ∀ M : RootedModel κ α, [M.IsFiniteGL] → M.root.1 ⊩[_] A,
  ∀ {κ : Type u}, [Nonempty κ] → ∀ M : RootedModel κ α, [M.IsFiniteGLTree] → M.root.1 ⊩[_] A,
  ∀ (n : ℕ) [NeZero n] (M : Model (Fin n) α), [M.IsFiniteGL] → M ⊧ A,
  ∀ (n : ℕ) [NeZero n] (M : RootedModel (Fin n) α), [M.IsFiniteGL] → M.root.1 ⊩[_] A
].TFAE
```

SOURCE: 1

Here are a few remarks. The Kripke completeness of GL (the equivalence of 1 and 6) is due to Segerberg [127]. This is the usual Kripke completeness with respect to the class of finite GL-models, and has already been mechanized in HOL Light by Maggesi and Perini Brogi [95, 96]. However, the proof of the arithmetical completeness theorem described later requires not the mere Kripke completeness, but the completeness with respect to rooted models (7, and furthermore 8). The transformation of a rooted model into a tree model is done by the technique known as tree unraveling (cf. [36:Theorem 3.18]). The equivalence of 3 and 4 is the special case of the cut-elimination theorem (Theorem 3.3) for sequents of the form $\Rightarrow A$. The equivalence with 9 and 10 is provided for the sake of *concrete* (or useful) countermodels. Due to universe issues, the models in the completeness clauses range over types `κ` in the same universe level as the one that `α` belongs to. Hence, to construct a countermodel, one would have to define it over a type lifted to the matching universe level, such as `PUnit` or `PLift (Fin n)` [16]. However, dealing with such universe issues every time is quite tedious. Thus we prepared some lemmas that internally dispose of the universe issues via a suitable type equivalence, so that countermodels can be constructed concretely over `Fin n`.

[16] https://leanprover-community.github.io/mathlib4_docs/Init/Prelude.html#PLift

Next, we introduce the modal logics S (due to Solovay [142]) and D (due to Japaridze (Dzhaparidze) [68]), which play important roles in provability logic.

**Definition 3.8** For logics $L_1, L_2$, we write $L_1 + L_2$ for the logic obtained by closing the union of $L_1$ and $L_2$ under MP and substitution.

Then Solovay's provability logic and Japaridze's provability logic are defined as follows.

- $\mathsf{S} := \mathsf{GL} + \{\Box A \to A \mid A\}$.
- $\mathsf{D} := \mathsf{GL} + (\{\neg\Box\bot\} \cup \{\Box(\Box A \vee \Box B) \to \Box A \vee \Box B \mid A, B\})$.

```
inductive Logic.sumQuasiNormal (L₁ L₂ : Logic α) : Logic α
  | mem₁ {A}    : L₁ A → sumQuasiNormal L₁ L₂ A
  | mem₂ {A}    : L₂ A → sumQuasiNormal L₁ L₂ A
  | mdp  {A B}  : sumQuasiNormal L₁ L₂ (A ➝ B) → sumQuasiNormal L₁ L₂ A → sumQuasiNormal L₁ L₂ B
  | subst {A s} : sumQuasiNormal L₁ L₂ A → sumQuasiNormal L₁ L₂ (A⟦s⟧)
infix:50 " +ᴸ " => Logic.sumQuasiNormal

abbrev LogicS {α} : Logic α := LogicGL +ᴸ ({ □A ➝ A | A })

abbrev LogicD {α} : Logic α := LogicGL +ᴸ (insert (∼□⊥) { □(□A ⋎ □B) ➝ (□A ⋎ □B) | (A) (B) })
```

SOURCE: 1 2 3

These logics are non-normal, i.e., not closed under the necessitation rule, so Kripke semantics cannot be applied directly. However, they are known to be sound and complete with respect to classes of infinite models obtained by suitably extending finite GL-models. Such models for S are called tail models [155], and for D the so-called pseudo tail models (cf. [19]) are used. We omit the details of these constructions.

Both logics also admit Gentzen-style sequent calculi, but in the calculi, sequents have levels. Kushida [86] gave such a calculus for S with two levels of sequents, and Kashima et al. [76] extended his approach to a calculus for D with three levels.

**Definition 3.9 (Sequent calculi for S and D [75, 76, 86])** A *layered sequent* is a sequent $\Gamma \Rightarrow \Delta$ together with a level. The calculus $\mathcal{G}_{\mathsf{S}}$ uses the two levels $\Rightarrow^1$ and $\Rightarrow^2$, and the calculus $\mathcal{G}_{\mathsf{D}}$ uses the three levels $\Rightarrow^1$, $\Rightarrow^2$, and $\Rightarrow^3$. At each level $l$ separately, both systems contain the propositional rules (Ax), ($\bot$L), (WL), (WR), ($\to$L), and ($\to$R) of $\mathcal{G}_{\mathsf{GL}}$, with $\Rightarrow$ replaced by $\Rightarrow^l$. In addition, both systems contain the following rules, except that (Lift $^2_3$) belongs to $\mathcal{G}_{\mathsf{D}}$ only.

$$\frac{\Box A, \Gamma, \Box\Gamma \Rightarrow^1 A}{\Box\Gamma \Rightarrow^1 \Box A}\,(\Box_{\mathsf{GL}}) \qquad \frac{\Gamma \Rightarrow^1 \Delta}{\Gamma \Rightarrow^2 \Delta}\,(\mathrm{Lift}^1_2)$$

$$\frac{A, \Gamma \Rightarrow^2 \Delta}{\Box A, \Gamma \Rightarrow^2 \Delta}\,(\Box\mathrm{L}) \qquad \frac{\Box\Gamma \Rightarrow^2 \Box\Delta}{\Box\Gamma \Rightarrow^3 \Box\Delta}\,(\mathrm{Lift}^2_3)$$

Neither system contains a cut rule; $\mathcal{G}_{\mathsf{S}}$ + (Cut) and $\mathcal{G}_{\mathsf{D}}$ + (Cut) denote the systems extended with the cut rule, which is level-preserving.

```
structure TwoLayeredSequent (α : Type u) extends Sequent α where
  level : Fin 2
notation:50 Γ:51 " ⇒[" l "] " Δ:51 => TwoLayeredSequent.mk (Γ ⇒ Δ) l

inductive LogicS.ProofGentzen : TwoLayeredSequent α → Type u
| ...
| boxGL   {Γ A}   : ProofGentzen ((insert (□A) (Γ ∪ Γ.box)) ⇒[0] {A}) →
                    ProofGentzen (Γ.box ⇒[0] {□A})
| liftUp {Γ Δ}    : ProofGentzen (Γ ⇒[0] Δ) → ProofGentzen (Γ ⇒[1] Δ)
| boxL    {Γ Δ A} : ProofGentzen (insert A Γ ⇒[1] Δ) →
```

```
                   ProofGentzen (insert (□A) Γ ⇒[1] Δ)
scoped prefix:120 "⊢ᵍ[S]! " => LogicS.ProofGentzen

abbrev LogicS.ProvableGentzen (S : TwoLayeredSequent α) : Prop := Nonempty (⊢ᵍ[S]! S)
scoped prefix:120 "⊢ᵍ[S] " => LogicS.ProvableGentzen

structure ThreeLayeredSequent (α : Type u) extends Sequent α where
  level : Fin 3
notation:50 Γ:51 " ⇒[" l "] " Δ:51 => ThreeLayeredSequent.mk (Γ ⇒ Δ) l

inductive LogicD.ProofGentzen : ThreeLayeredSequent α → Type u
| ...
| boxGL {Γ : FormulaFinset α} {A}   : ProofGentzen ((insert (□A) (Γ ∪ □Γ)) ⇒[0] {A}) →
                                      ProofGentzen (□Γ ⇒[0] {□A})
| liftUp₀₁ {Γ Δ}                    : ProofGentzen (Γ ⇒[0] Δ) → ProofGentzen (Γ ⇒[1] Δ)
| boxL {Γ Δ A}                      : ProofGentzen (insert A Γ ⇒[1] Δ) →
                                      ProofGentzen (insert (□A) Γ ⇒[1] Δ)
| liftUp₁₂ {Γ Δ : FormulaFinset α}  : ProofGentzen (□Γ ⇒[1] □Δ) →
                                      ProofGentzen (□Γ ⇒[2] □Δ)
scoped prefix:120 "⊢ᵍ[D]! " => LogicD.ProofGentzen

abbrev LogicD.ProvableGentzen (S : ThreeLayeredSequent α) : Prop := Nonempty (⊢ᵍ[D]! S)
scoped prefix:120 "⊢ᵍ[D] " => LogicD.ProvableGentzen
```

NOTE: Levels are implemented as elements of `Fin 2` and `Fin 3`, hence are numbered from 0 in the mechanization: the levels $\Rightarrow^1$, $\Rightarrow^2$, and $\Rightarrow^3$ above correspond to `⇒[0]`, `⇒[1]`, and `⇒[2]` respectively.

SOURCE: 1 2 3

By construction, the $\Rightarrow^1$-fragment of both systems is exactly $\mathcal{G}_{\mathsf{GL}}$, and the $\Rightarrow^1$ and $\Rightarrow^2$ fragments of $\mathcal{G}_{\mathsf{D}}$ are exactly $\mathcal{G}_{\mathsf{S}}$; these embeddings are mechanized as well, and are what lets the mechanization of $\mathcal{G}_{\mathsf{D}}$ reuse that of $\mathcal{G}_{\mathsf{S}}$.

As for $\mathcal{G}_{\mathsf{GL}}$, we can prove the cut-elimination theorem semantically.

**Theorem 3.10 (Cut elimination for $\mathcal{G}_{\mathsf{S}}$ and $\mathcal{G}_{\mathsf{D}}$ [75, 76])**
- If $\mathcal{G}_{\mathsf{S}} + (\mathrm{Cut}) \vdash \Gamma \Rightarrow^2 \Delta$, then $\mathcal{G}_{\mathsf{S}} \vdash \Gamma \Rightarrow^2 \Delta$.
- If $\mathcal{G}_{\mathsf{D}} + (\mathrm{Cut}) \vdash \Gamma \Rightarrow^3 \Delta$, then $\mathcal{G}_{\mathsf{D}} \vdash \Gamma \Rightarrow^3 \Delta$.

```
theorem LogicS.ProvableGentzen.of_with_cut {Γ Δ : FormulaFinset α}
  (h : ⊢ᵍᶜ[S] (Γ ⇒[1] Δ)) : ⊢ᵍ[S] (Γ ⇒[1] Δ)

theorem LogicD.ProvableGentzen.of_with_cut {Γ Δ : FormulaFinset α}
  (h : ⊢ᵍᶜ[D] (Γ ⇒[2] Δ)) : ⊢ᵍ[D] (Γ ⇒[2] Δ)
```

SOURCE: 1 2

With this result, the characterizations of $\mathsf{S}$ and $\mathsf{D}$ can be stated as follows. First, the following holds for $\mathsf{S}$.

**Proposition 3.11 (cf. [155])** The following are equivalent.
1. $\mathsf{S} \vdash A$
2. $\mathcal{G}_{\mathsf{S}} \vdash \Rightarrow^2 A$
3. On the chain of the tail model constructed from any finite $\mathsf{GL}$-model and any point $t$ of it, $A$ is eventually always forced.
4. $\bigwedge_{\Box B \in \mathrm{Sub}(A)} (\Box B \to B) \to A$ is forced at the root of every rooted finite $\mathsf{GL}$-model.
5. Same as 3, but only for the finite $\mathsf{GL}$-models whose set of points is $\{0, 1, \cdots, n-1\}$ for some $n \geq 1$.
6. $\mathsf{GL} \vdash \bigwedge_{\Box B \in \mathrm{Sub}(A)} (\Box B \to B) \to A$

```
theorem LogicS.provability_TFAE [DecidableEq α] : [
  A ∈ LogicS,
```

```
    ⊢ᵍ[S] (∅ ⇒[1] {A}),
    ∀ {κ : Type u}, [Nonempty κ] → ∀ (M : Model κ α), [M.IsFiniteGL] → ∀ (tail : M.World),
      ∃ k : ℕ, ∀ n : ℕ, k ≤ n → toTail.chainPoint n ⊩[(M.toTail tail).toModel] A,
    ∀ {κ : Type u}, [Nonempty κ] → ∀ (M : RootedModel κ α), [M.IsFiniteGL] →
      M.root.1 ⊩[_] (⋀A.subfmlsS ➝ A),
    ∀ (n : ℕ) [NeZero n] (M : Model (Fin n) α), [M.IsFiniteGL] → ∀ (tail : M.World),
      ∃ k : ℕ, ∀ m : ℕ, k ≤ m → toTail.chainPoint m ⊩[(M.toTail tail).toModel] A,
    (⋀A.subfmlsS ➝ A) ∈ LogicGL
  ].TFAE
```

Source: 1

Using this equivalence, we can show the following fact about the formulas obtained by replacing every $\Box$ with $\boxdot$.

**Definition 3.12 (Boxdot translation)** The *boxdot translation* $A^{\boxdot}$ of a formula $A$ is obtained by replacing every occurrence of $\Box$ with $\boxdot$, i.e., it is defined recursively as follows.

- $p^{\boxdot} = p$
- $\bot^{\boxdot} = \bot$
- $(A \to B)^{\boxdot} = A^{\boxdot} \to B^{\boxdot}$
- $(\Box A)^{\boxdot} = \boxdot(A^{\boxdot})$

```
def Formula.boxdotTranslate : Formula α → Formula α
  | #a     ⇒ #a
  | ⊥      ⇒ ⊥
  | A ➝ B ⇒ (boxdotTranslate A) ➝ (boxdotTranslate B)
  | □A     ⇒ ⊡(boxdotTranslate A)
postfix:90 "ᵇ" ⇒ Formula.boxdotTranslate
```

Source: 1

**Proposition 3.13** For every formula $A$, $\mathsf{GL} \vdash A^{\boxdot}$ if and only if $\mathsf{S} \vdash A^{\boxdot}$.

```
theorem LogicS.iff_provable_boxdot_GL_provable_boxdot_S [DecidableEq α] :
  (Aᵇ) ∈ LogicGL ↔ (Aᵇ) ∈ LogicS
```

Source: 1

The boxdot translation and the equivalence of $\mathsf{GL}$ and $\mathsf{S}$ on boxdot-translated formulas are also important in the connection with $\mathsf{Grz}$, which we discuss later in Section 3.7.

Next, we turn to $\mathsf{D}$.

**Proposition 3.14 (cf. [19, 76])** The following are equivalent, where $\Box^{-1}X = \{B \mid \Box B \in X\}$ for a set of formulas $X$.

1. $\mathsf{D} \vdash A$
2. $\mathcal{G}_{\mathsf{D}} \vdash\Rightarrow^3 A$
3. $A$ is forced at the root of the pseudo tail model constructed from any finite $\mathsf{GL}$-model.
4. $\bigwedge_{\Gamma \subseteq \Box^{-1}\,\mathrm{Sub}(A)} (\Box(\bigvee \Box\Gamma) \to \bigvee \Box\Gamma) \to A$ is forced at the root of every rooted finite $\mathsf{GL}$-model.
5. Same as 3, but only for the finite $\mathsf{GL}$-models whose set of points is $\{0, 1, \cdots, n-1\}$ for some $n \geq 1$.
6. $\mathsf{GL} \vdash \bigwedge_{\Gamma \subseteq \Box^{-1}\,\mathrm{Sub}(A)} (\Box(\bigvee \Box\Gamma) \to \bigvee \Box\Gamma) \to A$

```
theorem LogicD.provability_TFAE [DecidableEq α] : [
    A ∈ LogicD,
    ⊢ᵍ[D] (∅ ⇒[2] {A}),
```

```
  ∀ {κ : Type u}, [Nonempty κ] → ∀ (M : Model κ α), [M.IsFiniteGL] → ∀ r o,
    (M.toPseudoTail r o).root.1 ⊩[_] A,
  ∀ {κ : Type u}, [Nonempty κ] → ∀ (M : RootedModel κ α), [M.IsFiniteGL] →
    M.root.1 ⊩[_] (⋀A.subfmlsD → A),
  ∀ (n : ℕ) [NeZero n] (M : Model (Fin n) α), [M.IsFiniteGL] → ∀ r o,
    (M.toPseudoTail r o).root.1 ⊩[_] A,
  (⋀A.subfmlsD → A) ∈ LogicGL
].TFAE
```

Source: 1 2

The inclusions $\mathsf{GL} \subseteq \mathsf{D} \subseteq \mathsf{S}$ hold trivially by definition. Moreover, constructing countermodels via the semantics shows that these inclusions are proper.

**Proposition 3.15** $\mathsf{GL} \subsetneq \mathsf{D} \subsetneq \mathsf{S}$

```
lemma LogicGL_ssubset_LogicD [DecidableEq α] : (LogicGL : Logic α) ⊂ LogicD

lemma LogicD_ssubset_LogicS [Inhabited α] [DecidableEq α] : (LogicD : Logic α) ⊂ LogicS
```

Source: 1 2

### 3.2 Applications of the sequent calculus

Sambin and Valentini [122] give several further applications of the sequent calculus for GL. First, since it is a pure sequent calculus, the Craig interpolation property (CIP) can be shown straightforwardly by Maehara's method [93] (cf. [145]).

**Theorem 3.16 (Craig interpolation property for GL)** If $\mathsf{GL} \vdash A \to B$, then there exists a formula $C$ such that $\mathsf{GL} \vdash A \to C$ and $\mathsf{GL} \vdash C \to B$, and every propositional variable of $C$ occurs in both $A$ and $B$.

```
theorem LogicGL.CIP (h : (A → B) ∈ LogicGL) :
  ∃ C : Formula α, (A → C) ∈ LogicGL ∧ (C → B) ∈ LogicGL ∧ C.atoms ⊆ A.atoms ∩ B.atoms
```

Source: 1 2

The CIP of GL is important in particular because it yields the fixpoint theorem of GL [25, 140]. We have also mechanized the fixpoint theorem of GL via the sequent calculus.

**Definition 3.17** A propositional variable $p$ is *modalized* in a formula $A$ if every occurrence of $p$ in $A$ is within the scope of $\Box$.

```
def ModalizedIn (p : α) : Formula α → Prop
  | #a    ⇒ a ≠ p
  | ⊥     ⇒ True
  | A → B ⇒ A.ModalizedIn p ∧ B.ModalizedIn p
  | □_    ⇒ True
```

Source: 1

**Theorem 3.18 (Fixpoint theorem of GL [122])** Suppose that $p$ is modalized in $A$. Then there exists a formula $D$ not containing $p$ and consisting only of propositional variables of $A$ such that

$$\mathsf{GL} \vdash A[p := D] \leftrightarrow D$$

Moreover, such a fixpoint is unique up to provable equivalence: for any formula $E$ such that $\mathsf{GL} \vdash A[p := E] \leftrightarrow E$, we have $\mathsf{GL} \vdash D \leftrightarrow E$.

```
theorem LogicGL.fixpointTheorem
  (hpq : p ≠ q) (hA : A.ModalizedIn p) (hq : q ∉ A.atoms) :
  ∃ D : Formula α, D.atoms ⊆ A.atoms \ {p} ∧ ((A⟦p ↦ D⟧) ↔ D) ∈ LogicGL ∧
    ∀ E : Formula α, ((A⟦p ↦ E⟧) ↔ E) ∈ LogicGL → (D ↔ E) ∈ LogicGL
```

NOTE: The fresh propositional variable `q` serves only as a placeholder in the construction of the fixpoint.

SOURCE: 1

A special case of the fixpoint theorem of $\mathsf{GL}$ has also been mechanized in Lean by Gignoux [45], as a supporting lemma for a mechanization of the CIP of $\mathsf{GL}$ based on non-wellfounded proof systems formulated coalgebraically. We note that Gignoux's mechanization treats formulas of the form $\Box A$ and $\Diamond A$ and proves the fixpoint equivalence semantically over Kripke frames, and its interpolants are given by a noncomputable function, whereas in our mechanization the interpolants and the fixpoints can be computed constructively inside Lean from derivation trees of the sequent calculus. However, at present derivation trees of the sequent calculus cannot be constructed automatically by proof search or the like, so concrete derivation trees have to be input by hand. Also, when $\mathsf{GL} \vdash A$ is proved non-constructively, e.g., via Kripke semantics, the interpolants and the fixpoints are of course not computable in Lean.

Finally, we have also mechanized facts on the CIP of $\mathsf{S}$ and $\mathsf{D}$, which we briefly mention.

**Theorem 3.19 ([17, 18])** $\mathsf{S}$ has the CIP, but $\mathsf{D}$ does not.

```
theorem LogicS.CIP (h : (A ➝ B) ∈ LogicS) :
  ∃ C : Formula α, (A ➝ C) ∈ LogicS ∧ (C ➝ B) ∈ LogicS ∧ C.atoms ⊆ A.atoms ∩ B.atoms

theorem LogicD.notCIP {a b c : α} (hab : a ≠ b) (hac : a ≠ c) (hbc : b ≠ c) :
  ∃ A B : Formula α, (A ➝ B) ∈ LogicD ∧
    ¬ ∃ C : Formula α, (A ➝ C) ∈ LogicD ∧ (C ➝ B) ∈ LogicD ∧
      C.atoms ⊆ A.atoms ∩ B.atoms
```

SOURCE: 1 2

### 3.3 On the labelled sequent calculus

We have also mechanized the labelled sequent calculus for $\mathsf{GL}$ by Negri [104, 105]. As for prior work, our mechanization is almost the same in its method as the mechanization of the labelled sequent calculus for $\mathsf{GL}$ in HOL Light by Maggesi and Perini Brogi [95, 96], and in that sense it has little novelty. We nevertheless touch on some differences between the implementations.

The termination of Maggesi and Perini Brogi's *implementation* of the calculus is guaranteed as a mathematical fact on the meta-level. That is, by the mathematical fact proved in [105], the computation is guaranteed to terminate under the assumption that their implementation is correct. On the other hand, when defining the proof search, we guarantee its termination inside the theorem prover itself, since Lean requires the definition to be well-founded. In this respect, our mechanization gives a stronger guarantee. However, Maggesi and Perini Brogi's mechanization has practical utility: although the termination is not guaranteed, it can actually be executed and used as a tactic, which automates simple proofs of $\mathsf{GL}$ appearing in their mechanization. In contrast, due to the implementation constraints in the well-foundedness proof, our labelled sequent calculus cannot be used, e.g., as a Lean tactic, and its properties are mechanized purely as mathematical facts. Hence, regarding the practical utility, Maggesi and Perini Brogi's mechanization has the advantage.

Moreover, we remark that interpolation for labelled sequent calculi in general is discussed, e.g., in [44:Section 5], but whether it is possible for the labelled sequent calculus for $\mathsf{GL}$ seems to be open at present. This suggests that labelled calculi are less suitable for mechanizing the properties of $\mathsf{GL}$ described in Section 3.2.

### 3.4 Arithmetical completeness theorems

In this section, we describe the main results of our mechanization of provability logic: the mechanization of Solovay's arithmetical completeness theorem [142] and its generalization.

First, we define arithmetical interpretations, which translate modal formulas into arithmetic sentences. In what follows, $T$ is an arithmetic theory with a $\Delta_1$-definable axiomatization extending $\mathsf{I}\Sigma_1$. Moreover, $\mathfrak{B}$ denotes a provability in the sense of Section 2.4. Although the definition allows $\mathfrak{B}$ to be arbitrary, we mainly consider the standard provability $\mathfrak{B}_T$ of $T$.

**Definition 3.20** A map $f : \mathrm{Var} \to \mathrm{Sent}_A$, where $\mathrm{Sent}_A$ denotes the set of arithmetic sentences, is called an *arithmetical realization* (or simply a *realization*). Given a realization $f$ and a provability $\mathfrak{B}$, the *arithmetical interpretation* of $A$ by $\mathfrak{B}$, denoted $f_{\mathfrak{B}}(A)$, is the extension of $f$ translating each modal formula $A$ into an arithmetic sentence as follows.

$$\begin{aligned} f_{\mathfrak{B}}(p) &= f(p) \\ f_{\mathfrak{B}}(\bot) &= \bot \\ f_{\mathfrak{B}}(A \to B) &= f_{\mathfrak{B}}(A) \to f_{\mathfrak{B}}(B) \\ f_{\mathfrak{B}}(\Box A) &= \mathfrak{B}(f_{\mathfrak{B}}(A)) \end{aligned}$$

In particular, the interpretation $f_{\mathfrak{B}_T}(A)$ by $\mathfrak{B}_T$ is called the *standard interpretation* of $A$ and is written $f_T(A)$.

```
structure Realization (α : Type*) (L : FirstOrder.Language) where
  val : α → FirstOrder.Sentence L

def Formula.interpret (f : Realization α L) {T₀ T : FirstOrder.Theory L} (𝔅 : Provability T₀ T) :
  Formula α → FirstOrder.Sentence L
  | #a     ⇒ f.val a
  | ⊥      ⇒ ⊥
  | A ➝ B ⇒ (A.interpret f 𝔅) ➝ (B.interpret f 𝔅)
  | □A     ⇒ 𝔅 (A.interpret f 𝔅)

noncomputable abbrev Formula.standardInterpret (f : Realization α _)
  (T : FirstOrder.ArithmeticTheory) [T.Δ₁] := Formula.interpret f T.standardProvability

noncomputable instance : CoeFun (Realization α ℒₒᵣ)
  (fun _ ↦ (T : FirstOrder.ArithmeticTheory) → [T.Δ₁] → Formula α → FirstOrder.Sentence ℒₒᵣ) :=
  ⟨Formula.standardInterpret⟩
```

Note: The interpretation `A.interpret f 𝔅` corresponds to $f_{\mathfrak{B}}(A)$. By the coercion, the standard interpretation `A.standardInterpret f T` can be written as `f T A`, which corresponds to $f_T(A)$.

Source: 1

**Definition 3.21** The *(standard) provability logic of $T$ relative to $U$*, written $\mathrm{PL}_T(U)$, is defined as follows.

$$\mathrm{PL}_T(U) = \left\{ A \mid U \vdash f_{\mathfrak{B}_T}(A) \text{ for every realization } f \right\}$$

```
def FirstOrder.ArithmeticTheory.provabilityLogicRelativeTo
  (T U : FirstOrder.ArithmeticTheory) [T.Δ₁] : Logic α :=
  {A | ∀ f : Realization α ℒₒᵣ, U ⊢ f T A}

abbrev FirstOrder.ArithmeticTheory.provabilityLogic
  (T : FirstOrder.ArithmeticTheory) [T.Δ₁] : Logic α := T.provabilityLogicRelativeTo T
```

Source: 1

Solovay's arithmetical completeness theorem states that the behavior of the standard provability predicate, viewed as a modal operator, is captured exactly by the modal logic $\mathsf{GL}$. That is, for appropriate choices of $T$ and $U$, the provability logic $\mathrm{PL}_T(U)$ coincides with $\mathsf{GL}$. Here we present the generalized version (Theorem 3.24) using the notion of the *height* of a theory due to Visser [154].

**Definition 3.22 (Height of a theory)** For $n \geq 0$, $\mathfrak{B}^n$ denotes the $n$-fold iteration of the provability $\mathfrak{B}$ (where $\mathfrak{B}^0\sigma \equiv \sigma$). The *height* $\mathrm{hgt}(T) \leq \omega$ of a theory $T$ is the least $n \in \omega$ such that $T \vdash \mathfrak{B}^n_T\bot$; if no such $n$ exists, we set $\mathrm{hgt}(T) = \omega$.

```
noncomputable def Provability.height (𝔅 : Provability T₀ T) : ENat := ENat.find (T ⊢ 𝔅^[·] ⊥)

noncomputable abbrev ArithmeticTheory.height (T : ArithmeticTheory) [T.Δ₁] : ℕ∞ :=
  T.standardProvability.height
```

Source: 1

Note that if $T$ is $\Sigma_1$-sound, then $T \nvdash \mathfrak{B}^n_T\bot$ for every $n \in \omega$, and hence $\mathrm{hgt}(T) = \omega$.

**Definition 3.23** For $n \leq \omega$, abusing notation, we define the logic $\mathsf{GL} + \Box^n\bot$ as follows: for $n < \omega$ it is $\mathsf{GL} + \{\Box^n\bot\}$, and for $n = \omega$ it is $\mathsf{GL}$ itself.

```
def LogicGLPlusBoxBot {α} : ℕ∞ → Logic α
  | .some n ⇒ LogicGL +ᴸ □^[n]⊥
  | .none   ⇒ LogicGL
```

Source: 1

The main result of our mechanization of provability logic is the following.

**Theorem 3.24 ([154])** $\mathrm{PL}_T(T) = \mathsf{GL} + \Box^{\mathrm{hgt}(T)}\bot$

```
lemma LogicGLPlusBoxBot.eq_provabilityLogic :
  LogicGLPlusBoxBot (α := α) T.height = T.provabilityLogic
```

Source: 1

Theorem 3.24 is proved by embedding into arithmetic a rooted finite $\mathsf{GL}$-model of an appropriate height, obtained as a countermodel when $\mathsf{GL} + \Box^{\mathrm{hgt}(T)}\bot \nvdash A$ (the construction of Solovay sentences). This is where the completeness with respect to rooted models stated in Theorem 3.7 is needed. As a corollary, we obtain Solovay's original statement.

**Corollary 3.25 (Solovay's (first) arithmetical completeness theorem [142])** If $T$ is $\Sigma_1$-sound, then $\mathrm{PL}_T(T) = \mathsf{GL}$. In particular, $\mathrm{PL}_{\mathsf{PA}}(\mathsf{PA}) = \mathsf{GL}$.

```
theorem LogicGL.eq_provabilityLogic_sigma1_sound [T.SoundOnHierarchy 𝚺 1] :
  @LogicGL α = T.provabilityLogic

theorem LogicGL.eq_provabilityLogic_peano_arithmetic : @LogicGL α = (𝐏𝐀.provabilityLogic)
```

Source: 1

Furthermore, Solovay also proved that $\mathsf{S}$ is arithmetically complete with respect to the true arithmetic $\mathsf{TA}$. The reduction of $\mathsf{S}$ to $\mathsf{GL}$ stated in Proposition 3.11 is used in an essential way in this proof.

**Theorem 3.26 (Solovay's (second) arithmetical completeness theorem [142])** Let $T$ be a sound theory. For every formula $A$, $\mathsf{S} \vdash A$ if and only if $\mathbb{N} \models f_{\mathfrak{B}_T}(A)$ for every realization $f$. That is, $\mathrm{PL}_T(\mathsf{TA}) = \mathsf{S}$.

```
variable {T : FirstOrder.ArithmeticTheory} [T.Δ₁] [IΣ₁ ≼ T] [ℕ↓[ℒₒᵣ] ⊧* T]

theorem LogicS.arithmetical_completeness_iff [DecidableEq α] :
  A ∈ LogicS ↔ (∀ f : Realization α ℒₒᵣ, ℕ↓[ℒₒᵣ] ⊧ f T A)

theorem LogicS.eq_provabilityLogicRelativeTo_TA [DecidableEq α] :
  @LogicS α = T.provabilityLogicRelativeTo TA
```

Source: 1

### 3.5 The classification theorem of provability logics

The classification of the provability logics obtained as $\mathrm{PL}_T(U)$, where $T$ is as in Section 3.4 and $U$ is an arbitrary arithmetic theory, was studied by Artemov, Beklemishev, Visser, Japaridze, and others, and was finally completed by Beklemishev [19]. We have also mechanized this classification theorem. Since its proof involves difficult arguments in both arithmetic and Kripke semantics, we again omit the details of the proofs and list the main results. For the details, see [10, 19].

**Definition 3.27** The *trace* $\mathrm{tr}(A) \subseteq \omega$ of a formula $A$ is the set of all $n$ such that there exists a rooted finite $\mathsf{GL}$ -model of height $n$ whose root does not force $A$. The *trace* of a logic $L$ is defined by $\mathrm{tr}(L) := \bigcup_{A \in L} \mathrm{tr}(A)$.

```
def trace (A : Formula α) : Set ℕ := { n |
  ∃ κ : Type u, ∃ _ : Nonempty κ, ∃ M : RootedModel κ α, ∃ _ : Fintype M.World, ∃ _ : M.IsGL,
  (M.height = n ∧ M.root.1 ⊮[_] A) }

abbrev Logic.trace (L : Logic α) : Set ℕ := ⋃ A ∈ L, A.trace
```

Source: 1

**Definition 3.28** For $n \in \omega$, define the formula $F_n := \Box^{n+1}\bot \to \Box^n\bot$. For $\alpha \subseteq \omega$ and cofinite $\beta \subseteq \omega$, we define the following non-normal modal logics.
- $\mathsf{GL}_\alpha := \mathsf{GL} + \{F_n : n \in \alpha\}$
- $\mathsf{GL}_\beta^- := \mathsf{GL} + \left\{\neg \bigwedge_{n \in \omega \setminus \beta} F_n\right\}$

In particular, we call $\mathsf{GL}_\omega$ simply $\mathsf{A}$[17].

```
def TBB (n : ℕ) : Formula α := (□^[(n + 1)]⊥) ➝ (□^[n]⊥)

abbrev LogicGLAlpha {α} (X : Set ℕ) : Logic α := (@LogicGL α) +ᴸ ↑(X.image $ TBB (α := Empty))

abbrev LogicA {α} : Logic α := LogicGLAlpha Set.univ

noncomputable abbrev TBBMinus [DecidableEq α] (X : Set ℕ) (X_finite : X.Finite) : Formula α :=
  ∼⋀(X_finite.toFinset.image TBB)

abbrev LogicGLBetaMinus {α} [DecidableEq α] (X : Set ℕ) (X_cofinite : Xᶜ.Finite) : Logic α :=
  (@LogicGL α) +ᴸ (LetterlessFormulaSet.lift { TBBMinus _ X_cofinite })
```

Note: Since $F_n$ is hard to use as an identifier, the mechanization names $F_n$ as `TBB` (axiom T for Box Bot).

Source: 1 2 3 4

[17] We follow the naming of [70]; it presumably stands for Artemov.

The most essential lemmas in the proof of the classification theorem are the following two facts. One is proved by an arithmetical argument, and the other by a Kripke-semantical argument.

**Lemma 3.29 ([10:Corollary 55, Corollary 58])** Let $L$ be a provability logic with $\mathrm{tr}(L) = \omega$.
1. It is impossible that $\mathsf{A} \subsetneq L \subsetneq \mathsf{D}$.
2. It is impossible that $\mathsf{D} \subsetneq L \subsetneq \mathsf{S}$.

```
theorem no_logic_between_LogicA_LogicD :
  letI L : Logic α := T.provabilityLogicRelativeTo U;
  L.trace = Set.univ → ¬((LogicA ⊂ L) ∧ (L ⊂ LogicD))

theorem no_logic_between_LogicD_LogicS :
  letI L : Logic α := T.provabilityLogicRelativeTo U;
  L.trace = Set.univ → ¬((LogicD ⊂ L) ∧ (L ⊂ LogicS))
```

Source: 1 2

The classification theorem is stated as follows.

**Theorem 3.30 (Classification theorem of provability logics [10:Theorem 40, 19])** Let $L = \mathrm{PL}_T(U)$, where $T$ is a $\Delta_1$-definable arithmetic theory extending $\mathsf{I}\Sigma_1$ and $U$ is an arbitrary arithmetic theory. Then $L$ is classified as follows:
1. If $\mathrm{tr}(L)$ is coinfinite, then $L = \mathsf{GL}_{\mathrm{tr}(L)}$.
2. If $\mathrm{tr}(L)$ is cofinite and $L \nsubseteq \mathsf{S}$, then $L = \mathsf{GL}^-_{\mathrm{tr}(L)}$.
3. If $\mathrm{tr}(L)$ is cofinite and $L \subseteq \mathsf{S}$, then $L$ is one of $\mathsf{GL}_{\mathrm{tr}(L)}$, $\mathsf{D} \cap \mathsf{GL}^-_{\mathrm{tr}(L)}$, and $\mathsf{S} \cap \mathsf{GL}^-_{\mathrm{tr}(L)}$.

```
theorem classification_provability_logics [DecidableEq α] :
  letI L : Logic α := T.provabilityLogicRelativeTo U;
  if h_coinfinite : L.traceᶜ.Infinite then
    L = LogicGLAlpha L.trace
  else
    haveI h_cofinite : L.traceᶜ.Finite := Set.not_infinite.mp h_coinfinite;
    if ¬(L ⊆ LogicS) then
      L = LogicGLBetaMinus L.trace h_cofinite
    else
      L = LogicGLAlpha L.trace ∨
      L = LogicD ∩ LogicGLBetaMinus L.trace h_cofinite ∨
      L = LogicS ∩ LogicGLBetaMinus L.trace h_cofinite
```

Source: 1

Furthermore, [19] also proved the classification theorem for the *truth provability logics*, i.e., the logics of the form $\mathrm{PL}_T(\mathsf{TA})$. We have mechanized this fact as well.

**Theorem 3.31 (Classification theorem of truth provability logics [10:Corollary 41, 19])** Let $L = \mathrm{PL}_T(\mathsf{TA})$. Then exactly one of the following four cases holds.
1. $T$ is sound, and $L = \mathsf{S}$.
2. $T$ is $\Sigma_1$-sound but not sound, and $L = \mathsf{D}$.
3. $T$ is not $\Sigma_1$-sound, $\mathrm{hgt}(T) = \omega$, and $L = \mathsf{A}$.
4. $\mathrm{hgt}(T) = n < \omega$ for some $n$, and $L = \mathsf{GL}^-_{\omega \setminus \{n\}}$.

In particular, $L$ is determined by the properties of $T$ listed above.

```
theorem classification_provabilityLogic_TA [DecidableEq α] [Nonempty α] :
  letI L : Logic α := T.provabilityLogicRelativeTo TA;
  (ℕ↓[ℒₒᵣ] ⊧* T ∧ L = LogicS) ∨
  (T.SoundOnHierarchy Σ 1 ∧ ¬(ℕ↓[ℒₒᵣ] ⊧* T) ∧ L = LogicD) ∨
```

```
(¬(T.SoundOnHierarchy Σ 1) ∧ T.height = (⊤ : ℕ∞) ∧ L = LogicA) ∨
∃ n : ℕ, T.height = n ∧ L = LogicGLBetaMinus {n}ᶜ (by simp)
```

SOURCE: 1

### 3.6 On some remaining `sorry` s

Although the mechanization of the classification theorem itself does not depend on them, the mechanizations of some facts of provability logic still contain `sorry` s. We note them here.

The first is the statement that D is indeed a provability logic, which is currently not `sorry` -free.

**Theorem 3.32 ([10:Example 60, 68])** Let $T$ be $\Sigma_1$-sound. Then $\mathsf{D} = \mathrm{PL}_T\big(T + \mathrm{Rfn}_{\Sigma_1}(T)\big)$, where $\mathrm{Rfn}_{\Sigma_1}(T)$ is the (local) reflection principle for $\Sigma_1$ formulas of $T$.

This is because the following fact has not been mechanized in our development. Proving it requires arguments involving partial truth definitions, which we have not yet completed.

**Theorem 3.33 (Unboundedness [10:Theorem 23, 82])** For $T$ as above, $\mathrm{Rfn}_{\Sigma_n}(T)$ is not provable in any consistent r.e. extension of $T$ by $\Pi_n$ sentences.

The other is the uniform arithmetical completeness theorem.

**Theorem 3.34 (Uniform arithmetical completeness theorem)** For every $\Sigma_1$-sound theory $T$, there exists a uniform realization $f$ such that for every formula $A$, $\mathsf{GL} \vdash A$ if and only if $T \vdash f_{\mathfrak{B}_T}(A)$.

### 3.7 On Grz

The Grzegorczyk logic Grz is also closely related to GL. Unlike GL, it is an extension of S4, so that $\Box$ behaves reflexively; nevertheless, as we describe below, it is tightly connected to GL and S through the boxdot translation, and this connection yields an arithmetical completeness theorem for Grz with respect to a *strong* arithmetical interpretation. We also mention that Grz has been mechanized in HOL Light by Bilotta's HOLMS project [24]. For instance, what they call the Kuznetsov–Goldblatt–Boolos theorem [24:Theorem 2] is mechanized in our development as Theorem 3.39.

We first introduce the Hilbert-style proof system, which is the usual definition of Grz.

**Definition 3.35** The Hilbert-style proof system $\mathcal{H}_{\mathsf{Grz}}$ for Grz is obtained from $\mathcal{H}_{\mathsf{GL}}$ by replacing the axiom Löb with the following two axioms.

1. Axiom T: $\Box A \to A$
2. Axiom Grz: $\Box(\Box(A \to \Box A) \to A) \to A$

As for GL, we define the logic $\mathsf{Grz} := \{A \mid \mathcal{H}_{\mathsf{Grz}} \vdash A\}$.

```
inductive LogicGrz.ProofHilbert : Formula α → Type u
| ...
| modalK   {A B} : ProofHilbert $ □(A ➝ B) ➝ (□A ➝ □B)
| modal4   {A}   : ProofHilbert $ □A ➝ □□A
| modalT   {A}   : ProofHilbert $ □A ➝ A
| modalGrz {A}   : ProofHilbert $ □(□(A ➝ □A) ➝ A) ➝ A
| mdp      {A B} : ProofHilbert (A ➝ B) → ProofHilbert A → ProofHilbert B
| nec      {A}   : ProofHilbert A → ProofHilbert (□A)
notation:50 "⊢ʰ[Grz]! " A:51 => LogicGrz.ProofHilbert A

abbrev LogicGrz.ProvableHilbert (A : Formula α) := Nonempty (⊢ʰ[Grz]! A)
notation:50 "⊢ʰ[Grz] " A:51 => LogicGrz.ProvableHilbert A

abbrev LogicGrz {α} : Logic α := { A | ⊢ʰ[Grz] A }
```

NOTE: As in $\mathcal{H}_{\mathsf{GL}}$, the propositional part is taken axiomatically, and the axiom 4 is adopted as an axiom although it is derivable from T and Grz.

SOURCE: 1 2

Next we introduce the Kripke semantics.

**Definition 3.36 (Grz-model)** Let $R$ be a binary relation on $W$, and let $R^{\neq} = \{(x, y) \mid xRy \text{ and } x \neq y\}$ be the irreflexivization of $R$. $R$ is *weakly converse well-founded* if $R^{\neq}$ is conversely well-founded. If $R$ is transitive, this is equivalent to saying that $R$ admits no infinite ascending chain $x_0 R x_1 R \cdots$ consisting of pairwise distinct points.

Using this notion, we define the following.

- A model is a Grz-*model* if $R$ is reflexive, transitive, and weakly converse well-founded.
- A finite model whose $R$ is reflexive, transitive, and antisymmetric (i.e., a finite partial order) is called a *finite* Grz-*model*.

We note that every finite Grz-model is a Grz-model.

```
def Rel.IrreflGen (r : Rel α α) : Rel α α := fun x y ⇒ r x y ∧ x ≠ y

abbrev WeaklyConverseWellFounded {α} (rel : Rel α α) := ConverseWellFounded rel.IrreflGen

class IsWeaklyConverseWellFounded (α) (rel : Rel α α) : Prop where
  wcwf : WeaklyConverseWellFounded rel

class Model.IsGrz (M : Model κ α) extends
  Std.Refl M.Rel, IsTrans _ M.Rel, IsWeaklyConverseWellFounded _ M.Rel

class Model.IsFiniteGrz (M : Model κ α) extends
    Std.Refl M.Rel, IsTrans _ M.Rel, Std.Antisymm M.Rel where
  [finite : Finite M.World]

instance [M.IsFiniteGrz] : M.IsGrz
```

SOURCE: 1 2

Finally we introduce the sequent calculus. Sequent calculi for Grz were formulated by Avron [14] and by Borga and Gentilini [29]; the former gives a semantic cut elimination, the latter a syntactic one. Non-wellfounded proof systems for Grz are studied by Savateev and Shamkanov [126].

**Definition 3.37** The sequent calculus $\mathcal{G}_{\mathsf{Grz}}$ for Grz is obtained from $\mathcal{G}_{\mathsf{GL}}$ by replacing the rule $(\Box_{\mathsf{GL}})$ with the following two rules.

$$\frac{B, \Gamma \Rightarrow \Delta}{\Box B, \Gamma \Rightarrow \Delta}\,(\Box\mathrm{T}) \qquad \frac{\Box(A \to \Box A), \Box\Gamma \Rightarrow A}{\Box\Gamma \Rightarrow \Box A}\,(\Box_{\mathsf{Grz}})$$

As for $\mathcal{G}_{\mathsf{GL}}$, this system contains no cut rule, and $\mathcal{G}_{\mathsf{Grz}} + (\mathrm{Cut})$ denotes the system extended with the cut rule.

```
inductive LogicGrz.ProofGentzen : Sequent α → Type u
| ...
| boxT   {Γ Δ : FormulaFinset α} {B} :
    ProofGentzen (insert B Γ ⇒ Δ) → ProofGentzen (insert (□B) Γ ⇒ Δ)
| boxGrz {Γ : FormulaFinset α} {A}   :
    ProofGentzen (insert (□(A → □A)) (□Γ) ⇒ {A}) → ProofGentzen (□Γ ⇒ {□A})
notation:120 "⊢ᵍ[Grz]! " S:121 ⇒ LogicGrz.ProofGentzen S

abbrev LogicGrz.ProvableGentzen (S : Sequent α) : Prop := Nonempty (⊢ᵍ[Grz]! S)
notation:120 "⊢ᵍ[Grz] " S:121 ⇒ LogicGrz.ProvableGentzen S

inductive LogicGrz.GentzenWithCutProof : Sequent α → Type u
| ...
```

```
  | cut {Γ₁ Γ₂ Δ₁ Δ₂ A} :
      GentzenWithCutProof (Γ₁ ⇒ insert A Δ₁) → GentzenWithCutProof (insert A Γ₂ ⇒ Δ₂) →
      GentzenWithCutProof (Γ₁ ∪ Γ₂ ⇒ Δ₁ ∪ Δ₂)
notation:120 "⊢ᵍᶜ[Grz]! " S:121 => LogicGrz.GentzenWithCutProof S

abbrev LogicGrz.GentzenWithCutProvable (S : Sequent α) : Prop := Nonempty (⊢ᵍᶜ[Grz]! S)
notation:120 "⊢ᵍᶜ[Grz] " S:121 => LogicGrz.GentzenWithCutProvable S
```

NOTE: In [14], the rule $(\Box_{\mathsf{Grz}})$ carries arbitrary side formulas. As with $(\Box_{\mathsf{GL}})$, we adopt the more economical presentation in which the conclusion is exactly $\Box\Gamma \Rightarrow \Box A$, and recover the side formulas afterwards by the weakening rules.

SOURCE: 1

We mechanized the finite model property of Grz with respect to the Kripke semantics, and as its corollaries we mechanized the cut elimination for $\mathcal{G}_{\mathsf{Grz}}$ and its equivalence with the Hilbert-style system.

**Theorem 3.38 (Characterization of Grz)** The following are equivalent.

1. $\mathsf{Grz} \vdash A$.
2. $\mathcal{H}_{\mathsf{Grz}} \vdash A$.
3. $\mathcal{G}_{\mathsf{Grz}} \vdash \Rightarrow A$.
4. $\mathcal{G}_{\mathsf{Grz}} + (\mathrm{Cut}) \vdash \Rightarrow A$.
5. $A$ is forced at every point of every finite Grz-model.
6. $A$ is forced at the root of every rooted finite Grz-model.

```
theorem LogicGrz.provability_TFAE [DecidableEq α] {A : Formula α} : [
  A ∈ LogicGrz,
  ⊢ʰ[Grz] A,
  ⊢ᵍ[Grz] (∅ ⇒ {A}),
  ⊢ᵍᶜ[Grz] (∅ ⇒ {A}),
  ∀ {κ : Type u}, [Nonempty κ] → ∀ M : Model κ α, [M.IsFiniteGrz] → M ⊨ A,
  ∀ {κ : Type u}, [Nonempty κ] → ∀ M : RootedModel κ α, [M.IsFiniteGrz] → M.root.1 ⊩[_] A
].TFAE
```

SOURCE: 1 2

Furthermore, Grz is related to GL and S through the boxdot translation as follows.

**Theorem 3.39** For every formula $A$, the following hold.
1. $\mathsf{Grz} \vdash A$ if and only if $\mathsf{GL} \vdash A^{\boxdot}$.
2. $\mathsf{Grz} \vdash A$ if and only if $\mathsf{S} \vdash A^{\boxdot}$ (cf. Proposition 3.13).

```
theorem iff_provable_boxdot_GL_provable_Grz : Aᵇ ∈ LogicGL ↔ A ∈ LogicGrz

theorem iff_provable_boxdot_S_provable_Grz : Aᵇ ∈ LogicS ↔ A ∈ LogicGrz
```

SOURCE: 1

Using this fact, Goldblatt [49] and Boolos [26] showed that Grz is arithmetically complete with respect to the *strong* arithmetical interpretation, in which $\Box$ is read as “provable and true” rather than merely “provable”.

**Definition 3.40 (Strong interpretation)** Given a realization $f$ and a provability $\mathfrak{B}$, the *strong (arithmetical) interpretation* $f^{\mathsf{s}}_{\mathfrak{B}}(A)$ is defined exactly as the interpretation $f_{\mathfrak{B}}(A)$ of Definition 3.20 except for the modal clause, which reads

$$f^{\mathsf{s}}_{\mathfrak{B}}(\Box A) = f^{\mathsf{s}}_{\mathfrak{B}}(A) \land \mathfrak{B}(f^{\mathsf{s}}_{\mathfrak{B}}(A)).$$

Then $T \vdash f^{\mathrm{s}}_{\mathfrak{B}}(A)$ if and only if $T \vdash f_{\mathfrak{B}}(A^{\boxdot})$ (and similarly for truth in a model of $T$ on which $\mathfrak{B}$ is sound), and this is how the arithmetical completeness of Grz is reduced to that of GL and S.

```
def Formula.strongInterpret (f : Realization α L) {T₀ T : FirstOrder.Theory L}
  (𝔅 : Provability T₀ T) : Formula α → FirstOrder.Sentence L
  | #a     ⇒ f.val a
  | ⊥      ⇒ ⊥
  | A ➝ B ⇒ (A.strongInterpret f 𝔅) ➝ (B.strongInterpret f 𝔅)
  | □A     ⇒ (A.strongInterpret f 𝔅) ⋏ 𝔅 (A.strongInterpret f 𝔅)

lemma Formula.iff_interpret_boxdot_strongInterpret [𝔅.HBL2] :
  T ⊢ (Aᵇ).interpret f 𝔅 ⭤ T ⊢ A.strongInterpret f 𝔅

lemma Formula.iff_models_interpret_boxdot_strongInterpret
  {M} [Nonempty M] [Structure L M] [M↓[L] ⊧* T] [𝔅.HBL2] [𝔅.SoundOn M] :
  M↓[L] ⊧ (Aᵇ).interpret f 𝔅 ⭤ M↓[L] ⊧ A.strongInterpret f 𝔅
```

Source: 1

**Theorem 3.41 (Arithmetical completeness of Grz [26, 49])** Let $T$ be a theory with $\mathrm{hgt}(T) = \omega$ (in particular, any $\Sigma_1$-sound $T$). Then $\mathsf{Grz} \vdash A$ if and only if $T \vdash f^{\mathrm{s}}_{\mathfrak{B}_T}(A)$ for every realization $f$. Moreover, if $T$ is sound, then $\mathsf{Grz} \vdash A$ if and only if $\mathbb{N} \models f^{\mathrm{s}}_{\mathfrak{B}_T}(A)$ for every realization $f$.

```
variable {T : FirstOrder.ArithmeticTheory} [T.Δ₁] [𝐈𝚺₁ ⪯ T]

theorem LogicGrz.arithmetical_completeness_iff_of_infinity_height
  (height : T.height = (⊤ : ℕ∞)) [DecidableEq α] :
  A ∈ LogicGrz ⭤ (∀ f : Realization α ℒₒᵣ, T ⊢ A.strongInterpret f T.standardProvability)

theorem LogicGrz.arithmetical_completeness_iff_of_sigma1_sound
  [T.SoundOnHierarchy 𝚺 1] [DecidableEq α] :
  A ∈ LogicGrz ⭤ (∀ f : Realization α ℒₒᵣ, T ⊢ A.strongInterpret f T.standardProvability)

variable [ℕ↓[ℒₒᵣ] ⊧* T]

theorem LogicGrz.arithmetical_completeness_model_iff [DecidableEq α] :
  A ∈ LogicGrz ⭤ (∀ f : Realization α ℒₒᵣ, ℕ↓[ℒₒᵣ] ⊧ A.strongInterpret f T.standardProvability)
```

Source: 1

### 3.8 On GL.3

A sequent calculus for GL.3 was given by Valentini and Solitro [151] and Valentini [150]. In particular, [151] shows that GL.3 enjoys a certain arithmetical completeness with respect to the class of arithmetic sentences called consistency assertions. We briefly describe these results.

**Definition 3.42** GL.3 is the normal modal logic obtained from GL by adding the weak linearity axiom $\Box(\boxdot A \to B) \lor \Box(\boxdot B \to A)$[18].

```
inductive Logic.sumNormal (L₁ L₂ : Logic α) : Logic α
  | mem₁ {A}     : L₁ A → sumNormal L₁ L₂ A
  | mem₂ {A}     : L₂ A → sumNormal L₁ L₂ A
  | mdp  {A B}   : sumNormal L₁ L₂ (A ➝ B) → sumNormal L₁ L₂ A → sumNormal L₁ L₂ B
  | subst {A s}  : sumNormal L₁ L₂ A → sumNormal L₁ L₂ (A⟦s⟧)
  | nec  {A}     : sumNormal L₁ L₂ A → sumNormal L₁ L₂ (□A)
infix:50 " ⊕ᴸ " ⇒ Logic.sumNormal
```

[18] In older literature, it was also written as GLLin [150, 151] or K4.3W [127].

```
abbrev LogicGLPoint3 {α} : Logic α := LogicGL ⊕ᴸ { (□((⊡A) ➝ B)) ⋎ (□((⊡B) ➝ A)) | (A) (B) }
```

SOURCE: 1 2

**Definition 3.43** A finite GL-model is called a *finite* GL.3*-model* when $\prec$ is linear, i.e., $x \prec y$ and $x \prec z$ imply $y \prec z$ or $y = z$ or $z \prec y$.

```
class IsFiniteGLPoint3 (M : Model κ α) extends Model.IsFiniteGL M where
  linear : ∀ {x y z : M.World}, x ≺ y → x ≺ z → y ≺ z ∨ y = z ∨ z ≺ y
```

SOURCE: 1

**Definition 3.44** The sequent calculus $\mathcal{G}_{\mathsf{GL.3}}$ for GL.3 is obtained from the sequent calculus for GL by replacing the $(\Box_{\mathsf{GL}})$ rule with the following rule, where $\Delta \neq \varnothing$ and $\{S_1, \ldots, S_m\} = \mathcal{P}(\Delta) \setminus \{\varnothing\}$ (hence $m = 2^{|\Delta|} - 1$).

$$\frac{\Gamma, \Box\Gamma, \Box S_1 \Rightarrow S_1, \Box(\Delta \setminus S_1) \quad \cdots \quad \Gamma, \Box\Gamma, \Box S_m \Rightarrow S_m, \Box(\Delta \setminus S_m)}{\Box\Gamma \Rightarrow \Box\Delta}\,(\Box_{\mathsf{GL.3}})$$

Note that the case $\Delta = \{A\}$ is exactly the $(\Box_{\mathsf{GL}})$ rule.

```
inductive LogicGLPoint3.ProofGentzen : Sequent α → Type u
| axm (A) : ProofGentzen ({A} ⇒ {A})
| botL : ProofGentzen ({⊥} ⇒ ∅)
| wkL  {Γ Γ' Δ}  : ProofGentzen (Γ ⇒ Δ) → Γ ⊆ Γ' → ProofGentzen (Γ' ⇒ Δ)
| wkR  {Γ Δ Δ'}  : ProofGentzen (Γ ⇒ Δ) → Δ ⊆ Δ' → ProofGentzen (Γ ⇒ Δ')
| impL {Γ Δ A B} : ProofGentzen (Γ ⇒ (insert A Δ)) → ProofGentzen (insert B Γ ⇒ Δ) →
                   ProofGentzen ((insert (A ➝ B) Γ) ⇒ Δ)
| impR {Γ Δ A B} : ProofGentzen ((insert A Γ) ⇒ (insert B Δ)) →
                   ProofGentzen (Γ ⇒ (insert (A ➝ B) Δ))
| boxGLPoint3 {Γ Δ} (hΔ : Δ.Nonempty) :
    (∀ S : FormulaFinset α, S ⊆ Δ → S.Nonempty →
      ProofGentzen ((Γ.box ∪ Γ ∪ S.box) ⇒ (S ∪ (Δ \ S).box))) →
    ProofGentzen (Γ.box ⇒ Δ.box)
notation:120 "⊢ᵍ[GLPoint3]! " S:121 => LogicGLPoint3.ProofGentzen S

abbrev LogicGLPoint3.ProvableGentzen (S : Sequent α) : Prop :=
  Nonempty (⊢ᵍ[GLPoint3]! S)
notation:120 "⊢ᵍ[GLPoint3] " S:121 => LogicGLPoint3.ProvableGentzen S
```

SOURCE: 1

For these characterizations, equivalences analogous to those for GL hold.

**Theorem 3.45 ([151])** The following are equivalent.

1. $\mathsf{GL.3} \vdash A$.
2. $\mathcal{G}_{\mathsf{GL.3}} \vdash\Rightarrow A$.
3. $A$ is forced at every point of every finite GL.3-model.
4. $A$ is forced at the root of every rooted finite GL.3-model.
5. For every $n \geq 1$, $A$ is forced at every point of every finite GL.3-model whose set of points is $\{0, 1, \cdots, n-1\}$.
6. For every $n \geq 1$, $A$ is forced at the root of every rooted finite GL.3-model whose set of points is $\{0, 1, \cdots, n-1\}$.

```
theorem LogicGLPoint3.provability_TFAE [DecidableEq α] {A : Formula α} : [
  A ∈ LogicGLPoint3,
  ⊢ᵍ[GLPoint3] (∅ ⇒ {A}),
  ∀ {κ : Type u}, [Nonempty κ] → ∀ M : Model κ α, [M.IsFiniteGLPoint3] → M ⊧ A,
```

```
  ∀ {κ : Type u}, [Nonempty κ] → ∀ M : RootedModel κ α, [M.IsFiniteGLPoint3] → M.root.1 ⊩[_] A,
  ∀ (n : ℕ) [NeZero n] (M : Model (Fin n) α), [M.IsFiniteGLPoint3] → M ⊧ A,
  ∀ (n : ℕ) [NeZero n] (M : RootedModel (Fin n) α), [M.IsFiniteGLPoint3] → M.root.1 ⊩[_] A
].TFAE
```

Source: 1

In particular, on closed formulas GL.3 and GL do not differ; this can be shown using the trace for closed formulas due to [9] and the sequent calculus.

**Theorem 3.46 ([151:Theorem 2])** For a closed formula $A$, $\mathsf{GL.3} \vdash A$ if and only if $\mathsf{GL} \vdash A$.

```
theorem LogicGLPoint3.eq_LogicGL_on_letterless : @LogicGLPoint3 Empty = @LogicGL Empty
```

Source: 1

Finally, we state the arithmetical completeness of GL.3 with respect to consistency assertions.

**Definition 3.47**
- A sentence $\sigma$ is a *consistency assertion* if it is generated from $\neg\mathfrak{B}\bot$ and $\mathfrak{B}\bot$ by closing under $\mathfrak{B}$, $\neg$, $\wedge$, $\vee$, and $\to$.
- A realization $f$ is a *consistency realization* if $f$ sends every propositional variable to a consistency assertion.

```
inductive Provability.IsConsistencyAssertion (𝔅 : Provability T₀ T) : FirstOrder.Sentence L → Prop
  | con       : IsConsistencyAssertion 𝔅 (∼(𝔅 ⊥))
  | incon     : IsConsistencyAssertion 𝔅 (𝔅 ⊥)
  | prov {σ}  : IsConsistencyAssertion 𝔅 σ → IsConsistencyAssertion 𝔅 (𝔅 σ)
  | neg {σ}   : IsConsistencyAssertion 𝔅 σ → IsConsistencyAssertion 𝔅 (∼σ)
  | and {σ τ} : IsConsistencyAssertion 𝔅 σ → IsConsistencyAssertion 𝔅 τ → IsConsistencyAssertion 𝔅
(σ ⋏ τ)
  | or {σ τ}  : IsConsistencyAssertion 𝔅 σ → IsConsistencyAssertion 𝔅 τ → IsConsistencyAssertion 𝔅
(σ ⋎ τ)
  | imp {σ τ} : IsConsistencyAssertion 𝔅 σ → IsConsistencyAssertion 𝔅 τ → IsConsistencyAssertion 𝔅
(σ ➝ τ)

def Realization.IsConsistencyRealization (f : Realization α L) (𝔅 : Provability T₀ T) : Prop :=
  ∀ a, 𝔅.IsConsistencyAssertion (f.val a)

abbrev ConsistencyRealization (α : Type*) (𝔅 : Provability T₀ T) :=
  {f : Realization α L // f.IsConsistencyRealization 𝔅}

instance {𝔅 : Provability T₀ T} :
  CoeFun (ConsistencyRealization α 𝔅) (fun _ ↦ Formula α → FirstOrder.Sentence L) :=
  ⟨fun f ↦ Formula.interpret f.1 𝔅⟩

abbrev StandardConsistencyRealization (α : Type*) (T : FirstOrder.ArithmeticTheory) [T.Δ₁] :=
  ConsistencyRealization α T.standardProvability
```

Source: 1

**Theorem 3.48 ([151:Theorem 1])** $\mathsf{GL.3} \vdash A$ if and only if $\mathsf{PA} \vdash f_{\mathfrak{B}_{\mathsf{PA}}}(A)$ for every consistency realization $f$ over PA.

```
theorem LogicGLPoint3.arithmetical_completeness_iff_peano_arithmetic [DecidableEq α] :
  A ∈ LogicGLPoint3 ↔ ∀ f : StandardConsistencyRealization α PA, PA ⊢ f A
```

Source: 1

## 4 Related work and future work

Finally, in this section, we mention some prior work related to our mechanization, that is, mechanizations of the incompleteness theorems and of facts concerning provability logic in proof assistants. Moreover, on that basis, we indicate several directions in which we plan to proceed. For facts that we have not mechanized, see also Section 2.6 and Section 3.6. Concerning provability logic, there is much prior work on mechanizations in the broader area of modal logic in general (e.g., tense logic and epistemic logic), but since these are outside the scope of the present report, we omit them.

### 4.1 Further metamathematical topics

Our mechanization of the incompleteness theorems is an achievement, but it is a start rather than a goal. Section 2.6 collects several further results, but many metamathematical facts about arithmetic and the incompleteness theorems remain unmechanized. For example, there are many important tools for the metamathematical analysis of arithmetic, such as reflection principles, partial truth definitions, arguments about nonstandard models of arithmetic, and the arithmetized completeness theorem. Our development does not contain these tools at present. Without them, we cannot prove facts such as Ryll-Nardzewski's theorem [120], which states that $\mathsf{PA}$ is not finitely axiomatizable. We also cannot fill some of the `sorry` s that we left in Section 3.6. For these topics, we plan to mechanize the arguments of the standard textbooks [57, 89].

Proof-theoretic analysis is another direction. As for prior work, Hydras & Co. [34, 35] is a Rocq mechanization of the termination (in Rocq) of the hydra game [77], and of related arguments about the ordinals that proof theory frequently uses. We have mechanized almost no proof-theoretic result so far, so we plan to work on proof-theoretic analysis and ordinal analysis in the future (see also Section 5 for an experiment on autoformalization in this direction).

These tools are also necessary for the polymodal provability logics of Section 4.7, because they give the arithmetical meaning of these logics.

### 4.2 Interpretability

We mechanized the incompleteness theorems above in arithmetic, that is, in theories of the language $\mathcal{L}_{\mathrm{OR}}$. They depend on the choice of the language and on the details of the coding. Thus, even if we mechanize set theory in our framework, we cannot conclude the incompleteness theorems for it immediately. The mechanization of interpretability is important for this problem. Let $T$ be a theory of the language $\mathcal{L}_T$, and let $U$ be a theory of the language $\mathcal{L}_U$. Roughly speaking, $T$ is interpretable in $U$ if there is a suitable translation $t$ such that $T \vdash \varphi \Longrightarrow U \vdash t(\varphi)$ for every $\mathcal{L}_T$-sentence $\varphi$; we write $U \rhd T$. Interpretability is a tool for the comparison of theories: if $U \rhd T$ and $T$ is essentially undecidable, then $U$ is also essentially undecidable (cf. [147]). See, e.g., Lindström [89:Section 4] for a further discussion. As far as we know, no prior work mechanized interpretability itself in a proof assistant.

Section 4.4 discusses the set theories $\mathsf{ZF}$ and $\mathsf{ZFC}$. If we mechanize the fact that $\mathsf{ZF} \rhd \mathsf{PA}$, we can also mechanize the incompleteness theorems for these set theories. Then we do not have to repeat inside set theory the arguments that we carried out for arithmetic, and we expect that this skips a large part of the proof. In another direction, we can consider other theories, because the analysis of the incompleteness phenomena is not restricted to arithmetic. The theory of concatenation $\mathsf{TC}$ is a first-order theory that directly axiomatizes the concatenation of strings, and Grzegorczyk initiated its study [55, 56]. In particular, $\mathsf{TC} \rhd \mathsf{Q}$ holds [42, 143, 144, 158]. Such a minimal system can be easier to mechanize than arithmetic itself.

Interpretability logic develops provability logic further and treats interpretability itself as a modality. We discuss it in Section 4.7.

### 4.3 Intuitionistic first-order logic and arithmetic

Intuitionistic logic is classical logic without the law of excluded middle, and the corresponding predicate logic is intuitionistic first-order logic $\mathsf{IQL}$. $\mathsf{IQL}$ satisfies several constructive principles. It has the disjunction property: if $\mathsf{IQL} \vdash \varphi \vee \psi$, then $\mathsf{IQL} \vdash \varphi$ or $\mathsf{IQL} \vdash \psi$. It also has the existence property: if $\mathsf{IQL} \vdash \exists x\, \varphi(x)$, then there is a term $t$, possibly containing free variables, such that $\mathsf{IQL} \vdash \varphi(t)$. Intuitionistic predicate logic also has connections to other fields, for example to dependent type theory via the Curry–Howard correspondence.

At present, our mechanization of intuitionistic predicate logic is not far advanced: it contains the syntax, a Hilbert-style deduction system, and Kripke semantics. Nevertheless, we have already used it to mechanize the cut-elimination theorem for classical first-order logic semantically, following Avigad [11]: a classical derivation is translated by the Gödel–Gentzen negative translation into a derivation in minimal logic, and the soundness of minimal logic with respect to a Kripke-style forcing relation yields a cut-free classical derivation[19]. Cut elimination for the intuitionistic sequent calculus itself has not been mechanized yet; we plan to prove it semantically in the same way, by developing the Kripke semantics of intuitionistic predicate logic further. As prior work, Herbelin and Lee [63][20] already mechanized cut elimination for intuitionistic logic in Rocq.

Forster, Kirst, Wehr, and their colleagues carried out a series of mechanizations in Rocq [40, 79][21]. This prior work has a wider scope than ours. Their design is notable: they take intuitionistic logic as the base, and they obtain classical logic as the extension by Peirce's law, controlled by a flag. They give Tarski, Kripke, algebraic, and game semantics, and for each one they analyze which non-constructive principles the completeness theorem requires in the constructive type theory of Rocq. Our implementation is specific to classical logic: it defines dual connectives as primitives, and it uses a Tait calculus.

Heyting arithmetic $\mathsf{HA}$ is intuitionistic logic together with the axioms of Peano arithmetic. The library of Forster et al. discusses $\mathsf{Q}$ and $\mathsf{PA}$ over intuitionistic natural deduction, so that provability in the latter is exactly provability in $\mathsf{HA}$. Kirst and Hermes [78] proved that these systems are undecidable, through a reduction from Hilbert's tenth problem (the MRDP theorem), and that every axiomatization that is sound in the standard model is incomplete. The same authors also analyzed, in the same setting, Tennenbaum's theorem, which states that no nonstandard model of $\mathsf{PA}$ has computable addition and multiplication [64]. The same library mechanizes the Friedman translation, which transforms a proof in classical logic into a proof in minimal logic, and thus shows that $\mathsf{Q}$ and $\mathsf{PA}$ over minimal or intuitionistic logic are also undecidable. Our framework already covers the classical side, so the mechanization of such translations is a practical route to $\mathsf{HA}$.

The exact axiomatization of the provability logic of Heyting arithmetic has remained a difficult open problem for a long time. We mention it again in Section 4.6.

### 4.4 Set theory and forcing

One of our current goals is to mechanize a general framework for forcing and to establish foundational results such as the independence of the continuum hypothesis. Han and van Doorn [59] have already mechanized the latter result, but their approach is based on Boolean-valued models and has more limited applicability than forcing.

There are several possible ways to mechanize forcing. Two basic approaches are as follows:

1. The standard textbook model-theoretic approach: As in, for example, Kunen [83], one begins with a countable transitive model $M$ of $\mathsf{ZFC}$ and a forcing poset $\mathbb{P}$ with generic filter $G \subseteq \mathbb{P}$, and constructs a new model by the forcing extension $M[G]$.
2. An approach using proof-theoretic forcing: One constructs a kind of interpretation between theories $T_1$ and $T_2$, called a *forcing interpretation*, and establishes an appropriate conservativity result $T_1 \subseteq_\Gamma T_2$ [12]. For example, let $T_1 := \mathsf{ZFC} + \neg\mathsf{CH}$ (where $\mathsf{CH}$ is the continuum hypothesis), $T_2 := \mathsf{ZFC}$, and $\Gamma := \{\bot\}$. Suppose that one can construct a $\Gamma$-conservative forcing interpretation of $T_1$ in $T_2$. A proof of $\mathsf{ZFC} \vdash \mathsf{CH}$ easily yields a proof of $\mathsf{ZFC} + \neg\mathsf{CH} \vdash \bot$, from which the forcing interpretation would in turn yield $\mathsf{ZFC} \vdash \bot$.

The first approach is the standard choice. A frequently noted drawback is that the existence of a countable transitive model of $\mathsf{ZFC}$ (or of a similar set theory) is strictly stronger than the mere consistency of $\mathsf{ZFC}$. Given the strength of Lean as an ambient formal system, however, this is unlikely to pose a serious problem.

The second approach is attractive in several respects. First, it yields a stronger result than the first approach: as the preceding example illustrates, proving the independence of $\mathsf{CH}$ requires only the consistency of $\mathsf{ZFC}$, with no need for any additional stronger assumption. It also has constructive and finitistic advantages, since the translation of proofs induced by a forcing interpretation is essentially syntactic and finitary, and can moreover be computed by a polynomial-time function. This is a substantively stronger result than mere independence.

Although we have not yet undertaken a mechanization of forcing for set theory, we have already used a highly simplified version of forcing to mechanize the completeness theorem for first-order logic. The idea underlying this proof is due to Avigad [11]. We briefly describe it here, assuming that the language is countable.

Let $\mathbb{P}$ be the set of **LK**-sequents $\Gamma$ such that $\mathbf{LK} \nvdash \neg\Gamma$. Endow $\mathbb{P}$ with the relation inductively defined by the following rules. This relation is a preorder whose greatest element is the empty sequent:

[19] See https://github.com/FormalizedFormalLogic/Foundation/blob/v1/Foundation/FirstOrder/Hauptsatz.lean. The underlying forcing argument is described in Section 4.4.
[20] Implementation: https://formal.hknu.ac.kr/Kripke.
[21] See https://github.com/uds-psl/coq-library-fol.

$$
\begin{gathered}
\Xi \preceq \Xi \\
\varphi, \psi, \Gamma \preceq \Xi \Rightarrow \varphi \wedge \psi, \Gamma \preceq \Xi \\
\varphi(t) \preceq \Xi \Rightarrow \forall x\, \varphi(x), \Gamma \preceq \Xi \\
\Delta \preceq \Xi \text{ and } \Delta \subseteq \Gamma \Rightarrow \Gamma \preceq \Xi
\end{gathered}
$$

If $\mathbf{LK} \vdash \varphi$, then the Gödel–Gentzen translation gives $\mathbf{LJ} \vdash \varphi^{\mathrm{GG}}$. Since Kripke semantics is sound for **LJ**, we have $p \Vdash \varphi^{\mathrm{GG}}$ for every $p \in \mathbb{P}$. Viewing $p \Vdash \varphi^{\mathrm{GG}}$ as a *weak forcing* relation $p \Vdash_{\mathrm{w}} \varphi$ yields a sound Kripke model for **LK**. Moreover, this model is canonical in the following sense: for every formula $\varphi$,

$$\mathbf{LK} \vdash \varphi \quad \text{iff} \quad \forall p \in \mathbb{P}\, (p \Vdash_{\mathrm{w}} \varphi)$$

```
lemma complete {φ : Proposition L} : ℙ⁻ ∀⊩ᶜ φ ↔ LK¹ ⊢ φ
```

SOURCE: 1

Now suppose that $\mathbf{LK} \nvdash \neg\sigma$. Since $p := \{\sigma\} \in \mathbb{P}$, we can construct a filter $G \subseteq \mathbb{P}$ that contains $p$ and is generic with respect to the following two countable families of dense sets:

$$
\begin{aligned}
\mathcal{D}_{\varphi} &:= \{p \in \mathbb{P} \mid p \Vdash_{\mathrm{w}} \varphi \vee p \Vdash_{\mathrm{w}} \neg\varphi\} \\
\mathcal{H}_{\psi} &:= \{p \in \mathbb{P} \mid \forall q \preceq p\, (q \Vdash_{\mathrm{w}} \exists x\, \psi(x) \Longrightarrow \exists t : \text{term } \ q \Vdash_{\mathrm{w}} \psi(t))\}
\end{aligned}
$$

Let term model $\mathfrak{T}$ be a model by defining $\mathfrak{T} \vDash \alpha :\Leftrightarrow \exists p \in G\, (p \Vdash_{\mathrm{w}} \alpha)$ for atomic formulas $\alpha$, then the forcing lemma can be proved:

$$\mathfrak{T} \vDash \varphi \quad \text{iff} \quad \exists p \in G\, (p \Vdash_{\mathrm{w}} \varphi)$$

```
def GenericForces (p : ℙ⁻) (φ : Proposition K) : Prop := ∃ q ∈ genericFilter p, q ⊩ᶜ φ

local infix: 60 " ⊫ " ⇒ GenericForces

lemma forcing_lemma (φ : Semiformula K ξ n) {fv : ξ → 𝕿} {bv : Fin n → 𝕿} :
    φ.Eval (s := termModelOf p) bv fv ↔ p ⊫ Rew.bind bv fv ▹ φ
```

SOURCE: 1

Since $G$ contains $\{\sigma\}$ and $\{\sigma\} \Vdash_{\mathrm{w}} \sigma$, it follows that $\mathfrak{T} \vDash \sigma$. This proves the completeness theorem for **LK**.

```
lemma satisfiable_of_irrefutable (σ : Sentence L) (h : LK¹ ⊬ ∼(σ : Proposition L)) :
    Satisfiable {σ}
```

SOURCE: 1

To develop forcing interpretations for set theory, an argument of the kind just described must be carried out *internally* to the set theory. As discussed in the section of incompleteness theorems (Section 2.2), a direct syntactic treatment is likely to be too complex.

The same remedy may be applicable here: one can instead proceed model-theoretically via the completeness theorem. This may make it possible to reuse the externally defined weak forcing relation $\Vdash_{\mathrm{w}}$ and the general theory of Kripke models. Moreover, such an external argument may be technically close to the forcing arguments ordinarily employed by set theorists.

### 4.5 Proof theory of provability logics

The proof theory of GL has been studied extensively. First, Gentzen-style sequent calculi have been investigated in numerous works [14, 28, 30, 52, 87, 99, 121, 122, 125, 149]. In particular, as a syntactic issue, whether the termination of the cut-elimination algorithm holds for sequent calculi based on multisets had long been a matter of debate, and the issue is considered to have been resolved by [52]. On the other hand, Brighton [30] gave an alternative proof of the termination of the cut-elimination algorithm using the technique called *regression trees*, and this argument has been mechanized in Rocq by Goré, Ramanayake, and Shillito [53]. Furthermore, Férée et al. [38] mechanized in Rocq the uniform interpolation theorem [21] for GL via sequent calculi. In particular, although their proof is based on Bílková [21], we mention that in the course of the mechanization they discovered an error in [21] and were able to correct it[22]. These mechanizations can be regarded as significant results in that they settled a debate over ambiguous pen-and-paper arguments by strict computer verification.

In addition, there are also many approaches to non-Gentzen-style proof systems for GL, i.e., systems obtained by adding further machinery to ordinary sequent calculi: e.g., the *labelled sequent calculi* by Negri [104, 105], the *tree-hypersequent calculus* by Poggiolesi [113], the *nested sequent calculi* by Maniwa and Kashima [97], and the *non-wellfounded proofs* (or *circular proofs*) by Shamkanov [129][23]. For discussions on the equivalence of the provability of several of these sequent systems, including the Gentzen-style ones, see Goré and Ramanayake [51] and Lyon [92]. In particular, Shamkanov's non-wellfounded proofs have the advantage that the Lyndon interpolation theorem can be proved syntactically [129:Chapter 4][24]. As far as we know, the only mechanizations of the proof theory of sequent calculi equipped with such additional machinery are the mechanization of the labelled sequent calculus in HOL Light by Maggesi and Perini Brogi [95, 96] and, along the same line, Bilotta's HOLMS project [22–24], and the recent mechanization in Lean by Gignoux [45] of non-wellfounded proof systems for GL, formulated coalgebraically, through which the CIP of GL is proved.

Tableau methods for GL are discussed in [27:Chapter 10] for instance. A tableau-based automated theorem prover for GL was implemented by Goré and Kelly [50], where the efficiency of the implementation is also discussed.

The proof theory of S and D has been developed only recently. Sierra Miranda and Studer [138] proved the Lyndon interpolation property of S using non-wellfounded proofs. As a different approach, Kushida [86] proposed a sequent calculus for S with two levels of sequents, and Kashima et al. [75, 76] developed this approach further, obtaining a sequent calculus for D with three levels of sequents; these are the systems we have mechanized (see Definition 3.9).

In the present work, we have mechanized the Gentzen-style sequent calculus for GL, the labelled sequent calculus for GL (see Section 3.3), and the two-level sequent calculus for S together with the three-level sequent calculus for D (see Definition 3.9). For future work, we plan to mechanize sequent calculi with other machinery as well, together with the equivalence of their provability. In particular, although Shamkanov's circular proofs involve infinitary structures, the studies by Sierra Miranda et al. [66, 137–139] have revealed that they have many applications, so their mechanization seems to be a technically challenging but worthwhile task. Gignoux's coalgebraic mechanization of non-wellfounded proof systems for GL [45] mentioned above can be regarded as a first step in this direction.

### 4.6 Provability logic of Heyting arithmetic

The provability logic of intuitionistic or constructive arithmetic, in particular, Heyting arithmetic HA, has been a subject of study for a long time (see [10:Section 9, 20:Section 4]). Even among the recent developments alone, there is prior work such as [7, 8, 100, 101, 136].

Here, we define iK and iGL. Intuitionistic modal logic iK is obtained from intuitionistic propositional logic by adding the axiom K for $\Box$ and the necessitation rule (note that the language does not contain $\Diamond$), and intuitionistic Gödel–Löb logic iGL is obtained by adding Löb's axiom $\Box(\Box A \to A) \to \Box A$ to iK. It is known that the provability logic of HA contains at least iGL, that is, iGL is arithmetically sound with respect to HA. For purely logical studies of iGL, consult [43, 90, 148]. As for mechanization, Goré and Shillito [54] gave a refined version of the proof of cut elimination for the sequent calculus for iGL due to van der Giessen and Iemhoff [43], and this proof has been mechanized in Rocq (see also [134]).

On the other hand, the logic called the intuitionistic strong Löb logic iSL, obtained by adding the strong Löb axiom $(\Box A \to A) \to A$ to iK, is also important. For a survey of iSL itself as a logic, see, e.g., [161]. Shillito et al. [135] gave a new sequent calculus for iSL admitting cut elimination, and mechanized it in Rocq. Férée et al. [38] mechanized the uniform interpolation theorem for iSL in Rocq (see also Section 4.5).

Finally, the provability logic of Heyting arithmetic has been announced in Mojtahedi's preprint [101]. However, at the time of writing, this preprint is still under review[25]. In the future, we plan to mechanize these arguments, which will make it possible to verify them rigorously and thus to settle this problem in a more reliable way.

[22] Quoted from [38:p.2]: During our work on formalising this proof in Coq, we uncovered an incompleteness in it ([21]), and our formalisation contains a corrected version of the construction of…

[23] Here we mention only the systems for GL. For general discussions of each formalism, we refer the reader to the references of the respective papers.

[24] This fact itself is also proved in [128], but the proof there relies on Kripke-semantical techniques.

[25] The first version was submitted to arXiv in 2022.

### 4.7 Enriched modalities

There are also extensions in the direction of adding further modal operators in order to express various notions related to provability. Here we mention two directions for which mechanizations can be found: polymodal provability logic and interpretability logic.

Japaridze [68, 69] extended the modality of $\mathsf{GL}$ to infinitely many modal operators $[1], [2], \ldots$ together with their duals $\langle 1\rangle, \langle 2\rangle, \ldots$, and introduced the logic $\mathsf{GLP}$. For the meaning of these modal operators, we may consult [10:Chapter 8.3]. $\mathsf{GLP}$ is useful in the proof-theoretic analysis of arithmetic and is moreover decidable, but it is also known to be Kripke incomplete. It is complete with respect to topological semantics, but that semantics has the drawback of being technically hard to work with. It turned out that the strictly positive fragment of $\mathsf{GLP}$ admits a technically much simpler formulation without losing much expressive power, and nowadays such a system is called *reflection calculus* $\mathsf{RC}$ (see [16]). At the time of writing, prior work on the mechanization of reflection calculi has been carried out mainly by Joosten's group. Together with Joosten, de Almeida Borges proposed the *worm calculus* $\mathsf{WC}$ [5], a variable-free subsystem of $\mathsf{RC}$ built up solely from $\top$ and modal operators indexed by ordinals, and mechanized it in Rocq [2]. They further proposed the *quantified reflection calculus with one modality* $\mathsf{QRC}_1$ [6], a system that admits quantifiers while remaining reasonably tractable, and mechanized its soundness and completeness in Rocq [3, 4][26]. On the other hand, Santiago-Fernández et al. [124] formulated a term-rewriting-like system (a tree rewriting system) for derivations of $\mathsf{RC}$, and its mechanization in Rocq appears to be in progress in [123].

As another extension of provability logic, there is the *interpretability logic* proposed by Visser [156]. Interpretability logic is the extension of provability logic with an additional binary modal operator $\rhd$ representing interpretability (informally, $A \rhd B$ means that the extended theory $T + f(A)$ interprets $T + f(B)$; see also Section 4.2). There are several semantics for interpretability logic, including *de Jongh–Veltman semantics* [73] and *Verbrugge semantics*, also known as *generalized Veltman semantics* (cf. [74]). The latter can handle completeness and definability for more axioms, but it has the drawback that the arguments become very involved. As prior work, mechanization of frame definability for Verbrugge semantics has been carried out in Agda by Rovira [119].

As for our own progress, we have mechanized syntactic proofs and frame definability for some additional axioms and weak interpretability logics based on work by Kurahashi and Okawa [85][27]. However, we have not yet established modal completeness with respect to frames, and as for the arithmetical completeness theorem, we have not been able to mechanize it at all.

## 5 Appendix: Mechanizing logics with AI

We describe how the AI is used in our development. Claude does not mechanize everything autonomously: the author first fixes the overall strategy for proving the main theorems and writes their formal statements, and only then delegates the actual proofs to Claude. Within the proofs as well, the author gives appropriate directions and tactics, e.g., to proceed by induction on the structure of formulas or on the rules of a sequent calculus. As a rule of thumb in pure mathematical logic, a fact proved by such an induction requires no special idea: one simply carries out the calculation, and on paper one typically works out a few representative cases and omits the rest. In a mechanization, every case must be treated without omission; we saw little value in a human spending time on such code, so we actively delegated it to the AI. In practice, the overall refactoring of FormalizedFormalLogic/ProvabilityLogic and the mechanization of the classification theorem were mostly completed in about three weeks of actual work, which the second author regards as a substantial gain in speed and efficiency.

Apart from such delegation of proofs, we also experimented with autoformalization by an LLM, in the direction of proof theory. We tried to mechanize the sequent calculus for $\mathsf{PA}$ with the $\omega$-rule, its cut-elimination theorem, and the fact that $\mathsf{PA}$ does not prove the termination of Goodstein sequences [77]. The proofs in the generated code[28] contain no `sorry` and no additional axiom. However, a human check of its definitions and statements is still in progress, and we therefore do not count this experiment among the results reported in this paper.

[26] Only the mechanization of soundness direction is reported in [3], but as far as we can tell from de Almeida Borges' doctoral thesis [4], completeness and decidability have been mechanized since then.
[27] See: https://github.com/FormalizedFormalLogic/InterpretabilityLogic
[28] For more details, see https://github.com/FormalizedFormalLogic/goodstein-independence.